\documentclass[12pt]{article}
\usepackage{amssymb}
\usepackage{graphics}
\usepackage{psfrag}

\DeclareFontFamily{U}{rsf}{}
\DeclareFontShape{U}{rsf}{m}{n}{
  <5> <6> rsfs5 <7> <8> <9> rsfs7 <10-> rsfs10}{}
\DeclareMathAlphabet\Scr{U}{rsf}{m}{n}

\catcode`\@=11
\def\@citex[#1]#2{%
\if@filesw \immediate \write \@auxout {\string \citation {#2}}\fi
\@tempcntb\m@ne \let\@h@ld\relax \def\@citea{}%
\@cite{%
  \@for \@citeb:=#2\do {%
    \@ifundefined {b@\@citeb}%
      {\@h@ld\@citea\@tempcntb\m@ne{\bf ?}%
      \@warning {Citation `\@citeb ' on page \thepage \space undefined}}%
      {\@tempcnta\@tempcntb \advance\@tempcnta\@ne%
      \@tempcntb\number\csname b@\@citeb \endcsname \relax%
      \ifnum\@tempcnta=\@tempcntb 
        \ifx\@h@ld\relax%
          \edef \@h@ld{\@citea\csname b@\@citeb\endcsname}%
        \else%
          \edef\@h@ld{\ifmmode{-}\else--\fi\csname b@\@citeb\endcsname}%
        \fi%
      \else
        \@h@ld\@citea\csname b@\@citeb \endcsname%
        \let\@h@ld\relax%
      \fi}%
    \def\@citea{,\penalty\@highpenalty\,}%
  }\@h@ld
}{#1}}

\def\@citeb#1#2{{[#1]\if@tempswa , #2\fi}}
\def\@citeu#1#2{{$^{#1}$\if@tempswa , #2\fi }}
\def\@citep#1#2{{#1\if@tempswa , #2\fi}}

\def\bcites{         
        \catcode`\@=11
        \let\@cite=\@citeb
        \catcode`\@=12
}

\def\upcites{         
        \catcode`\@=11
        \let\@cite=\@citeu
        \catcode`\@=12
}

\def\plaincites{      
        \catcode`\@=11
        \let\@cite=\@citep
        \catcode`\@=12
}

\newcount\hour
\newcount\minute
\newtoks\amorpm
\hour=\time\divide\hour by 60
\minute=\time{\multiply\hour by 60 \global\advance\minute by-\hour}
\edef\standardtime{{\ifnum\hour<12 \global\amorpm={am}%
        \else\global\amorpm={pm}\advance\hour by-12 \fi
        \ifnum\hour=0 \hour=12 \fi
        \number\hour:\ifnum\minute<10 0\fi\number\minute\the\amorpm}}
\edef\militarytime{\number\hour:\ifnum\minute<10 0\fi\number\minute}

\def\draftlabel#1{{\@bsphack\if@filesw {\let\thepage\relax
   \xdef\@gtempa{\write\@auxout{\string
      \newlabel{#1}{{\@currentlabel}{\thepage}}}}}\@gtempa
   \if@nobreak \ifvmode\nobreak\fi\fi\fi\@esphack}
        \gdef\@eqnlabel{#1}}
\def\@eqnlabel{}
\def\@vacuum{}
\def\marginnote#1{}
\def\draftmarginnote#1{\marginpar{\raggedright\scriptsize\tt#1}}
\def\draft{
        \pagestyle{plain}
        \overfullrule=2pt
        \oddsidemargin -.5truein
        \def\@oddhead{\sl \phantom{\today\quad\militarytime} \hfil
        \smash{\Large\sl DRAFT} \hfil \today\quad\militarytime}
        \let\@evenhead\@oddhead
        \let\label=\draftlabel
        \let\marginnote=\draftmarginnote
        \def\ps@empty{\let\@mkboth\@gobbletwo
        \def\@oddfoot{\hfil \smash{\Large\sl DRAFT} \hfil}
        \let\@evenfoot\@oddhead}
        \def\@eqnnum{(\theequation)\rlap{\kern\marginparsep\tt\@eqnlabel}%
        \global\let\@eqnlabel\@vacuum}  }

\def\section{\@startsection {section}{1}{\z@}{3.ex plus 1ex minus
 .2ex}{2.ex plus .2ex}{\large\bf}}
\def\subsection{\@startsection{subsection}{2}{\z@}{2.75ex plus 1ex minus
 .2ex}{1.5ex plus .2ex}{\bf}}        

\def\appendix{{\newpage\section*{Appendix}}\let\appendix\section%
        {\setcounter{section}{0}
        \gdef\thesection{\Alph{section}}}\section}

\def\abstract{\if@twocolumn
\section*{Abstract}
\else 
\begin{center}
{\bf Abstract\vspace{-.5em}\vspace{0pt}}
\end{center}
\quotation
\fi}

\catcode`\@=12

\newcommand{\beq}{\begin{equation}}
\newcommand{\eeq}{\end{equation}}
\newcommand{\beqa}{\begin{eqnarray}}
\newcommand{\eeqa}{\end{eqnarray}}
\newcommand{\dd}{{\rm d}}

\newcommand{\Z}{{\bf Z}}
\newcommand{\ZZ}{{\mathbb Z}}

\newcommand{\R}{{\bf R}}

\newcommand{\C}{{\bf C}}
\newcommand{\CC}{{\mathbb C}}

\newcommand{\e}{\,{\rm e}}
\newcommand{\CP}{{\bf CP}}

\newcommand{\be}{\begin{equation}}
\newcommand{\ee}{\end{equation}}
\newcommand{\bea}{\begin{eqnarray}}
\newcommand{\eea}{\end{eqnarray}}

\def\to{\rightarrow}

\def\lae{\mathrel{\mathop{\smash{\lower .5 ex \hbox{$\stackrel<\sim$}}}}}
\def\lae{\mathrel{\mathop{\smash{\lower .5 ex \hbox{$\stackrel>\sim$}}}}}

\def\l:{\mathopen{:}\,}
\def\r:{\,\mathclose{:}}

\catcode`\@=11
\def\theequation{\arabic{equation}}
\catcode`\@=12

\bcites

\catcode`\@=11
\def\theequation{\thesection.\arabic{equation}}
\@addtoreset{equation}{section}
\@addtoreset{footnote}{section}
\@addtoreset{footnote}{subsection}
\catcode`\@=12

\newcommand{\oQ}{\overline{Q}}

\newcommand{\bpsi}{\overline{\psi}}

\newcommand{\bx}{\overline{x}}

\newcommand{\bz}{\overline{z}}

\newcommand{\blambda}{\overline{\lambda}}
\newcommand{\bsigma}{\overline{\sigma}}

\newcommand{\wta}{\widetilde{a}}
\newcommand{\wtb}{\widetilde{b}}
\newcommand{\wtc}{\widetilde{c}}
\newcommand{\wtx}{\widetilde{x}}
\newcommand{\wty}{\widetilde{y}}
\newcommand{\wtz}{\widetilde{z}}
\newcommand{\wtX}{\widetilde{X}}
\newcommand{\wtY}{\widetilde{Y}}
\newcommand{\wtm}{\widetilde{m}}
\newcommand{\wtr}{\widetilde{r}}
\newcommand{\wttheta}{\widetilde{\theta}}
\newcommand{\wthu}{\widetilde{\theta}_{\mathfrak{u}(1)}}
\newcommand{\wtt}{\widetilde{t}}
\newcommand{\wts}{\widetilde{s}}
\newcommand{\wtk}{\widetilde{k}}
\newcommand{\whc}{\widehat{c}}

\newcommand{\Ost}{O_+}
\newcommand{\Ons}{O_-}
\newcommand{\Ostns}{O_{\pm}}
\newcommand{\Vct}{V}
\newcommand{\nn}{\nonumber}

\begin{document}

\begin{titlepage}

\begin{center}

\hfill\today

\vskip 2.5 cm
{\large \bf Duality In Two-Dimensional 
$(2,2)$ Supersymmetric\\[0.2cm] Non-Abelian
Gauge Theories}
\vskip 1 cm

{Kentaro Hori}\\
\vskip 0.5cm
{\sl IPMU, The University of Tokyo, Kashiwa, Japan}

\end{center}

\vskip 0.5 cm
\begin{abstract}
We study the low energy behaviour of
${\mathcal N}=(2,2)$ supersymmetric gauge theories
in $1+1$ dimensions, with orthogonal and symplectic gauge groups
and matters in the fundamental representation.
We observe supersymmetry breaking in super-Yang-Mills theory 
and in theories with small numbers of flavors.
For larger numbers of flavors,
we discover duality between regular theories with different gauge groups
and matter contents, 
where regularity refers 
to absence of quantum Coulomb branch.
The result is applied to study families of superconformal field theories
that can be used for superstring compactifications, with
corners corresponding to three-dimensional Calabi-Yau manifolds. 
This work is motivated by recent development in mathematics concerning 
equivalences of derived categories.
\end{abstract}

\end{titlepage}

\newpage

\thispagestyle{empty}

\tableofcontents
\thispagestyle{empty}

\newpage

\setcounter{page}{1}

\section{Introduction}

What is the space of all $(2,2)$ superconformal field theories in
$1+1$ dimensions with $c=9$ and integral R-charges? 
It is an interesting question in its own right and is also believed
to be important. 
Related questions are: What is the space of all
three-dimensional Calabi-Yau manifolds? and
What is the space of all string compactifications to $3+1$ dimensions?

Linear sigma models provide a useful tool to find 
$(2,2)$ superconformal field theories, 
analyze them, and study their quantum moduli spaces \cite{Wphases}.
Investigation so far had been centered on models
with Abelian gauge groups.
The low energy theory in a typical ``phase'' has simple interpretation
as an orbifold of non-linear sigma model with a superpotential.
The geometric phase is related to toric geometry, where well developed
techniques are available.
 Models with non-Abelian gauge groups, on the other hand,
had not yet been studied much. One reason is that such models
typically have phases where a simple gauge group is unbroken,
but the nature of the low energy behaviour of such theories 
had not been well understood.
In \cite{HoTo}, David Tong and the author attempted to understand 
the nature of a class of theories with special unitary gauge groups,
and the result is applied to linear sigma models.
In the present work, we study more about
non-Abelian gauge interaction, with emphasis on orthogonal and symplectic
gauge groups, and apply our findings to linear sigma models
relevant for string compactifications.
We hope that this work will eventually lead us to
expand our perspective concerning the above questions.

One problem in theories with simple gauge groups
is that there is often a non-compact Coulomb branch
which makes it impossible to have a sensible conformal field theory
with discrete spectrum in the infra-red limit.
Coulomb branch is interesting in its own right, especially
in relation to six dimensional theories associated with Neveu-Schwarz 
fivebranes. However, its presence is problematic
in order to obtain
theories relevant for string compactifications.
This motivates us to look at {\it regular} theories, where
Coulomb branch is lifted by quantum correction.

Let us describe the main results of this paper.

We first describe the result for
the $O(k)$ or $SO(k)$ gauge theory with $N$ massless fields in the vector
representation, $x_1,\ldots, x_N$, with vanishing superpotential
($k=1,2,3,\ldots$ and $N=0,1,2,3,\ldots,$).
For $k\geq 3$, since these groups have $\Z_2$ fundamental group,
we need to specify the mod 2 theta angle \cite{WTheta}.
Also, an $O(k)$ theory can be regarded as a $\Z_2$ orbifold of 
an $SO(k)$ theory, and there are two possibilities additionally,
denoted by $\Ost(k)$ and $\Ons(k)$.
The theory with $k\geq 2$ is regular 
when $N-k$ is odd and the mod 2 theta angle is turned off,
or when $N-k$ is even and the mod 2 theta angle is turned on.
For $N\leq k-2$,  whether regular or not, the theory
has no normalizable supersymmetric ground state.
That is, the supersymmetry is spontaneously broken.
The rest applies only to regular theories.
For $N=k-1$, the $SO(k)$ and $\Ons(k)$ 
theories flow in the infra-red limit to
the free theory of the scalar products, 
$(x_ix_j)=\sum_{a=1}^kx_i^ax_j^a$, the ``mesons''.
The $\Ost(k)$ theory flows to two copies of such free theory.
For $N\geq k$, we propose that there is a duality:
\beqa
~~~\Ost(k)&\longleftrightarrow& SO(N-k+1)\nn\\
~~~SO(k)&\longleftrightarrow& \Ost(N-k+1)\\
~~~\Ons(k)&\longleftrightarrow& \Ons(N-k+1).\nn
\eeqa
The theory with gauge group on the left hand side
flows to the same infra-red fixed point as
the theory with gauge group on the right hand side
with $N$ vectors
$\wtx^1,\ldots,\wtx^N$ and 
${N(N+1)\over 2}$ singlets $s_{ij}=s_{ji}$ with the superpotential
\beq
W=\sum_{i,j=1}^Ns_{ij}(\wtx^i\wtx^j).
\label{Oduality}
\eeq
The mesons in the original theory correspond to the singlets in 
the dual, $(x_ix_j)=s_{ij}$.
The symmetry $\Z_2=O(k)/SO(k)$ in the $SO(k)$ theory corresponds to
the quantum $\Z_2$ symmetry of the dual $\Ost(N-k+1)$ theory (regarded as
a $\Z_2$ orbifold).
In particular, the ``baryons''
$[x_{i_1}\cdots x_{i_k}]=\det(x^a_{i_b})$ in the $SO(k)$ theory
correspond to twist operators in
the dual $O(N-k+1)$ theory.
There are similar correspondences between $\Z_2$ symmetries 
in the other dual pairs.
The duality is tested against non-trivial checks, including
't Hooft anomaly matching, flow by complex mass deformation, 
vacuum counting with twisted mass deformation,
comparison of  $(c,c)$ and $(a,c)$ chiral rings.

We next describe the results for the $USp(k)$ gauge theory
with $N$ massless fundamentals, $x_1,\ldots,x_N$, with vanishing 
superpotential  ($k=2,4,6,\ldots$ and $N=0,1,2,3,\ldots$). 
It is regular if and only if $N$ is odd.
For $N\leq k$, there is no normalizable supersymmetric ground state
in both regular ($N$ odd) and irregular ($N$ even) theories.
For $N=k+1$, the low energy theory is the free conformal field theory
of the mesons, $[x_ix_j]=\sum_{a,b=1}^kx_i^aJ_{ab}x_j^b$,
where $J_{ab}$ is the symplectic structure defining the gauge group.
For higher odd $N\geq k+3$, there is a duality:
\beq
~~~USp(k)~~\longleftrightarrow~~ USp(N-k-1).
\eeq
That is, the theory is dual to
the $USp(N-k-1)$ gauge theory with $N$ fundamentals $\wtx^1,\ldots,\wtx^N$ and 
${N(N-1)\over 2}$ singlets $a_{ij}=-a_{ji}$ with the superpotential
\beq
W=\sum_{i,j=1}^Na_{ij}[\wtx^i\wtx^j].
\label{USpduality}
\eeq
The mesons in the original theory
correspond to the singlets in the dual, $[x_ix_j]=a_{ij}$.
Again, the duality is tested against non-trivial checks.

For completeness, we record the results obtained in \cite{HoTo}
for the $SU(k)$  gauge theory
with $N$ massless fundamentals, $x_1,\ldots,x_N$, with vanishing 
superpotential  ($k=2,3,4,\ldots$ and $N=0,1,2,\ldots$). 
The theory is regular when there is no
$k$ distinct $N$-th roots of unity that sum to zero.
For $N\leq k$, there is no normalizable supersymmetric ground state.
For $N=k+1$, the low energy theory is the free conformal field theory
of the baryons $[x_{i_1}\cdots x_{i_k}]$.
For $N\geq k+2$, there is a duality between regular theories:
\beq
~~~SU(k)~~\longleftrightarrow~~ SU(N-k).
\eeq
That is, the theory is dual to the $SU(N-k)$ gauge theory with
$N$ fundamentals $\wtx^1,\ldots,\wtx^N$ and vanishing superpotential.
$[x_{i_1}\cdots x_{i_k}]=\epsilon_{i_1\cdots i_kj_1\cdots j_{N-k}}
[\wtx^{j_1}\cdots \wtx^{j_{N-k}}]$ is the relation of the variables.
This duality itself was not explicitly stated in \cite{HoTo}
but can be proven rather trivially based on the relation $G(k,N)\cong G(N-k,N)$ 
of Grassmannians, just as in the series of dualities found in \cite{HoTo}.
In a way, this is the most
fundamental case from which the members in the series follows by addition
of superpotential.
It would be interesting to study
$SU(k)$ gauge theories with $N$ fundamentals and $M$ anti-fundamentals.
We postpone the discussion of such theories for future works.

This pattern looks strikingly similar to the results found
in ${\mathcal N}=1$ gauge theories in $3+1$ dimensions
\cite{ADS,SeibergExact,Seiberg} and \cite{SOSeiberg,USpSeiberg};
supersymmetry breaking for low (but non-zero) flavors,
quantum deformed moduli space for a ``critical'' flavor,
and duality for supercritical flavors:
\beqa
~~SU(k)&\longleftrightarrow& SU(N-k)\nn\\
~~~SO(k)&\longleftrightarrow& SO(N-k+4)\\
~~~USp(k)&\longleftrightarrow& USp(N-k-4).\nn
\eeqa
For special unitary groups, the number of fundamentals must be equal to the 
number of antifundamentals to avoid the gauge anomaly.
For symplectic groups, $N$ must be even to avoid the global anomaly.
There is also a similar pattern in ${\mathcal N}=2$ gauge theories in
$2+1$ dimensions \cite{AHW,DHO,AHISS,Karch,Aharony,Kapustin}.
In particular, for supercritical
flavors, there is a duality between the following gauge groups:
\beqa
~~U(k)&\longleftrightarrow& U(N-k)\nn\\
~~~O(k)&\longleftrightarrow& O(N-k+2)\\
~~~USp(k)&\longleftrightarrow& USp(N-k-2).\nn
\eeqa
There are also versions of duality between theories
with Chern-Simons terms.
The shift in rank for the orthogonal and symplectic groups
decreases as $\pm 4,\pm 2,\pm 1$ as the dimension 
is reduced from four to two. This suggests an interpretation of our duality
in terms of brane construction, as
the charge of orientifold planes decreases by a factor of $2$
for each reduction of dimension by $1$.

The present work is motivated by recent development
in mathematics concerning equivalences of derived categories
(see for example \cite{BOrev} for an introduction).
In fact, the direct motivation came from the paper
by Hosono and Takagi \cite{HoTa} in which the authors suggested 
an equivalence of the derived categories of two distinct Calabi-Yau 
manifolds, $X$ and $Y$, of dimension three.
$X$ is the intersection of five symmetric quadrics in $\CP^4\times \CP^4$
divided by the exchange involution, while
$Y$ is a double cover over the degeneration locus
of the associated quadratic form on $\CP^4$
(a determinantal quintic in $\CP^4$) which is
ramified along a curve of higher degeneration.
This suggests existence of a quantum K\"ahler moduli space
of $(2,2)$ superconformal field theories,
which has a corner corresponding to $X$
and another corner corresponding to $Y$.
$X$ naturally leads us to consider a linear sigma model
with a non-Abelian gauge group
\beq
{U(1)\times O(2)\over \{(\pm 1,\pm {\bf 1}_2)\}}.
\eeq
In the opposite phase, an $O(2)$ subgroup of the gauge group is entirely
unbroken, and we are forced to understand the low energy behaviour of 
such a quantum gauge theory.
We managed to understand it to some extent, and the result
tells us that we indeed have a ramified 
double cover over the determinantal quintic, 
but does not tell us how to construct it globally.
This again forced us to have a better understanding and led us
to discover the non-Abelian duality. In the dual model with gauge group
$(U(1)\times SO(4))/\{(\pm 1,\pm {\bf 1}_4)\}$, the global ramified
double cover $Y$ emerges naturally and completely classically.
The dual pair of linear sigma models play complementary r\^oles
at both of the two corners corresponding to $X$ and $Y$.
If the gauge symmetry is unbroken and a non-trivial quantum analysis
is necessary in one theory, the gauge symmetry is completely Higgsed
and purely classical analysis suffices in the dual theory.

\begin{figure}[htb]
\centerline{\includegraphics{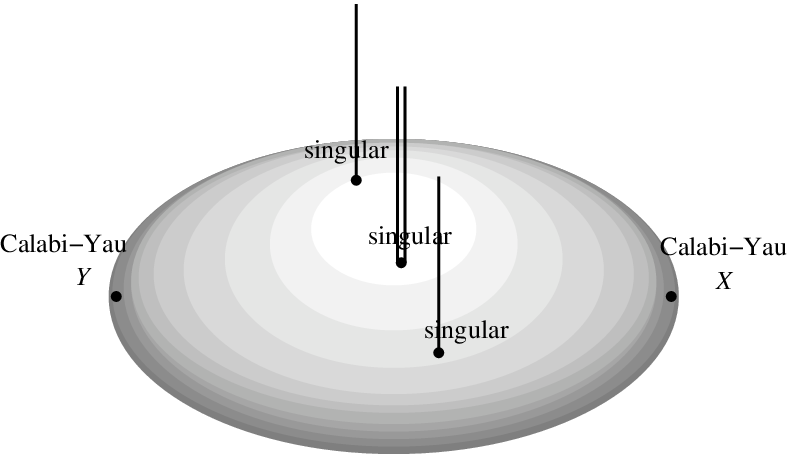}}
\caption{The quantum K\"ahler moduli space for Hosono-Takagi example.}
\label{fig:HoTa}
\end{figure}

Figure~\ref{fig:HoTa} shows the structure of the quantum K\"ahler moduli space
obtained by the dual pair of linear sigma models.
There are three singular points at which Coulomb branch emerges.
One point is special in that there are {\it two copies} of a
one-dimensional Coulomb branch. Comparing with the result of \cite{HoTa},
this seems to be related to having {\it two} BPS particles that become massless
at this point.

The example of Hosono-Takagi can be regarded as a sister of the example of 
R\o dland \cite{Rodland}
which was the motivation of the work \cite{HoTo}. In that work, we 
studied a linear sigma model with gauge group $U(2)$, that is,\footnote{It
is interesting to see that the two simplest non-Abelian groups,
$USp(2)$ and $O(2)$, appeared in this way. This reminds us of
their r\^ole in discovering the M-theory lift 
of orientifold six planes \cite{Sp1O2}.}
\beq
{U(1)\times USp(2)\over \{(\pm 1,\pm {\bf 1}_2)\}}.
\eeq
In this paper, we apply the duality also to this model and find
a new picture. We will also apply the duality to
linear sigma models relevant for intersection of quadrics,
which was studied in mathematics \cite{alggeom,BOso,BOrev,KiQ}
as the basic example of non-trivial equivalence of derived categories.
This exercise was also useful in refining our understanding
of $\Z_2$ orbifolds and that resulted
in shaping up the duality in the present form.

\section{$O(1)$ Theories}
\label{sec:O1}

In this section, we study several two-dimensional 
$(2,2)$ supersymmetric gauge theories with
gauge group $O(1)=\{1,-1\}$, i.e., orbifolds by the group
$\Z_2=\Z/2\Z$. 

We start with making general remarks concerning the definition of
$\Z_2$ orbifolds in supersymmetric field theories.

\subsection{General Remarks On $\ZZ_2$ Orbifolds}
\label{subsec:grZ2}

Let us consider a $(2,2)$ supersymmetric quantum field theory in $1+1$
dimensions with an involutive symmetry $\tau$
commuting with the supercharges. We would like to define an orbifold 
with respect to $\Z_2=\{1,\tau\}$.

For $g,h\in \{1,\tau\}$, we denote by
${\displaystyle {}^g\mathop{\mbox{\large $\Box$}}_h}$ 
the $g$-twisted Witten index on the unprojected $h$-twisted 
Ramond-Ramond (RR) sector.
It is equal to the partition function on a torus
with $h$-twist in the ``space'' direction and $g$-twist in the ``time''
direction. As usual, it receives
contribution only from supersymmetric ground states,
and does not depend on the metric of the torus nor on the distinction between
the space and time directions. In particular, we have
\beq
{}^{\tau}\mathop{\mbox{\huge $\Box$}}_{\tau}
~~~{}^{\displaystyle =}~~~
{}^{\tau\tau=1}\mathop{\mbox{\huge $\Box$}}_{\tau}
~~~{}^{\displaystyle =}~~~
{}^{\tau}\mathop{\mbox{\huge $\Box$}}_{1}.
\label{moinv}
\eeq
This provides constraints and relations concerning the action of
$\tau$ and $(-1)^F$ on the twisted and untwisted sectors.

As an important example, let us consider a theory in which there is exactly
a single supersymmetric ground state in each of the untwisted and the twisted
sectors.
The first equality (\ref{moinv}) means that the twisted ground state
must be $\tau$ invariant and hence survives the orbifold projection.
The second equality means that, if $\tau$ acts as the sign
$\varepsilon_{\tau}$ on the untwisted ground state, then the action of
$(-1)^F$ on the twisted and untwisted sectors are related by
\beq
(-1)^F|_{\tau}=\varepsilon_{\tau}(-1)^F|_1.
\label{Fn}
\eeq
The untwisted ground state survive the projection if $\varepsilon_{\tau}=1$
but is projected out if $\varepsilon_{\tau}=-1$.
The sign $\varepsilon_{\tau}$ is a part of the data of the
orbifold which has such a significant effect.

In general, $\Z_2$ orbifolds of theories with spinors
always come in pairs --- if we can define an orbifold
by a symmetry $\tau$ then we can also consider the orbifold by
$(-1)^{F_s}\tau$, where $(-1)^{F_s}$ is the operator that acts
as the sign flip of all states in the untwisted RR sector. This was emphasized
for superconformal field theories in \cite{DGH,IV}.
It follows from the operator product rule
\beqa
&&{\rm NSNS}^{}_g\times {\rm NSNS}^{}_h~\longrightarrow~{\rm NSNS}^{}_{gh}
\nn\\
&&{\rm NSNS}^{}_g\times {\rm RR}^{}_h~\longrightarrow~{\rm RR}^{}_{gh}
\\
&&{\rm RR}^{}_g\times {\rm RR}^{}_h~\longrightarrow~{\rm NSNS}^{}_{gh}
\nn
\eeqa
that the switch from $\tau$ to 
$(-1)^{F_s}\tau$ does not change the orbifold projection 
in the untwisted NSNS sector 
but reverses the one in the twisted NSNS sector.
For later use we record the effect:
\beq
(-1)^{F_s}=\left\{\begin{array}{ll}
1&\mbox{~in untwisted NSNS and twisted RR}\\
-1&\mbox{~in twisted NSNS and untwisted RR}.
\end{array}\right.
\eeq
Dressing by $(-1)^{F_s}$ is different from the discrete torsion, 
which is absent 
for the orbifold group $\Z_2$ as ${\rm H}^2(\Z_2,U(1))$ is trivial.

There is a ``canonical'' choice of orbifold in a class of theories.
Let us consider the supersymmetric
non-linear sigma model with a target K\"ahler manifold $X$.
The spectral flow, or A-twist, provides a
linear isomorphism between the space of RR ground states
and the underlying space of the $(a,c)$ ring, which is
a deformation of the de Rham cohomology ring ${\rm H}_{\rm dR}(X,\C)$.
Suppose $X$ has an involutive holomorphic isometry $\tau$, with which
we would like to define an orbifold.
In the untwisted sector, there is a canonical choice of orbifold projection:
the identity operator must be kept and hence,
in the the untwisted $(a,c)$ ring, those corresponding to
the $\tau$-invariant cohomology classes must remain.
Now, the ``canonical'' choice would be the one that keeps
the linear isomorphism between the untwisted RR ground states
and the space of untwisted $(a,c)$ ring elements.
The non-canonical one would select the RR ground states corresponding
to the anti-invariants in ${\rm H}_{\rm dR}(X,\C)$.
We shall sometimes denote the orbifold group
by $\Z_2(-1)^{F_s}$ for the non-canonical choice.
As an important example, let us consider the case where $\tau$ is 
the identity map of $X$. 
In this case, the unprojected twisted sector is isomorphic to the
space of states of the original sigma model.
For either choice of orbifold, 
the untwisted NSNS and twisted RR sector must survive the projection 
entirely.
The untwisted RR as well as twisted NSNS sector survive entirely
for the canonical choice, while they are all projected out for the 
non-canonical choice.
Thus, the canonical orbifold $X/\Z_2$
is isomorphic to the sigma model
whose target space is the disjoint union of two copies of $X$, 
while the non-canonical one $X/\Z_2(-1)^{F_s}$
is isomorphic to the original sigma model on $X$.

In the course of the paper, we shall introduce the notion
of ``canonical'' or ``standard'' $\Z_2$ orbifold for other type of
theories. This will also be extended to the definition of
gauge theories with $O(k)$ gauge group, which can be regarded as 
$\Z_2$ orbifolds of $SO(k)$ gauge theories.

\subsection{Massive Fields}
\label{subsec:massive}

As the first example, we study an orbifold of the
theory of massive chiral multiplets with respect to the sign flip.
Our main interest will be the spectrum of supersymmetric ground states.

One way to give a mass to a chiral multiplet $(x,\psi_{\pm})$ is to
introduce a superpotential
\beq
W={m\over 2}x^2.
\label{cmass}
\eeq
An alternative is {\it twisted mass} \cite{HH} which is given by 
the following procedure: gauge the phase rotation symmetry of $x$,
give a value $-\wtm$ to the scalar component of the gauge multiplet,
and then turn off the gauge interaction. 
A superpotential mass shall be called a {\it complex mass}.
Note that a twisted mass is possible only when the phase rotation is
a symmetry. In particular, we cannot give both complex and twisted masses
at the same time, since the phase rotation is not a symmetry of (\ref{cmass}).
To be more explicit,
the complex mass term reads
\beq
{\mathcal L}_{m}=-|m|^2|x|^2-m\psi_+\psi_--\overline{m}\bpsi_-\bpsi_+,
\eeq
while the twisted mass term is
\beq
{\mathcal L}_{\wtm}=-|\wtm|^2|x|^2-\wtm\psi_+\bpsi_-
-\overline{\wtm}\psi_-\bpsi_+.
\eeq

Let us consider a $\Z_2$ orbifold of the theory of $(x,\psi_{\pm})$
with the usual kinetic term plus either of the two mass terms, by
the sign flip symmetry,
\beq
\tau:(x,\psi_{\pm})\to (-x,-\psi_{\pm}).
\eeq
At first sight, the two orbifold theories, one with a complex mass
and the other with a twisted mass, are isomorphic, 
as one can switch from one Lagrangian to the other by exchanging
$\psi_-$ and $\bpsi_-$. However, we would like to define the orbifolds
with respect to {\it a common $\tau$ action on 
the common space of states}. That is, we define them 
as two different mass deformations
of a given orbifold of a massless chiral multiplet.
Then, as we will see, the two are not isomorphic.

Each of the two theories have one untwisted and one twisted supersymmetric
ground states before the orbifold projection.
Let us compare the ground state wavefunctions in the two theories.
There is literally no difference in the dependence on the
bosonic field $x$ and hence our focus will be the fermionic fields
$\psi_{\pm}$.
The fields, both bosons and fermions, are integer ({\it resp}. half-integer) 
moded in the untwisted ({\it resp}. twisted) RR sector.
Let us first focus on the zero modes in the untwisted sector.
The (fermionic part of) Hamiltonians of the two systems read
\beqa
&&H_{m}=m\psi_{+0}\psi_{-0}+\overline{m}\bpsi_{-0}\bpsi_{+0},
\\
&&H_{\wtm}=\wtm\psi_{+0}\bpsi_{-0}+\overline{\wtm}\psi_{-0}\bpsi_{+0}.
\eeqa
The lowest energy states are respectively
\beqa
&&
|0\rangle_{0}+{\overline{m}\over |m|}\bpsi_{+0}\bpsi_{-0}
|0\rangle_{0},
\label{gscm}\\
&&
\bpsi_{+0}|0\rangle_{0}+{\wtm\over |\wtm|}\bpsi_{-0}
|0\rangle_{0},
\label{gstm}
\eeqa
where $|0\rangle_{0}$ is the state annihilated by $\psi_{+0}$
and $\psi_{-0}$. The state $|0\rangle_{(0)}$ is transformed by 
$\tau$ to itself up to a sign, since the defining property
is $\tau$ invariant.
Therefore, $\tau$ transforms 
the two ground states, (\ref{gscm}) and (\ref{gstm}), 
to themselves but {\it with opposite signs},
as it flips the sign of $\bpsi_{\pm 0}$.
Non-zero modes are decoupled into infinite sectors labeled
by the absolute value of the momentum.
The lowest energy states in each sector are again different between
the two theories
but the $\Z_2$ orbifold action on them are the same.
Therefore, the $\Z_2$ actions on the untwisted sector RR ground states
are opposite between the two theories, while the
actions on the twisted sector RR ground states are the same.
The same computation can be used to study the $\Z_2$ action on
NSNS sector states, where
the fermions are half-integer ({\it resp}. integer) moded
in the untwisted ({\it resp}. twisted) sector.
The $\Z_2$ actions on the untwisted sector NSNS ground states
are the same between the two theories, while the
actions on the twisted sector NSNS ground states are the opposite.

As remarked in the previous subsection, the twisted sector RR ground 
state must survive the orbifold projection, in each of the two theories. 
On the other hand,
whether the untwisted sector RR ground state survives or not
is up to our choice. 
There are two possibilities
(let $|\Omega\rangle^{}_{\rm RR}$ {\it resp}. 
$|\widetilde{\Omega}\rangle^{}_{\rm RR}$
be the untwisted ground state of the theory with a complex {\it resp}.
twisted mass): 
$|\Omega\rangle^{}_{\rm RR}$ is invariant and 
$|\widetilde{\Omega}\rangle^{}_{\rm RR}$ is 
anti-invariant, or
$|\Omega\rangle^{}_{\rm RR}$ is anti-invariant and 
$|\widetilde{\Omega}\rangle^{}_{\rm RR}$ 
is invariant.
In the rest of the paper, we shall take the latter 
as our ``standard'' convention
for the $\Z_2$ orbifold of a chiral multiplet by the sign flip.
Namely:
{\it $|\Omega\rangle^{}_{\rm RR}$ is anti-invariant and is projected out,
while $|\widetilde{\Omega}\rangle^{}_{\rm RR}$ 
is invariant and survives the projection.}
Note that
$|\Omega\rangle^{}_{\rm RR}$ and $|\widetilde{\Omega}\rangle^{}_{\rm RR}$
have opposite statistics, as can be seen from (\ref{gscm}) and (\ref{gstm}).
Let us assume that $|\Omega\rangle^{}_{\rm RR}$ is fermionic,
and hence $|\widetilde{\Omega}\rangle^{}_{\rm RR}$ is bosonic.
Then, it follows from the general constraint
(\ref{Fn}) that the twisted ground states
in the two theories are both bosonic.
Therefore, under this assignment, we have
\beq
{\rm Tr}(-1)^F=\left\{\begin{array}{ll}
1&\mbox{~for complex mass}\\
2&\mbox{~for twisted mass}.
\end{array}\right.
\eeq
What about the NSNS sector? 
As always, the NSNS ground state in the untwisted sector is
invariant and survive the projection in each of the two theories.
To determine the action in the twisted sector, we note that
the theory with a twisted mass has one twisted and one 
untwisted supersymmetric ground states. This requires existence of 
a twist operator in the infra-red limit. On the other hand, we do not 
need such an operator for the theory with a complex mass.
From this we conclude that
the twisted NSNS ground state is anti-invariant 
and is projected out in the theory with complex mass
while the one in the theory with twisted mass
is invariant and survives the projection.

The definition and the result for theories with several massive multiplets
is obtained simply by tensor product:
Let us consider the orbifold of $N$ fields (i.e. $N$ chiral multiplets)
with complex masses and $M$ fields 
with twisted masses, by the simultaneous sign flip
of all the $N+M$ fields.
Then it has one supersymmetric ground state from the twisted sector
if $N$ is odd, while it has two supersymmetric ground states,
one twisted and one untwisted, if $N$ is even.
The states are all bosonic, and in particular,
\beq
{\rm Tr}(-1)^F=\left\{\begin{array}{ll}
1&\mbox{~if $N$ is odd}\\
2&\mbox{~if $N$ is even}.
\end{array}\right.
\label{ind1}
\eeq

This result can be extended to the following periodicity phenomenon.
We know that simply adding a massive field to a system does not change
the infra-red behaviour. Does it hold also in orbifolds?
Suppose we have a $\Z_2$ orbifold of a $2d$ $(2,2)$ supersymmetric 
quantum field theory ${\mathcal A}$ by 
its involutive symmetry $\tau_{\mathcal A}$.
Let us add to ${\mathcal A}$ a single chiral multiplet $x$
with a complex or twisted mass and mod out the combined system by
$\tau=(\tau_{\mathcal A},\tau_x)$ where $\tau_x$ is the sign 
flip symmetry considered above.
At energies below the mass of $x$, 
relevant states in each sector
are the tensor product of states of the ${\mathcal A}$ system
with the ground state of the $x$ system in that sector.
By the transformation property of the ground states
learned above, we find that the orbifold projection
in this theory is the same as the one for ${\mathcal A}/\tau_{\mathcal A}$
if $x$ has a twisted mass. On the other hand, if $x$ has a complex mass,
the projection is the same as ${\mathcal A}/\tau_{\mathcal A}$
in the untwisted NSNS and twisted RR sectors
but is opposite to ${\mathcal A}/\tau_{\mathcal A}$
in the twisted NSNS and untwisted RR sectors.
This lead us to claim that (at low energies) {\it the combined orbifold
theory is equivalent to the original orbifold
${\mathcal A}/\tau_{\mathcal A}$
when $x$ has a twisted mass, while it is equivalent
to the other orbifold ${\mathcal A}/(-1)^{F_s}\tau_{\mathcal A}$
when $x$ has a complex mass.}
A similar conclusion holds if we combine ${\mathcal A}$
with $N$ fields with complex mass and $M$ fields with twisted mass:
The combined orbifold system is equivalent at low energies 
to ${\mathcal A}/\tau_{\mathcal A}$ if $N$ is even
and to ${\mathcal A}/(-1)^{F_s}\tau_{\mathcal A}$ if $N$ is odd.

\subsection*{\sl Embedding Into Linear Sigma Models}

We now show that the above ``standard'' choice of $\Z_2$ orbifold
appears naturally as a part of linear sigma models.
Let us consider a $U(1)$ gauge theory consisting of
a field $p$ of charge $-2$ and fields $x_1,\ldots,x_N$ of charge $1$.
First let us set $W=0$ and give no twisted mass.
When the Fayet-Iliopoulos (FI) parameter $r$ is negative,
the D-term equation $-2|p|^2+\sum_{i=1}^N|x_i|^2=r$ requires $p$ to have a
non-zero value, breaking the $U(1)$ gauge group to its $\Z_2$ subgroup.
In the limit $r\to -\infty$, only $x$'s remain as massless degrees of 
freedom and we obtain the free orbifold $\C^N/\Z_2$ by
the simultaneous sign flip of $x_1,\ldots, x_N$.
We study deformations of this linear sigma model
that correspond to giving complex and twisted masses to $x$'s
in the orbifold.

Let us first consider turning on the superpotential
\beq
W=p(x_1^2+\cdots+x_N^2).
\eeq
The theory at $r\to -\infty$ is now the Landau-Ginzburg (LG) orbifold
of the variables $x_1,\ldots,x_N$ with superpotential
$W=x_1^2+\cdots+x_N^2$ by the simultaneous sign flip.
That is, a $\Z_2$ orbifold theory where all $N$ variables 
have complex masses.
The main question is: Is it the ``standard'' 
one or the other one?
To see this, let us analyze this linear sigma model in detail.
In the regime $r\gg 0$, $x$'s have non-zero values, 
breaking the gauge group completely,
and the theory reduces to the non-linear sigma model on the quadric 
hypersurface $Q^{N-2}=\{x_1^2+\cdots+x_N^2=0\}$ of $\CP^{N-1}$.
We may also have Coulomb branch vacua.
The effective twisted superpotential for
the scalar component $\sigma$ of the $U(1)$ vector multiplet is
\beq
\widetilde{W}_{\it eff}=-(-2\sigma)(\log(-2\sigma)-1)-N\sigma(\log\sigma-1)
-t\sigma,
\eeq
for $t=r-i\theta$ where $\theta$ is the theta angle.
The Coulomb branch vacua are found by solving
$\partial_{\sigma}\widetilde{W}_{\it eff}\equiv
0$ (mod $2\pi i\Z$), i.e.,
\beq
\sigma^{N-2}=4\e^{-t}.
\label{Cv}
\eeq
The detail of the theory depends on $N$.
Let us begin with the case $N=2$, where the axial $U(1)$ R-symmetry is
anomaly free and the FI parameter $r$ does not run.
We expect that the Witten index is constant if we move
$r$ from $r\ll 0$ to $r\gg 0$ as long as we avoid the point
$\e^t=4$ (which supports a non-compact Coulomb branch).
At $r\gg 0$, we have the sigma model on the quadric $Q^0=\{x_1^2+x_2^2=0\}$,
which is the set of two points. Its Witten index is of course 2. Hence
our LG orbifold at $r\to -\infty$ should also have Witten index $2$.
Let us next discuss the case $N=1$, in which
the FI parameter runs from negative to positive under the
renormalization group (toward longer distances). The theory describes
a flow from the LG orbifold.
The quadric $Q^{-1}=\{x_1^2=0\}$ is empty, but
we have a single Coulomb branch vacuum at $\sigma=\e^t/4$.
Thus the theory has a unique supersymmetric ground state.
Finally, we discuss the case $N>2$ where the FI parameter runs from 
positive to negative. 
The theory is a flow from the sigma model on $Q^{N-2}$ to our LG orbifold
 or to one of the $N-2$ Coulomb branch vacua.
Hodge number $h^{i,j}$ of the quadric $Q^{N-2}$
is (see, for example \cite{alggeom})
\beq
\mbox{$N$ even:}~
\left\{
\begin{array}{ll}
1&\mbox{$i=j\ne {N-2\over 2}$}\\
2&\mbox{$i=j={N-2\over 2}$}\\
0&\mbox{otherwise,}
\end{array}\right.
\qquad
\mbox{$N$ odd:}~
\left\{
\begin{array}{ll}
1&\mbox{$i=j$}\\
0&\mbox{otherwise.}
\end{array}\right.
\eeq
In particular,
the total number of supersymmetric ground states
of the sigma model is 
$N$ for even $N$ and $N-1$ for odd $N$.
Subtracting the number $N-2$ of the Coulomb branch vacua,
we obtain $2$ for even $N$ and $1$ for odd $N$.
To summarize, for all $N$, the result of the linear sigma model
agrees with the result, e.g. (\ref{ind1}), for our $\Z_2$ orbifold.
This means that the $\Z_2$ orbifold that appears at the $r\to -\infty$
limit of the linear sigma model is the ``standard'' one in
our convention.

Next, let us consider another deformation. Instead of turning on the 
superpotential, we give twisted masses 
with twisted masses $0,\wtm_1,\ldots,\wtm_N$ to $p,x_1,\ldots,x_N$.
In the limit $r\to -\infty$, the theory
reduces to the $\Z_2$ orbifold where all $N$ fields have twisted masses.
In the $r\gg 0$ regime,
the classical vacuum equations read
\beqa
&&-2|p|^2+\sum_{i=1}^N|x_i|^2=r,\nn\\
&&\sigma p=(\sigma-\wtm_1)x_1=\cdots =(\sigma-\wtm_N)x_N=0,
\label{classV}\\
&&\bsigma p=(\bsigma-\overline{\wtm}_1)x_1=\cdots= 
(\bsigma-\overline{\wtm}_N)x_N=0.
\nn
\eeqa
 If the twisted masses are distinct,
there are $N$ solutions at $\sigma=\wtm_1,\ldots,\wtm_N$, each of
which breaks the gauge symmetry completely.
The vacuum equation on the Coulomb branch reads
$\partial_{\sigma}\widetilde{W}_{\it eff}=2\log(-2\sigma)-\sum_{i=1}^N
\log(\sigma-\wtm_i)-t
\equiv 0$, or
\beq
4\sigma^2=\e^t(\sigma-\wtm_1)\cdots (\sigma-\wtm_N).
\label{tmCvac}
\eeq
Let us first analyze the system with $N=1$, which describes a
flow from our $\Z_2$ orbifold.
The equation (\ref{tmCvac})
has two solutions, both of which go indeed to $\sigma=0$
in the ultra-violet limit $t\to -\infty$.
In the infra-red limit $t\to +\infty$, one solution goes to 
$\sigma=\wtm$
and the other solution goes away to infinity as $\sigma\sim \e^t/4$.
The former supports the vacuum at $r\gg 0$ corresponding to the single
solution to (\ref{classV}), 
while the latter is a Coulomb branch vacuum.
The theory indeed has two bosonic ground states.
Next, let us consider the case $N=2$ where $r$ does not run.
In the $r\gg 0$ regime, we observed two classical vacua solving (\ref{classV}).
Let us check if that is everything. 
The equation (\ref{tmCvac}) has two solutions (except at $\e^t=4$);
at $r\gg 0$ the two solutions are at $\sigma\sim \wtm_1$ and $\wtm_2$
and indeed correspond to the two classical vacua, and at
$r\to -\infty$ the two solutions both go to $\sigma=0$ which is the right
value for our $\Z_2$ orbifold theory.
Therefore, we did not miss anything and
can conclude that the Witten index is $2$ at any value of $t$ 
(except $\e^t=4$) and in particular at $t\to-\infty$.
For $N>2$, the $N$ solutions to (\ref{tmCvac}) are at $\sigma\sim \wtm_i$
in the ultra-violet limit $t\to +\infty$ and indeed correspond to the
$N$ classical vacua solving (\ref{classV}).
In the infra-red limit, two of them go to $\sigma\to 0$ while
other $(N-2)$ go away to infinity. 
This again confirms that our $\Z_2$ orbifold has two bosonic
supersymmetric ground states.
To summarize, for all $N$,
the result of the linear sigma model
agrees with the result for our $\Z_2$ orbifold.
This ought to be the case as we have already confirmed that
the $\Z_2$ orbifold that appears at $r\to -\infty$
limit of the linear sigma model is the ``standard'' one in
our convention.

\subsection{Corank 1 Degeneration --- Branched Double Cover Of $\CC$
Or Its Orbifolds}
\label{subsec:rr1}

Let us next study the LG orbifold of $(N+1)$ variables,
$x_1,\ldots,x_N$ and $z$,
 with the superpotential
\beq
W=zx_1^2+x_2^2+\cdots +x_N^2,
\label{W2}
\eeq
modulo the $\Z_2$ generated by
\beq
\tau:(z,x_1,\ldots, x_N)\longmapsto (z,-x_1,\ldots, -x_N).
\label{inv2}
\eeq
By the periodicity mentioned earlier, the low energy behaviour of
the theory depends only on $N$ mod $2$.
The superpotential is quadratic in $x_i$'s with coefficients depending on $z$.
At $z\ne 0$ it is non-degenerate, 
i.e. the Hessian matrix is of maximal rank,
but as $z$ approaches $0$ the rank goes down by $1$.
In the region where $|z|$ is large enough, 
the fields $x_i$ are massive and can be integrated out first.
This system of $x$'s, as we have learned, has a single ($N$ odd) 
or two ($N$ even) massive vacua.
Thus, in the region of $z$ away from the origin
$z=0$, we have the theory of the variable $z$ without potential 
if $N$ is odd. If $N$ is even, we have two copies of the 
free theory of $z$, that is, the sigma model
whose target space is a double cover of the $z$-plane.
Of course this argument does not tell anything about the
behaviour near $z=0$, which will be the main point of the discussion.

\subsection*{\sl Even $N$}

Let us first study the $N$ even case. We take $N=2$ for simplicity. 
We employ a certain deformation of
the linear sigma model introduced in the previous subsection:
the $U(1)$ gauge theory with four fields of the following charges
\beq
\begin{array}{cccc}
p&x_1&x_2&z\\
-2&1&1&0
\end{array}
\eeq
with the superpotential
\beq
W=p(zx_1^2+x_2^2).
\eeq
The vector and axial $U(1)$ R-symmetry exists and
the FI parameter $r$ does not run in this theory.
In the limit $r\to-\infty$, we recover the $\Z_2$ orbifold under discussion.
In the positive $r$ regime, we have a sigma model whose target space is 
the hypersurface 
\beq
zx_1^2+x_2^2=0
\label{eqn2}
\eeq
in $\CP^1\times \C=\{([x_1:x_2],z)\}$.
The equation (\ref{eqn2}) and the D-term constraint
$-2|p|^2+|x|^2=r>0$ requires that $x_1$ must be non-zero, and
we may use an inhomogeneous coordinate $\widetilde{z}=ix_2/x_1$.
The equation (\ref{eqn2}) then reads
\beq
z=\widetilde{z}^2.
\label{bdc}
\eeq
This means that 
$\widetilde{z}$ provides a global coordinate of the hypersurface.
In particular the hypersurface is the complex plane $\C$ as a complex manifold.
The metric is smooth everywhere and has an asymptotic form
$\dd s^2\propto |\wtz\dd\wtz |^2$ as $|\wtz|\to \infty$,
which shows that the Euler density integral is $-1$, i.e., 
the curvature is mostly negative.
We expect that the metric flattens under the renormalization group,
and the theory flows to the free conformal field theory of 
the variable $\wtz$.
This holds for any large positive values of $r$ and hence
for all values of $r$ by the absence of singularity except at a point
in the FI-theta parameter space.
Therefore we conclude that
{\it the LG orbifold for even $N$ flows in the infra-red limit 
to the free conformal field theory of a single complex variable $\widetilde{z}$
that is related to $z$ via (\ref{bdc}).}
Note that the $\wtz$-plane is indeed a double cover of the
$z$-plane in the region away from $z=0$. 
It is a branched double cover with the branch point $z=0$.
That there is a branched double cover was also argued in \cite{CDHPS}
using Berry's phase.

As in any other $\Z_2$ orbifold, 
our LG orbifold has the quantum $\Z_2$ symmetry.
If we take the orbifold by this symmetry, we must get back 
the LG model before the orbifold (see, for example \cite{Ginsparg}).
In that theory, the $(x_1,x_2)$ system for a given non-zero $z$
has a unique zero energy ground state with a mass gap. 
Thus, we expect a {\it single} 
cover of the $z$-plane at least away from $z=0$.
This is achieved only when the quantum $\Z_2$ acts on $\wtz$ as
\beq
\wtz\,\,\longmapsto\,\,-\wtz.
\label{qZ2}
\eeq
In particular,
{\it the variable $\wtz$ is a twist field of the orbifold theory}.

\subsection*{\sl Unfolding The $\Z_2$}

At this occasion,
let us discuss more about the LG model before the orbifold.
In the absence of orbifolding, 
the models for all $N$ are equivalent at low energies. 
Thus, we may assume $N=1$ where we write $x=x_1$.
From what we have just seen, we can say that it is dual to the orbifold 
of the free theory of $\wtz$ by (\ref{qZ2}).
However, as always, we need to specify
the orbifold action in the untwisted RR sector. 
We claim that it is the non-standard one:
{\it the LG model of variables $x$ and $z$ with superpotential
$W=zx^2$ flows in the infra-red limit to the free orbifold conformal
field theory $\C/\Z_2(-1)^{F_s}$}. Or equivalently, {\it it is dual to
the $\Z_2$ orbifold of the theory of two variables, 
$\wtz$ without mass and $\wty$ with a complex mass,
by the simultaneous sign flip $(\wtz,\wty)\mapsto (-\wtz,-\wty)$.}

This can be derived as follows.
We have seen that the LG orbifold
$(W=zx_1^2+x_2^2)/\Z_2$ is dual to the free theory of $\C=\{\wtz\}$
with the relation $z=\wtz^2$. Let us add one variable $\zeta$ and 
perturb the system by the superpotential $\Delta W=\zeta z=\zeta\wtz^2$.
This changes the free theory of $\wtz$ to the LG model with superpotential
$W=\zeta\wtz^2$.
In the LG orbifold side, we have the superpotential
$W=zx_1^2+x_2^2+z\zeta$. If we integrate out $z$, we obtain the constraint
$\zeta=-x_1^2$ and we are left with the orbifold theory of
$x_1$ and $x_2$ with superpotential $W=x_2^2$.
With the notation change $(\wtz,\zeta)\to (x,z)$ and
$(x_1,x_2)\to (i\wtz,\wty)$, this is the claimed duality.

Let us do some consistency checks.
First, let us give a complex mass to $\wtz$.
This changes the dual theory to the LG orbifold $(W=\wtz^2+\wty^2)/\Z_2$
which has {\it two} supersymmetric ground states.
In the LG side, this corresponds, under $z=\wtz^2$, 
to deforming the superpotential to $W=zx^2+z$.
We find two critical points, $(x,z)=(0,i)$ and $(0,-i)$,
which means that there are {\it two} supersymmetric ground states,
agreeing with the dual result.
Next, let us give a twisted mass to $\wtz$.
The dual orbifold theory, 
which has one field with a twisted mass and another with
a complex mass, has {\it one} supersymmetric ground state.
In the LG side, we give twisted masses associated with the symmetry
where $z$ has charge $2$ and $x$ has charge $-1$. 
The scalar potential is
\beq
U=\left|2zx^{}_{}\right|^2+\left|x^2\right|^2
+\left|-\wtm x{}^{}_{}\right|^2+\left|2\wtm z{}^{}_{}\right|^2.
\eeq
It has a classical vacuum at the origin $(x,z)=(0,0)$.
Let us see what happens when we turn off the superpotential $W=zx^2$,
i.e., turn off the first two terms.
The potential still has just one classical vacuum at the origin
---
the vacuum at the origin before turning off $W$
stays there, and no other vacuum comes in from infinity.
Thus, we expect that the number of ground states does not change
if we set $W=0$. We know that the $W=0$ theory has a unique
RR ground state and hence we expect that  
the number of supersymmetric ground states is {\it one} in the 
theory with $W=zx^2$ as well.
This is confirmed by an exact analysis in Appendix~\ref{app:SUSYQM}. 
We again find that the result matches with
the one in the dual.
If we had chosen the dual to be the one without the massive field
$\wty$, we would have faced a problem: the number of ground states would be
{\it one} ({\it resp}. {\it two}) if we give a complex ({\it resp}. twisted)
mass to $\wtz$, which does not match with the LG result.

\subsection*{\sl Odd $N$}

Let us next discuss the $N$ odd case. We take $N=3$
and employ a chain of duality and standard relations as follows.
We denote by ${\mathcal A}_N$ the system of the variables $(z,x_1,\ldots,x_N)$
with the superpotential (\ref{W2}) equipped with the 
symmetry (\ref{inv2}), 
by ${\mathcal A}_N/\tau$ its orbifold equipped with the quantum symmetry
$\widehat{\tau}$,
by ${\mathcal B}$ the system of one massless variable $\wtz$ and
one variable $\wty$ with a complex mass equipped with the
symmetry $\tau:(\wtz,\wty)\to (-\wtz,-\wty)$, and
by ${\mathcal C}$ the system of one massless variable $\wtz$ 
equipped with the symmetry $\tau:\wtz\to -\wtz$.
We denote by
${\mathcal H}_{{{}_{\rm NSNS}}_{g}}^{\rm inv}({\mathcal A})$
{\it resp}. ${\mathcal H}_{{{}_{\rm NSNS}}_{g}}^{\rm anti}({\mathcal A})$
the space of invariants {\it resp}. anti-invariants 
in the $g$-twisted NSNS sector of a system ${\mathcal A}$,
and similarly for the RR sectors.
Then, we have the following equalities,
\beq
\begin{array}{ccccccccc}
{\mathcal H}_{{{}_{\rm NSNS}}_1}^{\rm inv}({\mathcal A}_3)&\!\!=\!\!&
{\mathcal H}_{{{}_{\rm NSNS}}_1}^{\rm inv}({\mathcal A}_2)&\!\!=\!\!&
{\mathcal H}_{{{}_{\rm NSNS}}_1}^{\rm inv}({\mathcal A}_2/\tau)&\!\!=\!\!&
{\mathcal H}_{{{}_{\rm NSNS}}_1}^{\rm inv}({\mathcal B})&\!\!=\!\!&
{\mathcal H}_{{{}_{\rm NSNS}}_1}^{\rm inv}({\mathcal C})\\[0.2cm]
{\mathcal H}_{{{}_{\rm NSNS}}_{\tau}}^{\rm inv}({\mathcal A}_3)&\!\!=\!\!&
{\mathcal H}_{{{}_{\rm NSNS}}_{\tau}}^{\rm anti}({\mathcal A}_2)&\!\!=\!\!&
{\mathcal H}_{{{}_{\rm NSNS}}_{\widehat{\tau}}}^{\rm anti}
({\mathcal A}_2/\tau)&\!\!=\!\!&
{\mathcal H}_{{{}_{\rm NSNS}}_{\tau}}^{\rm anti}({\mathcal B})&\!\!=\!\!&
{\mathcal H}_{{{}_{\rm NSNS}}_{\tau}}^{\rm inv}({\mathcal C})\\[0.2cm]
{\mathcal H}_{{{}_{\rm RR}}_1}^{\rm inv}({\mathcal A}_3)&\!\!=\!\!&
{\mathcal H}_{{{}_{\rm RR}}_1}^{\rm anti}({\mathcal A}_2)&\!\!=\!\!&
{\mathcal H}_{{{}_{\rm RR}}_{\widehat{\tau}}}^{\rm inv}
({\mathcal A}_2/\tau)&\!\!=\!\!&
{\mathcal H}_{{{}_{\rm RR}}_{\tau}}^{\rm inv}({\mathcal B})&\!\!=\!\!&
{\mathcal H}_{{{}_{\rm RR}}_{\tau}}^{\rm inv}({\mathcal C})\\[0.2cm]
{\mathcal H}_{{{}_{\rm RR}}_{\tau}}^{\rm inv}({\mathcal A}_3)&\!\!=\!\!&
{\mathcal H}_{{{}_{\rm RR}}_{\tau}}^{\rm inv}({\mathcal A}_2)&\!\!=\!\!&
{\mathcal H}_{{{}_{\rm RR}}_1}^{\rm anti}({\mathcal A}_2/\tau)&\!\!=\!\!&
{\mathcal H}_{{{}_{\rm RR}}_1}^{\rm anti}({\mathcal B})&\!\!=\!\!&
{\mathcal H}_{{{}_{\rm RR}}_1}^{\rm inv}({\mathcal C})
\end{array}
\label{chain}
\eeq
The first equality comes from the relation
$({\mathcal A}_3,\tau)\cong ({\mathcal A}_2,(-1)^{F_s}\tau)$.
The second equality is the standard relation between an orbifold 
$({\mathcal A},\tau)$ and its
quantum symmetry orbifold $({\mathcal A}/\tau,\widehat{\tau})$ 
(see, e.g. \cite{Ginsparg}).
The third equality follows from the duality found above,
$({\mathcal A}_2/\tau,\widehat{\tau})
\cong ({\mathcal B},\tau)$.
The fourth equality comes from the relation
$({\mathcal B},\tau)\cong ({\mathcal C},(-1)^{F_s}\tau)$.
The conclusion is that {\it our LG orbifold
with $N=3$ (and hence for any odd $N\geq 1$) flows in the infra-red limit to
the free orbifold conformal field theory $\C/\Z_2$} (the standard one).
It matches with the expectation that
the theory is a free theory of $z=\wtz^2$ in the region away from
$z=0$.
Note that the twisted and untwisted sectors are exchanged in the RR sector.
This means that the quantum symmetry of the LG orbifold
corresponds in the dual orbifold $\C/\Z_2$
to the quantum symmetry combined with
the symmetry $(-1)^{\bf F}$ which is defined by
\beq
(-1)^{\bf F}=\left\{\begin{array}{ll}
1&\mbox{~in the NSNS sector}\\
-1&\mbox{~in the RR sector}.
\end{array}\right.
\eeq

\subsection{Corank 2 Degeneration --- Conifold With $r=0$ And $\theta=\pi$}
\label{subsec:rr2}

As the final example in this section,
we consider the $\Z_2$ LG orbifold
\beq
W=ax^2+2cxy+by^2
\label{W3}
\eeq
\beq
(x,y,a,b,c)\longmapsto (-x,-y,a,b,c).
\label{orb3}
\eeq
As long as $(a,b,c)$ is away from the
degeneration locus
\beq
ab=c^2,
\label{degloc1}
\eeq
the fields $(x,y)$ are massive an can be integrated out:
the result of Section~\ref{subsec:massive} tells us
that the sector of $(x,y)$ mod $\Z_2$ has two massive vacua.
That is, we have a double cover of the open subset
$ab\ne c^2$ of the $(a,b,c)$-space.
Near the degeneration locus (\ref{degloc1}) but away from the origin
$(a,b,c)=(0,0,0)$, we may find a coordinate change,
$(x,y)\to (x',y')$, so that the superpotential is expressed as
\beq
W=(c^2-ab)x^{\prime 2}+y^{\prime 2}.
\eeq
The result of Section~\ref{subsec:rr1} then tells us that
the double cover is branched at the locus (\ref{degloc1}), i.e.,
of the form
\beq
c^2-ab=d^2.
\label{conifold}
\eeq
The main question is the behaviour of the theory near the origin.
Note that the equation (\ref{conifold}) is the one for the conifold.
Thus we expect that the theory is related in some way
to that of the conifold.
This is also what is observed in \cite{CDHPS}. We would now like to know
the precise relation to the conformal field theory
associated with resolution or deformation of the conifold.

We consider
a $U(1)$ gauge theory with six fields of the following charges
\beq
\begin{array}{cccccc}
p&x&y&a&b&c\\
-2&1&1&0&0&0
\label{LSM2}
\end{array}
\eeq
with the superpotential
\beq
W=p(ax^2+2cxy+by^2).
\eeq
In the $r\to-\infty$ limit,
we recover the LG orbifold under question.
The theory is singular at the value of $t=r-i\theta$ where there is
a non-compact Coulomb branch. 
The latter exists when 
$t_{\it eff}=\partial_{\sigma}\widetilde{W}_{\it eff}$ 
vanishes modulo $2\pi i \Z$,
where
$\widetilde{W}_{\it eff}(\sigma)=-(-2\sigma)(\log(-2\sigma)-1)
-2\sigma(\log\sigma -1)-t\sigma$.
The singular point is therefore
\beq
\e^t=4.
\eeq
In the $r\gg 0$ regime, we have a sigma model
whose target space is the hypersurface
\beq
ax^2+2cxy+by^2=0
\eeq
in $\CP^1\times \C^3$ where $x,y$ are the homogeneous coordinates of 
the first factor $\CP^1$ and $(a,b,c)$ are the coordinates of 
the second factor $\C^3$.
This is indeed a resolved conifold:
There are two solutions for $(x,y)$ if $(a,b,c)$ is away from
(\ref{degloc1}) and one solution if it is at (\ref{degloc1})
except at the origin $(a,b,c)=(0,0,0)$
where arbitrary $(x,y)$ solves the equation.
That is, the entire $\CP^1$ sits on the hypersurface at
the origin of $\C^3$.
More explicitly, if we set
\beq
d=\left\{\begin{array}{ll}
ax/y+c&y\ne 0\\
-by/x-c&x\ne 0,
\end{array}\right.
\eeq
then, $a,b,c,d$ satisfy the conifold equation (\ref{conifold}).
Thus, we conclude that the LG orbifold (\ref{W3})-(\ref{orb3})
belongs to a one parameter family of theories that also includes
a large volume limit of the resolved conifold.

At this point, we recall that there is another 
one parameter family that includes a large volume limit
of the resolved conifold --- in fact {\it two} large volume limits. 
It is obtained from the following $U(1)$ gauge theory with 
vanishing superpotential with the following matter content:
\beq
\begin{array}{cccc}
u_1&u_2&v_1&v_2\\
1&1&-1&-1
\end{array}
\label{LSM3}
\eeq
This one parameter family has one singular point
\beq
\e^t=1.
\label{sing2}
\eeq
$r\gg 0$ and $r\ll 0$ are the two large volume regimes.
If we set
\beq
a=u_1v_1,\quad
b=u_2v_2,\quad
c={u_1v_2+u_2v_1\over 2},\quad
d={u_1v_2-u_2v_1\over 2}
\label{u1inv}
\eeq
then, $a,b,c,d$ obey the relation (\ref{conifold}).

Now let us ask whether the LG orbifold (\ref{W3})-(\ref{orb3})
belongs to the second family. We propose that {\it it is
the theory at $\e^t=-1$.}
We give two evidences for this proposal. One is 
existence of a discrete symmetry. 
What is special about the theory at $\e^t=-1$ is that it has
an extra $\Z_2$ symmetry.
Let us consider the transformation
\beq
(u_1,u_2,v_1,v_2,\Vct)\longmapsto 
(v_1,v_2,u_1,u_2,-\Vct),
\label{Z2}
\eeq
where $\Vct$ is the vector superfield for the $U(1)$ gauge symmetry.
This reverses the FI-theta parameter,
$r\to- r$, $\theta\to -\theta$, and hence is
a symmetry of the theory only at $(r,\theta)=(0,0)$ and $(0,\pi)$.
But $(r,\theta)=(0,0)$ is the singular point, see (\ref{sing2}).
Thus, the theory at $(r,\theta)=(0,\pi)$ (i.e. $\e^t=-1$) is the only
regular theory that possesses the $\Z_2$ symmetry.
Note that it acts on the $U(1)$ invariants (\ref{u1inv}) as
\beq
a\to a, \quad
b\to b,\quad
c\to c,\quad
d\to -d.
\label{Z2p}
\eeq
We propose to identify this $\Z_2$ symmetry with the quantum $\Z_2$
symmetry of the LG orbifold (\ref{W3})-(\ref{orb3}).
Indeed, as discussed in Section~\ref{subsec:rr1},
the quantum symmetry acts as the exchange of the two 
sheets over the $(a,b,c)$ space
away from the degeneration locus (\ref{degloc1}). That is, it must 
exchange the two solutions for $d$ of the equation (\ref{conifold}),
which is nothing but the transformation (\ref{Z2p}).

To provide another evidence, let us discuss
the relation between the two families.
We write $t_1$ ({\it resp}. $t_2$) for the FI-theta parameter of the
first ({\it resp}. second) family. Recall that the singular points are at
$\e^{t_1}=4$ in the first family and $\e^{t_2}=1$ at the second family.
We first find the relation between $t_1$ and $t_2$ by
{\it assuming} the proposal. 
We expect that the relation is generically one to two, where
one value of $t_1$ corresponds to two values of $t_2$ related by
$t_2\to -t_2$. This requires the relation of the form 
$\e^{t_1}=f(\e^{t_2}+\e^{-t_2})$ for some rational function $f(x)$
of degree 1.
The relation between the singular points, large volume limits, and
the $\Z_2$ symmetric points requires the function $f(x)$ to satisfy
$f(2)=4$, $f(\infty)=\infty$ and $f(-2)=0$ respectively.
This fixes the relation as
\beq
\e^{t_1}=\e^{t_2}+\e^{-t_2}+2.
\label{t1t2}
\eeq
The main point of the second evidence is that this relation can be
supported by the Picard-Fuchs equation for the central charges of
B-type D-branes.
Recall that the central charge is expressed as the period integral 
of some differential form for the corresponding A-branes
in the mirror system, and satisfies Picard-Fuchs differential equation
\cite{Givental,HV}.
The mirror for the second family is known and the equation reads as
\beq
{\dd^2\over \dd t_2^2}\Pi=0.
\label{PF2}
\eeq
The mirror for the first family is not known but the dualization of
the charged sector as in \cite{HV} leads to the following equation
\beq
{\dd^2\over \dd t_1^2}\Pi=\e^{-t_1}\left(2{\dd\over \dd t_1}-1\right)
2{\dd \over\dd t_1}\Pi.
\label{PF1}
\eeq
The two equations (\ref{PF2}) and (\ref{PF1}) are 
equivalent provided that $t_1$ and $t_2$ are related by (\ref{t1t2}). 
This is a strong support for the relation (\ref{t1t2}), and in particular
the proposed identification of the LG orbifold 
 (\ref{W3})-(\ref{orb3}) as the theory of the second family at 
$\e^{t_2}=-1$.

\section{$O(2)$ Theories}
\label{sec:O2}

The group $O(2)$ is isomorphic to the semi-direct product
\beq
O(2)\cong SO(2)\rtimes \Z_2,
\eeq
where $\Z_2$ is generated by the reflection $\tau$ with respect to
the first axis,
\beq
\tau=\left(\begin{array}{cc}
1&0\\
0&-1
\end{array}\right)
\eeq
which acts on $SO(2)\cong U(1)$ 
by the group inversion.
By this, an $O(2)$ gauge theory can be regarded as
a $\Z_2$ orbifold of a $U(1)$ gauge theory.
For example, if the $O(2)$ theory consists of $N$ fields in the
fundamental representation 
(i.e. the doublet, ${\bf 2}$), $x_i=(x_i^1,x_i^2)^T$ 
($i=1,\ldots,N$), then,
$u_i=x_i^1+ix_i^2$ and $v_i=x_i^1-ix_i^2$ 
have $U(1)$ charges $+1$ and $-1$ respectively, 
and the generator $\tau$ of the orbifold group
acts on the fields as
\beq
\tau:(u_1,v_1,\ldots,u_N,v_N,\Vct)\longmapsto 
(v_1,u_1,\ldots,v_N,u_N,-\Vct).
\label{O2orb}
\eeq
To be precise, 
the $U(1)$ gauge theory is specified only when its FI-theta parameter
$t=r-i\theta$ is specified. 
But (\ref{O2orb}) reverses its sign, $t\to -t$.
Therefore, it is a symmetry only at $t=0$ or $t=\pi i$ (mod $2\pi i \Z$).
But one of them is a singular point in the one parameter family of
$U(1)$ theories because of the emergence of a non-compact Coulomb branch.
By looking at the effective twisted superpotential, we find that
the singular point is $t=\pi i N$ (mod $2\pi i \Z$).
We decide not to consider such a theory with a non-compact flat direction.
Thus, we take the other value
\beq
t=\pi i (N+1)\qquad\mbox{mod $2\pi i \Z$}.
\label{FIO2}
\eeq
This applies whether or not there is a superpotential for the
matter fields.
When $N$ is odd, we can take $t=0$ as the tree level FI-theta parameter.
When, $N$ is even, we should take $t=\pi i$ (mod $2\pi i\Z$).
Alternatively, in the latter case, we may take $t=0$
but introduce one additional doublet $x_{N+1}$ with a superpotential
$W={m\over 2}(x_{N+1}x_{N+1})$, as a regulator to prevent the
singularity due to Coulomb branch. 
This works no matter how the mass $m$ is large.

As always, we need to and will
specify the $\Z_2$ orbifold action on the space of states. 
As an important point, the $\Z_2$ action depends on the choice 
of ``regularizations'' for the even $N$ case
--- (i) setting $t=\pi i$ without introducing $x_{N+1}$
or (ii) setting $t=0$ while introducing $x_{N+1}$.
Switching from one to the other has an effect of dressing
the generator by $(-1)^{F_s}$ as we have learned.

\subsection{Yang-Mills Theory And QCD With Complex Mass
 --- Supersymmetry Breaking}
\label{subsec:O2reg}

Let us first study the theory without a matter field, 
i.e., the $O(2)$ ``Yang-Mills'' theory. According to our definition, it is
the orbifold of the Maxwell theory with $r=0$ and
$\theta=\pi$ by the $\Z_2$ that flips the sign of the vector multiplet
fields $\Vct=(v_{\mu},\sigma,\lambda)$. (Definition using a regulator field
will be included in the massive QCD below.)
We formulate the theory on the circle of length $L$.
Let us first consider the untwisted RR sector, where we impose the periodic
boundary condition on all fields.
The energy spectrum from the $U(1)$ gauge field $v_{\mu}$ is
\beq
E_n={e^2L\over 2}\left(-{1\over 2}+n\right)^2,\qquad
n\in\Z,
\eeq
where $e$ is the gauge coupling.
This is interpreted as the energy from the electric field 
$e^2\left(-{1\over 2}+n\right)$, where $-e^2/2$ 
is the background value associated 
with $\theta=\pi$ \cite{Coleman} and $e^2 n$ is from the conjugate
momentum for the Wilson line.
The remaining degrees of freedom, $\sigma$ and
$\lambda$, has the usual free massless Lagrangian.
Thus the states of the Maxwell theory is decomposed into sectors labeled by
$n\in \Z$.
The orbifold generator $\tau$ must reverse the electric field,
$e^2\left(-{1\over 2}+n\right)\to -e^2\left(-{1\over 2}+n\right)$.
Thus, it acts on the sectors as
\beq
\tau:(n,\sigma,\lambda)\longmapsto (-n+1,-\sigma,-\lambda).
\eeq
Note that the sectors are permuted and none of them is invariant.
In particular, there is no subtlety concerning the definition
of the $\Z_2$ orbifold action.
The orbifold theory may be identified as the sum of
sectors with the label $n$ running only over
\beq
n=0,1,2,3,\ldots.
\eeq
All the states have strictly positive energies.
Let us next consider the twisted RR sector, where we impose
the anti-periodic boundary condition for all fields.
Gauss law constraint requires that the electric field
is constant. Together with anti-periodicity, this means that
the electric field is constantly vanishing.
However, this is in conflict with the definition 
of $\theta=\pi$ as providing the background electric field
$e^2/2$ (mod $e^2\Z$). That is, there is no twisted
sector for this choice of the theta angle.
To summarize, the theory has no zero energy state. In particular
the supersymmetry is spontaneously broken.

\begin{figure}[htb]
\psfrag{a}{$-{L\over 2}$}
\psfrag{b}{${L\over 2}$}
\psfrag{e}{${e^2\over 2}$}
\psfrag{f}{$F_{01}$}
\psfrag{x}{$x^1$}
\centerline{\includegraphics{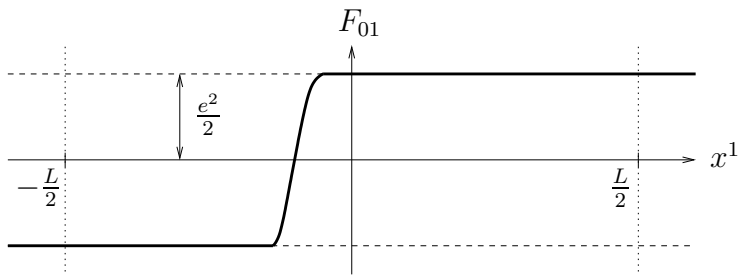}}
\caption{A consistent profile of the electric field in the twisted sector}
\label{fig:electric}
\end{figure}

Let us next consider ``QCD'' with quarks with complex masses, i.e.,
the theory with $N$ fundamental matter fields, $x_1,\ldots,x_N$,
with a non-degenerate quadratic superpotential, say,
\beq
W=(x_1x_1)+\cdots+(x_Nx_N).
\eeq
Since all fields are massive, they can be integrated out
and we are left with the $O(2)$ ``Yang-Mills'' theory.
Note that the theta angle for the $U(1)$ has changed from the
ultra-violet value (\ref{FIO2}) to the infra-red value
$\theta_{\it eff}=\pi$.
The treatment of the untwisted RR sector is as in the Yang-Mills theory.
In particular, there is no zero energy state there.
Unlike in the pure Yang-Mills,
the theory does have a twisted RR sector.
Gauss law constraint now takes the form
\beq
\partial_1\left({1\over e^2}F_{01}\right)=j^0
\eeq
where $j^0$ is the charge density. 
We may have a configuration 
as depicted in Figure~\ref{fig:electric}
that is compatible with Gauss law constraint, the anti-periodic
boundary condition, and $\theta_{\it eff}=\pi$.
All of such configurations have strictly positive energies.
Hence, the supersymmetry is spontaneously broken.

\subsection{Twisted Mass --- Defining The $\ZZ_2$ Orbifold}
\label{subsec:O2tmass}

Let us next discuss the $O(2)$ gauge theory with $N$ doublets
$x_1,\ldots, x_N$ now with twisted masses $\wtm_1,\ldots,\wtm_N$.
Our focus will be the spectrum of supersymmetric ground states.
Through the course of the analysis, we
specify the precise definition of the $\Z_2$ orbifold (for the massless
theory as well).

The classical vacuum equation reads as follows:
\beqa
&&\sum_{i=1}^N|u_i|^2=\sum_{i=1}^N|v_i|^2,\nn\\
&&(\sigma-\wtm_i)u_i=(-\sigma-\wtm_i)v_i=0,\quad\forall i,\\
&&(\bsigma-\overline{\wtm}_i)u_i=(-\bsigma-\overline{\wtm}_i)v_i
=0,\quad\forall i.\nn
\eeqa
We choose $\wtm_i$ to be generic. 
In particular, we assume $\wtm_i+\wtm_j\ne 0$ for all $i$ and $j$.
Then, there is no value of $\sigma$ at which 
both $u$ and $v$ can be non-zero, and hence $u=v=0$ is enforced by the first
equation.
In particular, the doublets are all massive at every value of $\sigma$.
Hence we can integrate them out and
study the effective theory for the vector multiplet.
The effective twisted superpotential for $\sigma$ is
\beqa
\widetilde{W}_{\it eff}
&=&
-\sum_{i=1}^N(\sigma-\wtm_i)(\log(\sigma-\wtm_i)-1)
-\sum_{i=1}^N(-\sigma-\wtm_i)(\log(-\sigma-\wtm_i)-1)\nn\\
&&+\pi i (N+1)\sigma,
\label{tweff}
\eeqa
and the vacuum equation reads
\beq
\prod_{i=1}^N(\sigma-\wtm_i)=(-1)^{N+1}\prod_{i=1}^N(-\sigma-\wtm_i).
\label{veqq}
\eeq
The equation is of order $N$ and is
symmetric under the $\Z_2$ orbifold action
\beq
\sigma~\longmapsto\,\, -\sigma.
\label{orbtmass}
\eeq
If $\wtm_i$ are generic and in particular $\wtm_i+\wtm_j\ne 0$,
the solutions are distinct and
are away from the forbidden region $\sigma=\pm \wtm_i$
where (\ref{tweff}) cannot be trusted.

When $N$ is even, there are ${N\over 2}$ pairs of non-zero solutions.
Since these solutions break the $\Z_2$ orbifold symmetry (\ref{orbtmass}),
there no need to consider the twisted sector, nor enters the subtlety of
defining the $\Z_2$ orbifold.
Thus, quite simply, 
there are ${N\over 2}$ supersymmetric ground states.

When $N$ is odd, there are 
${N-1\over 2}$ pairs of non-zero solutions, and
one solution at $\sigma=0$ at which the orbifold group is unbroken.
There are ${N-1\over 2}$ supersymmetric ground states from
the $\Z_2$ breaking solutions.
The main issue is the spectrum at the $\Z_2$ symmetric solution $\sigma=0$.
The effective superpotential near that solution is
\beq
\widetilde{W}_{\it eff}=\left({1\over \wtm_1}+\cdots+{1\over\wtm_N}\right)
\sigma^2+\cdots
\eeq
where the ellipses stand for a constant and higher order terms.
Even though we are considering the $U(1)$ gauge multiplet,
we can effectively treat the system as the $\Z_2$ orbifold of just
a single twisted chiral multiplet of this superpotential.
Before the orbifold projection there are two supersymmetric ground states,
one twisted and one untwisted.
As always, the one in the twisted sector is invariant and survives
the orbifold projection. On the other hand, we must make a choice
concerning the projection in the untwisted sector.
We shall call the orbifold ``standard'' if 
the untwisted RR ground state is invariant and survives
and ``non-standard'' if it is anti-invariant and is projected out.
We shall denote the corresponding gauge group
$\Ost(2)$ for the standard orbifold and $\Ons(2)$ for the non-standard one.
Under this definition, the total number of ground states is
\beq
\left\{\begin{array}{ll}
\displaystyle
{N-1\over 2}+2={N+3\over 2}&\mbox{in $\Ost(2)$ theory}\\[0.3cm]
\displaystyle
{N-1\over 2}+1={N+1\over 2}&\mbox{in $\Ons(2)$ theory}.
\end{array}\right.
\label{nO2t}
\eeq
This definition of orbifolds for odd $N$ can be extended, by continuity, 
to the theories where the twisted masses are turned off
and then the superpotential is turned on.
In particular, we have a definition for the theory with even $N$
plus an additional doublet with a complex mass (the ``regulator'').

\subsection*{\sl Embedding Into Linear Sigma Models}

This definition
can be compared with a geometrical setting where the notion 
of canonical or non-canonical orbifolds already exists 
(see Section~\ref{subsec:grZ2}).
This is done via a linear sigma model.
Let us consider for odd $N$ the theory with gauge group 
$(U(1)\times O(2))/\{(\pm 1,\pm{\bf 1}_2)\}$ consisting of
a field $p$ in the representation $(-2,{\bf 1})$ with zero twisted mass
and fields $x_1,\ldots, x_N$ in the representation $(1,{\bf 2})$
with twisted masses $\wtm_1,\ldots,\wtm_N$.
In the regime where the FI parameter $r_{U(1)}$ for the $U(1)$ factor
is negative, the field $p$ must have a non-zero value
and breaks the $U(1)$ to $\{\pm 1\}$, that is,
breaks the gauge group to simply $O(2)$.
In the limit $r_{U(1)}\to -\infty$, the theory reduces to
the $O(2)$ gauge theory we are discussing.
We would like to study this linear sigma model and in particular look
what we have at the other regime $r_{U(1)}\gg 0$.

We can write the gauge group as
$(U(1)\times U(1))/\{(\pm 1,\pm 1)\}\rtimes \Z_2$.
The FI-theta parameter of the first $U(1)$ is unconstrained 
but the one for the second
$U(1)$ must be zero since we have an odd number of doublets.
We shall reparametrize the continuous part of the group as
\beqa
{U(1)\times U(1)\over\{(\pm 1,\pm 1)\}}&\cong& U(1)_1\times U(1)_2.
\label{paramsigma}\\
{[}(g,h){]}~~&\longmapsto&~~(gh,gh^{-1})\nn
\eeqa
The FI-theta parameters must be equal between $U(1)_1$ and $U(1)_2$
and are denoted by $t=r-i\theta$;
$r$ is a free parameter of the theory when $N=1$ 
while it runs from positive to negative when $N\geq 2$.
The matter fields are $p$ of charge $(-1,-1)$,
$u_i$'s of charge $(1,0)$ and $v_i$'s of charge $(0,1)$ with respect to
$U(1)_1\times U(1)_2$. The symmetry $\tau$ 
acts as the exchange of $U(1)_1$ and $U(1)_2$
as well as $u$'s and $v$'s.

At $r\gg 0$, the classical vacuum equations for the scalar fields are
\beqa
&&-|p|^2+|u|^2=-|p|^2+|v|^2=r,\nn\\
&&(\sigma_1-\wtm_i)u_i=(\sigma_2-\wtm_i)v_i=0~\,\,\forall i,\quad
(\sigma_1+\sigma_2)p=0,\\
&&(\bsigma_1-\overline{\wtm}_i)u_i=(\bsigma_2-\overline{\wtm}_i)v_i=0
~\,\,\forall i,\quad
(\bsigma_1+\bsigma_2)p=0,
\eeqa
where $\sigma_1$ and $\sigma_2$ are the scalar components of 
the vector multiplets for $U(1)_1$ and $U(1)_2$.
If $\wtm_i$ are distinct, there are $N^2$ solutions:
$(\sigma_1,\sigma_2)=(\wtm_{i_1},\wtm_{i_2})$
with $|u_{i_1}|^2=|v_{i_2}|^2=r$ and all other $u$'s and $v$'s and $p$ are zero.
Note that the $\Z_2$ orbifold group
is broken at the $N^2-N$ solution with $i_1\ne i_2$
while unbroken at the $N$ solutions with $i_1=i_2$.
Each pair of broken solutions yields one supersymmetric ground state
while each unbroken solution yields
two vacua (one twisted and one untwisted) for the canonical
orbifold and one twisted vacuum for the non-canonical one,
in the sense of Section~\ref{subsec:grZ2}.
Thus, the total number of vacua is
\beq
\left\{\begin{array}{ll}
\displaystyle
{N^2-N\over 2}+2N&\mbox{for the canonical orbifold},\\[0.3cm]
\displaystyle
{N^2-N\over 2}+N&\mbox{for the non-canonical orbifold}.
\end{array}\right.
\label{result1}
\eeq
If $\wtm_i$ are not distinct, say all equal (write it $\wtm$),
then, there is a continuum of solutions at $\sigma_1=\sigma_2=\wtm$
with $|u|^2=|v|^2=r$ and $p=0$. The solution space is
$\CP^{N-1}\times\CP^{N-1}$ on which the $\Z_2$ orbifold group acts as
the exchange of the two $\CP^{N-1}$ factors.
The supersymmetric ground states in the untwisted sector are in one to one 
correspondence with invariant {\it resp}. anti-invariant
 cohomology classes of $\CP^{N-1}\times\CP^{N-1}$
for the canonical {\it resp}. non-canonical orbifold.
If $H_1$ and $H_2$ denote the hyperplane classes of 
the first and the second $\CP^{N-1}$ factor, invariant {\it resp}.
anti-invariant cohomology classes are
of the form $H_1^iH_2^j+H_1^jH_2^i$ {\it resp}. $H_1^iH_2^j-H_1^jH_2^i$, 
with $0\leq i,j\leq N-1$.
An elementary count finds that the total number of such classes is
${N^2+N\over 2}$ {\it resp}. $N^2-N\over 2$.
On the other hand, the twisted sector ground states
are in one to one correspondence with the cohomology classes of the diagonal
$\CP^{N-1}$ (there are $N$ of them), for both orbifolds.
Thus, the total number of ground states is
\beq
\left\{\begin{array}{ll}
\displaystyle
{N^2+N\over 2}+N&\mbox{for the canonical orbifold},\\[0.3cm]
\displaystyle
{N^2-N\over 2}+N&\mbox{for the non-canonical orbifold}.
\end{array}\right.
\label{result2}
\eeq
The result of course matches with (\ref{result1}), including the separation
into twisted and untwisted sectors.

In order to compare this result with the $O(2)$ gauge theory at $r\to-\infty$,
let us look into the Coulomb branch vacua.
The effective twisted superpotential is
$\widetilde{W}_{\it eff}=
-(-\sigma_1-\sigma_2)(\log(-\sigma_1-\sigma_2)-1)
-\sum_{a,i}(\sigma_a-\wtm_i)(\log(\sigma_a-\wtm_i)-1)
-t(\sigma_1+\sigma_2)$,
and the extremum equation reads,
\beq
\prod_{i=1}^N(\sigma_1-\wtm_i)
=\prod_{i=1}^N(\sigma_2-\wtm_i)=-\e^{-t}(\sigma_1+\sigma_2).
\eeq
The are $N^2$ solutions --- $N$ of them have $\sigma_1=\sigma_2$ 
($\Z_2$ preserving)
and $N^2-N$ of them have $\sigma_1\ne\sigma_2$ ($\Z_2$ breaking).
In the limit $r\to +\infty$, the solutions behave as
$\sigma_1\to\wtm_{i_1}$ and $\sigma_2\to \wtm_{i_2}$
for some $i_1$ and $i_2$. Thus, they all correspond to the classical
vacua at $r\gg 0$ studied above.
At $r\to -\infty$, some of the solutions have $\sigma_1+\sigma_2\to 0$
and correspond to vacua of the $O(2)$ gauge theory under consideration,
while the others have divergent values of $\sigma_1+\sigma_2$
and have nothing to do with the $O(2)$ gauge theory.
Among the $\Z_2$ preserving solutions,
one of them goes to $(0,0)$ as 
$\sigma_1=\sigma_2\sim -{1\over 2}\e^t\prod_{i=1}^N(-\wtm_i)$,
while the other $N-1$ diverge as $\sigma_a^{N-1}\sim -2\e^{-t}$.
Among the $\Z_2$ breaking solutions,
$(N-1)^2$ of them are divergent as
$\sigma_2\sim\omega\sigma_1$ ($\omega^N=1, \omega\ne 1$)
and $\sigma_1^{N-1}\sim -\e^{-t}(1+\omega)$, while the rest
($N-1$ solutions) are finite and hence has $\sigma_1+\sigma_2\to 0$
in the limit $r\to-\infty$. 
Among the ground states in (\ref{result1}) or (\ref{result2}), 
those from the solutions with $\sigma_1+\sigma_2\to 0$ 
in the limit $r\to -\infty$ are
\beq
\left\{\begin{array}{ll}
\displaystyle
{N-1\over 2}+2&\mbox{for the canonical orbifold},\\[0.3cm]
\displaystyle
{N-1\over 2}+1&\mbox{for the non-canonical orbifold}.
\end{array}\right.
\label{result3}
\eeq
This is in perfect agreement with (\ref{nO2t}), provided that
the ``standard'' $\Ost(2)$ theory corresponds to the canonical orbifold
in the geometric setting at $r\gg 0$
and ``non-standard'' $\Ons(2)$ theory corresponds to
the non-canonical orbifold.

\subsection{Massless QCD --- A Dual Description}
\label{subsec:O2dual}

\subsection*{\sl Single Flavor}

Next, let us consider the massless ``QCD'' with one flavor, i.e.,
the theory with a single fundamental matter field
$x=(x^1,x^2)^T$ with no superpotential nor twisted mass.
It is a $\Z_2$ orbifold of the $U(1)$ gauge theory with $r=\theta=0$
consisting of fields $u$, $v$ of charges $1$, $-1$.
We have a single generator of the $O(2)$ invariants
\beq
a=(xx):=(x^1)^2+(x^2)^2=uv,
\eeq
which is also the generator of the $U(1)$ invariants.

Let us first study the $U(1)$ theory before the $\Z_2$ orbifold.
It has no Coulomb branch since
the effective potential at large $|\sigma|$ is $e^2\pi^2/2$ by
the one loop theta angle.
Let us study the Higgs branch.
The D-term equation reads
\beq
|u|^2=|v|^2.
\eeq
We find a one-dimensional Higgs branch, parametrized by the invariant 
$a=uv$, whose metric is 
classically of the form $\dd s^2=|\dd a|^2/|a|$. It is the cone
$\C/\Z_2$ and has a conical singularity
at the origin $a=0$. The singularity appears because the metric 
is obtained by integrating out the gauge field 
which is classically massless at $a=0$.
However, the Coulomb branch is lifted by the non-zero value of the
effective theta angle and
the gauge field may not be massless at $a=0$.
We propose from this that the conical singularity is smeared 
in the quantum theory and that the theory 
flows in the infra-red limit to
the free conformal field theory of the single complex variable $a$.
That is, the sigma model whose target space is the complex $a$-plane $\C$.
Relevant mathematical fact is that the ring of $\C^{\times}$-invariant
polynomials of $u$ and $v$ is generated by the invariant $a=uv$, 
which obeys no relation,
\beq
\C[u,v]^{\C^{\times}}=\C[a].
\eeq
The situation is quite similar to the case of
$SU(k)$ QCD with $(k+1)$ flavors \cite{HoTo}, where
it was argued that it is the free theory of baryon variables.
Relevant mathematical fact there was again that
the the ring of $SL(k,\C)$-invariant polynomials of $(k+1)$ fundamentals
is the polynomial ring (with no relation) of the baryonic variables
\cite{Weyl}.

Let us come back to the $O(2)$ theory, which
is obtained by taking a $\Z_2$ orbifold of the $U(1)$ theory.
In the effective description obtained above,
the symmetry $\tau$ acts trivially on the variable $a$.
Thus, the orbifold is either the sigma model on the disjoint union
of two copies of $\C=\{a\}$ or $\C=\{a\}$ itself 
(see Section~\ref{subsec:grZ2}).
We claim that our standard choice yields the former
and the non-standard one the latter,
\beq
\begin{array}{ccc}
\mbox{\it $\Ost(2)$ with one massless doublet}&\longrightarrow&
\mbox{$\C\sqcup\C$}\\[0.2cm]
\mbox{\it $\Ons(2)$ with one massless doublet}&\longrightarrow&
\mbox{$\C$}.
\end{array}
\label{rg1}
\eeq
This can be shown by the embedding into the linear sigma model discussed in
Section~\ref{subsec:O2tmass}:
If we set $\wtm_1=0$ in the $N=1$ model, then the $r\gg 0$
theory is the $\Z_2$ orbifold of the sigma model whose target space is
the total space of the line bundle ${\mathcal O}(-1,-1)$ over
``$\CP^0\times \CP^0$'' (i.e. a complex line over a point),
where the orbifold acts trivially.
The $\Ost(2)$ theory corresponds, as we have learned, 
to the canonical orbifold
and hence to two copies of the line $\C$ while
the $\Ons(2)$ corresponds to the non-canonical one
and to a single copy of $\C$.

\subsection*{\sl Two Flavors}

Let us next consider the massless ``QCD'' with two flavors, i.e.,
the theory with two fundamentals, $x=(x^1,x^2)^T$ and $y=(y^1,y^2)^T$,
having no superpotential nor twisted mass.
It is defined as a $\Z_2$ orbifold of the $U(1)$ gauge theory
with $r=0$ and $\theta=\pi$ consisting of fields $u_1,u_2,v_1,v_2$ of charge 
$1,1,-1,-1$.
The theory before the orbifold is something
which we have already encountered in the present paper ---
it is the theory which is identified with the $\Z_2$ LG orbifold 
(\ref{W3})-(\ref{orb3}) discussed in Section~\ref{subsec:rr2}.
Furthermore, the $\Z_2$ orbifold symmmetry 
is identified as the quantum symmetry of the $\Z_2$ LG orbifold.
To be precise, there are two distinct orbifolds, corresponding to
$\Ost(2)$ or $\Ons(2)$, and only one of them has that
property. Without identifying which is the one,
let us proceed for now assuming that the $\Z_2$ orbifold symmetry
does correspond to the quantum $\Z_2$ of the LG orbifold 
(\ref{W3})-(\ref{orb3}).
Then, the orbifold theory is the LG model before the orbifold
i.e., (changing the notation to avoid possible confusion)
the LG model of five variables $\wtx,\wty,\wta,\wtb,\wtc$
with the superpotential
\beq
W=\wta\wtx^2+2\wtc\wtx\wty+\wtb\wty^2.
\label{dual1}
\eeq
The relation to the $O(2)$ invariants $(xx)$, $(xy)$, $(yy)$ are
\beq
\wta=u_1v_1=(xx),\qquad
\wtb=u_2v_2=(yy),\qquad
\wtc={u_1v_2+u_2v_1\over 2}=(xy).
\eeq
The quantum $\Z_2$ symmetry (for the orbifold by the $\Z_2=O(2)/SO(2)$) 
acts on the dual variables as
\beq
(\wtx,\wty,\wta,\wtb,\wtc)\longmapsto
(-\wtx,-\wty,\wta,\wtb,\wtc).
\label{qqact}
\eeq
This is because the quantum symmetry of the quantum symmetry orbifold
is the original orbifold symmetry, 
which was (\ref{orb3}) in the previous notation.
In particular, the variables $\wtx$ and $\wty$ are twist fields
with respect to the $\Z_2=O(2)/SO(2)$.

To find which of $\Ost(2)$ or $\Ons(2)$ we are discussing,
let us perturb the system by giving a mass $m$ to one of the two
fundamentals, say $y$, by the tree level superpotential
$W=m(yy)$. 
In the dual theory, this corresponds to deforming the superpotential to
\beq
W=\wta\wtx^2+2\wtc\wtx\wty+\wtb\wty^2+m\wtb.
\label{ttoo}
\eeq
If we integrate out $\wtb$, then we obtain the constraint
$\wty^2+m=0$, which has two solutions $\wty=\pm i\sqrt{m}$.
For each of them, plugging the value back to (\ref{ttoo}),
we may integrate out $\wtc$ yielding the constraint $\wtx=0$,
which leaves us with the free theory of the single variable
$\wta$ that corresponds to the invariant $(xx)$. 
That is, after the mass perturbation, we obtain the sigma model whose target
space is two copies of $\C$.
In view of (\ref{rg1}), we see that this result is consistent if our theory
was the $\Ost(2)$ theory, i.e., the $\Ost(2)$ theory with two massless
doublets, $x$ and $y$, and one regulator doublet $x_3$ with a complex mass.
Indeed, if that is the case, after the mass perturbation,
we have one massless doublet $x$ and two doublets $y$ and $x_3$
with complex masses. Addition of two fields with complex masses have
no effect whatsoever, both in the theta angle and in the
$\Z_2$ orbifold action. Thus, we are left with the
$\Ost(2)$ theory with a single  massless doublet $x$, whose low energy theory
is indeed two copies of the free theory of the singlet $(xx)$.

To summarize, we found that {\it the $\Ost(2)$ 
theory with
two massless doublets $x$ and $y$ is dual to the Landau-Ginzburg
model of five variables $\wtx,\wty,\wta,\wtb,\wtc$
with the superpotential (\ref{dual1}). $\wtx$ and $\wty$ are twist fields
with respect to the $\Z_2=\Ost(2)/SO(2)$ and the other variables are
the gauge invariant composites,
$\wta=(xx)$, $\wtb=(yy)$ and $\wtc=(xy)$.}

Using the chain of duality and standard relations as in (\ref{chain}),
we find the dual of the other theory as well:
{\it the $\Ons(2)$ theory with
two massless doublets is dual to the 
$\Z_2(-1)^{F_s}$ orbifold of the
Landau-Ginzburg model of five variables $\wtx,\wty,\wta,\wtb,\wtc$
with the superpotential (\ref{dual1}),
where the orbifold generator is $(\wtx,\wty,\wta,\wtb,\wtc)\mapsto
(-\wtx,-\wty,\wta,\wtb,\wtc)$ combined with $(-1)^{F_s}$.}

Let us draw some conclusions from these duality relations.

\subsection{Corank 1 Degeneration --- Two Or One Massive Vacua}
\label{subsec:O2rr1}

We consider the theory
with $N$ doublets $x_1,\ldots, x_N$ and a singlet $z$
with superpotential
\beq
W=z(x_1x_1)+\cdots+(x_Nx_N).
\eeq
By the definition of the $\Ostns(2)$ theory,
the low energy behaviour does not depend on $N$ 
as long as $N\geq 1$:
If $N$ is odd, the theory is defined as it is.
If $N$ is even, it is defined as the $\Ostns(2)$
theory with one additional massive field (a regulator). 
Changing $N$ by one either does not change anything
(i.e. the regulator is reinterpreted as a physical massive field,
 or vice versa)
or add/subtract two doublets with complex masses, which does not change
the low energy behaviour.

Let us take $N=1$ for simplicity.
If $z$ is fixed at a value away from zero, 
the superpotential for $x_1$ is regular and we have
supersymmetry breaking. The question is what happens in a 
neighborhood of $z=0$ and when the fluctuation of $z$ is taken into account.
To see this, we use the dual description.
For the theory with gauge group $\Ost(2)$ ({\it resp}. $\Ons(2)$),
the dual is two copies ({\it resp}. one copy)
 of the LG model of two variables, $z$ and $a$,
with the superpotential
\beq
W=za.
\eeq
This superpotential has a unique critical point $z=a=0$. 
Hence the theory has two ({\it resp}. one)
massive supersymmetric ground states,
with the expectation values
\beq
\langle z\rangle=0,\quad \langle(x_1x_1)\rangle=0.
\eeq

One may also consider the theory with $N=2$ and apply the duality 
obtained above.
The $\Ost(2)$ ({\it resp}. $\Ons(2)$) theory is dual to
the LG model ({\it resp}. $\Z_2(-1)^{F_s}$ LG orbifold)
of six variables,
$\wtx,\wty,\wta,\wtb,\wtc$ and $z$, with the superpotential
\beq
W=\wta\wtx^2+2\wtc\wtx\wty+\wtb\wty^2+z\wta+\wtb.
\eeq
There are two ({\it resp}. one) critical points,
$\wtx=\wta=\wtb=\wtc=z=0$ and $\wty=\pm i$. 
That is, there are two ({\it resp}. one) massive supersymmetric ground states,
with the expectation values
$\langle z\rangle=0$ and $\langle(x_1x_1)\rangle=\langle(x_2x_2)\rangle
=\langle(x_1x_2)\rangle=0$.

\subsection{Corank 2 Degeneration ---
Ramified Double Cover Of $\CC^2$ Or Its Orbifolds}
\label{subsec:O2rr2}

Let us now consider the theory of two doublets, $x$ and $y$,
and three singlets, $a$, $b$ and $c$, which are coupled
via the superpotential
\beq
W=a(xx)+2c (xy)+b(yy)
\label{W4}
\eeq
At values of $(a,b,c)$ away from the degeneration locus
\beq
ab=c^2,
\label{eqn4}
\eeq
the superpotential gives masses to both $x$ and $y$.
As we learned in Section~\ref{subsec:O2reg}, there is no 
zero energy state in such a theory.
Therefore, the low energy theory will concentrate near the
degeneration locus (\ref{eqn4}).
Near that locus but away from the origin,
$(a,b,c)=(0,0,0)$, we may change the variables to make
the superpotential into the form 
\beq
W=(c^2-ab)(x'x')+(y'y').
\eeq
The result of Section~\ref{subsec:O2rr1} then tells us that,
for the gauge group $\Ost(2)$ ({\it resp}. $\Ons(2)$)
we have two ({\it resp}. one) zero energy states
along the locus (\ref{eqn4}), as far as $(a,b,c)$ is away from the origin.
Thus, we expect to have some kind of double ({\it resp}. single)
cover over the degeneration locus (\ref{eqn4}).
We would like to find what really is the low energy theory.

Let us apply the dual description for the $\Ost(2)$ theory. 
It is simply the LG model of five plus three variables,
$\wtx,\wty,\wta,\wtb,\wtc$ and $a,b,c$,
with the superpotential
\beq
W=\wta\wtx^2+2\wtc\wtx\wty+\wtb\wty^2+a\wta+2c\wtc+b\wtb.
\label{W5}
\eeq
Integrating out $a,b,c$, we obtain the constraints
$\wta=\wtb=\wtc=0$
and we are left with the theory of $\wtx$
and $\wty$ only, with vanishing superpotential.
Thus, we obtain the conformal field theory of just two
variables $\wtx$ and $\wty$, with no constraint and no superpotential.
Extremizing $W$ with respect to $\wta,\wtb,\wtc$ finds the
relations
\beq
a=-\wtx^2,\qquad
b=-\wty^2,\qquad
c=-\wtx\wty.
\label{relabcxy}
\eeq
Such an $(a,b,c)$ indeed satisfies the equation (\ref{eqn4}). Conversely,
for each non-zero $(a,b,c)$ obeying (\ref{eqn4}),
the equation (\ref{relabcxy}) has two solutions for $(\wtx,\wty)$,
related by the sign flip $(\wtx,\wty)\to (-\wtx,-\wty)$.
That is, the $(\wtx,\wty)$ space is a double cover of the degeneration
locus $ab=c^2$, as expected.
Recall that $ab=c^2$ is the equation defining the $A_1$ surface
singularity, which is known to be realized by $\C^2/\Z_2$,
and the relations (\ref{relabcxy}) exhibit that realization,
\beq
\C^2\longrightarrow \C^2/\Z_2.
\label{A1quot}
\eeq
Recall that this $\Z_2$ symmetry is the quantum symmetry
of the orbifold by the $\Z_2=\Ost(2)/SO(2)$ (and hence
$\wtx$ and $\wty$ are twist fields). See  (\ref{qqact}).
Note also that $\wta=\wtb=\wtc=0$ means 
\beq
(xx)=(yy)=(xy)=0.
\label{singvanish}
\eeq

We conclude that {\it the $\Ost(2)$ theory flows in the infra-red limit to
to the free conformal field theory of two twist variables, $\wtx$ and $\wty$,
i.e., the sigma model with the target space $\C^2$.
The singlets $a$, $b$ and $c$ are related to $\wtx$ and $\wty$
by (\ref{relabcxy}). 
The $O(2)$ invariants, $(xx)$, $(yy)$ and $(xy)$,
vanish in the infra-red fixed point theory.}

We may unfold the $\Z_2=\Ost(2)/SO(2)$ by orbifolding
the associated quantum symmetry. 
Also,  we may use the chain of duality and standard relations as 
in (\ref{chain}).
These lead us to conclude that
{\it the $SO(2)$ ({\it resp}. $\Ons(2)$) 
theory flows in the infra-red limit to
the free orbifold conformal field theory $\C^2/\Z_2$ ({\it resp}. 
$\C^2/\Z_2(-1)^{F_s}$).}
Here, ``the $SO(2)$ theory'' of course stands for the $U(1)$ theory with
$r=0$ and $\theta=\pi$.

\section{Orthogonal Groups}
\label{sec:O}

In this section, we study low energy behaviour of theories with 
the orthogonal gauge group, $O(k)$ or $SO(k)$, with $N$ chiral multiplets
in the fundamental representation, i.e., the vector representation
${\bf k}$. 
We denote the chiral matter fields as
$x_1,\ldots, x_N$ where each $x_i$ is a column vector of length $k$,
$x_i=(x^a_i)_{a=1,\ldots,k}$. 
Our main focus will be the theory with vanishing superpotential
for $x_1,\ldots,x_N$.
The group
$O(k)$ is the semi-direct product $SO(k)\rtimes \Z_2$ for even $k$,
as in $O(2)$ discussed in the previous section, and the direct product
$SO(k)\times \Z_2$ for odd $k$, where $\{{\bf 1}_k\}\times
\Z_2$ corresponds to
the subgroup generated by the central element $-{\bf 1}_k$. In either case, 
the $O(k)$ gauge theory can be 
treated as a $\Z_2$ orbifold of the $SO(k)$ gauge theory.
As always, there are two versions of the orbifold,
related by $(-1)^{F_s}$. As the final important point, the groups
$O(k)$ and $SO(k)$ have a non-trivial fundamental group
\beq
~~~\pi_1(O(k))=\pi_1(SO(k))= \Z_2\qquad\mbox{for $k\geq 3$}.
\eeq
This means that there is a mod $2$ theta angle:
On a closed two-dimensional manifold,
there are two topological types of principal $G$ bundles for
$G=O(k)$ or $SO(k)$, 
the trivial and the non-trivial. And the
mod $2$ theta angle assigns a phase $(-1)$ to the path-integral weight
for the non-trivial $G$ bundle.

\subsection{The Space Of Classical Vacua}

Let us first describe the space of classical vacua --- 
the space of scalar fields that annihilate the classical potential.
We denote the scalar component of the vector multiplet
by $\sigma$. It is a $k\times k$ antisymmetric complex matrix.
We write $x$ for the $k\times N$ matrix $(x^a_i)$.
The vacuum equation reads
\beqa
&&[\sigma,\sigma^{\dag}]=0,\nn\\
&&xx^{\dag}=(xx^{\dag})^T,\label{OD}\\
&&\sigma x=\sigma^{\dag} x=0.\nn
\eeqa
The first equation means that $\sigma$ must lie in the complexification of
the Lie algebra of a maximal torus.
That is, up to gauge transformations, it is of the form
\beq
i\left(\begin{array}{ccccc}
&-\sigma_1&&&\\
\sigma_1&&&&\\
&&\ddots&&\\
&&&&-\sigma_{\ell}\\
&&&\sigma_{\ell}&
\end{array}\right)\quad\mbox{\it resp.}\quad
i\left(\begin{array}{ccccc|c}
&-\sigma_1&&&&\\
\sigma_1&&&&&\\
&&\ddots&&&\\
&&&&-\sigma_{\ell}&\\
&&&\sigma_{\ell}&&\\
\hline
&&&&&0
\end{array}\right),
\label{Omaxt}
\eeq
for $k=2\ell$ {\it resp.} $k=2\ell +1$.
The second equation means that $xx^{\dag}$ is a real symmetric matrix,
and hence it can be diagonalized using the gauge symmetry.
With an appropriate $U(N)$ flavor rotation, we can write the solution 
as
\beq
x=\left(\begin{array}{ccc}
a_1&&\\
&\ddots&\\
&&a_N\\
&&
\end{array}\right)\quad
\mbox{or}\quad
\left(\begin{array}{ccccc}
a_1&&&&\\
&\ddots&&&\\
&&\ddots&&\\
&&&a_k&
\end{array}\right)
\eeq
depending on $N\leq k$ or $N\geq k$.
The final equation requires that if the number of non-zero
$\sigma_a$'s is $s$, then the number of non-zero $a_i$'s is at most $k-2s$.
Let $C_s$ ({\it resp.} $H_r$)
be the set of gauge equivalence classes of solutions for
$\sigma$ of rank $2s$ or less ({\it resp}. $x$ of rank $r$ or less).
It has complex dimension $s$ ({\it resp.} $Nr-{r(r-1)\over 2}$).
The space of classical vacua is
\beq
{\mathcal M}=\bigcup_{s=s_{\rm min}}^{\ell}\Bigl(C_s\times H_{k-2s}\Bigr),
\label{OMc}
\eeq
where $s_{\rm min}=0$ if $N\geq k-1$
and $s_{\rm min}=[{k-N\over 2}]$ if $N\leq k-2$. 
When $N\geq k-1$, there is a Higgs branch $C_0\times H_k$ in which
$x$ is generically non-zero and breaks the gauge group
completely (or to a $\Z_2$ subgroup for the $O(k)$ theory with $N=k-1$). 
When $k$ is even, there is a Coulomb branch $C_{\ell}\times H_0$
in which $\sigma$ generically has the full rank $k$.
Other components are the mixed Coulomb-Higgs 
branches where both $x$ and $\sigma$ are generically non-zero.

\subsection{Regularity}
\label{subsec:Oreg}

Classically, the gauge theory reduces at low energies
to the non-linear sigma model whose target space is the classical vacuum
moduli space (\ref{OMc}). 
This space is singular and non-compact, and hence we do not know if
we have a sensible theory in the infra-red limit, or at least
the fixed point theory must be described by something
very much different from the sigma model for (\ref{OMc}) 
\cite{WittenH,AhaBer}.
Here we would like to discuss the possibility that
quantum corrections lift all non-compact flat directions in $\sigma$, 
i.e., Coulomb and all possible mixed branches.
We shall refer to such a theory as a {\it regular} theory.
Regularity is judged by the effective twisted superpotential
$\widetilde{W}_{\it eff}$ for $\sigma_a$'s.
By non-renormalization theorem, this is not affected even 
if the superpotential for the matter multiplets is introduced.
In particular, if the theory is regular, 
by introducing a superpotential that lifts the Higgs branch, we can
obtain a $(2,2)$ superconformal field theory with discrete spectrum
at the infra-red fixed point.
We would like to compute $\widetilde{W}_{\it eff}$
and find a criterion for regularity.

Let us first consider the Coulomb branch $C_{\ell}\times H_0$
for the $k$ even case ($k=2\ell$).
If $\sigma_a$'s are chosen so that $|\sigma_a|$
and $|\sigma_a\pm \sigma_b|$ for $a\ne b$ are all non-zero,
we have either massive multiplets or the massless vector
multiplets for the maximal torus $U(1)^{\ell}$. To
obtain the effective theory for the latter, we integrate out
the massive modes, consisting of the chiral multiplets
$x_i$ and the ``off-diagonal'' vector multiplets $\Vct^c_{\,\,\,\,d}$.
Let us first consider the vector multiplets.
The contribution to $\widetilde{W}_{\it eff}$ can be found 
\cite{WVerlinde} by
looking at the mass terms for the gaugino,
\beq
-{\rm tr}\left(\blambda_-[\sigma,\lambda_+]+
\blambda_+[\sigma^{\dag},\lambda_-]
\right).
\eeq
For $a\ne b$, those which are charged under $U(1)_a\times U(1)_b$
are in the $2\times 2$ block, $\Vct^{c}_{\,\,\,\,d}$ for
$c=2a-1,2a$ and $d=2b-1,2b$, and have masses
$\pm\sigma_a\pm\sigma_b$ (all the four possible sign combinations).
The contribution to $\widetilde{W}_{\it eff}$ of these four multiplets is
$\pi i (\sigma_a-\sigma_b)+\pi i (\sigma_a+\sigma_b)$ which vanishes
modulo $2\pi i$ times $\sigma_a$'s.
Computation of the contribution from the massive chiral multiplets
is standard,
$-\sum_{i,a}\sigma_a(\log\sigma_a-1)
-\sum_{i,a}(-\sigma_a)(\log(-\sigma_a)-1)$, which is 
$\pi i N\sum_a\sigma_a$, again modulo $2\pi i$ times $\sigma_a$'s.
 The total is 
\beq
\widetilde{W}_{\it eff}
= \pi i N\sum_{a=1}^{\ell}\sigma_a.
\label{Weffevenk}
\eeq

Let us next consider the $k$ odd case ($k=2\ell+1$) and look at
the mixed branch $C_{\ell}\times H_1$.
In this case, computation depends on the location of $x$ in $H_1$.
If it is zero, then, the entire gauge group, $O(k)$ or $SO(k)$,
is unbroken and we can do the computation as usual.
If it is non-zero, the gauge group is broken to its proper subgroup,
$O(k-1)$ or $SO(k-1)$, and we need to take into account the Higgs 
effect.
Let us first consider the former, expanding $x$ around $x=0$.
The last components $x^k_1,\ldots,x^k_N$ are massless and we leave
them in the effective theory.
Integration over massive modes 
can be done in the same way as above,
except that now, for odd $k$, we have the right-most off-diagonal components, 
$\Vct^c_{\,\,\,\,k}$ for $c=1,\ldots, k-1$. 
For $c=2a-1,2a$, they are charged only under $U(1)_a$ and
have masses $\pm\sigma_a$. They yield the non-trivial contribution
$\pi i\sigma_a$ to $\widetilde{W}_{\it eff}$.
Contribution from other massive modes are the same, and the total is
\beq
\widetilde{W}_{\it eff}=\pi i (N+1)\sum_{a=1}^{\ell}\sigma_a.
\label{Weffoddk}
\eeq
Let us next perform computation around a Higgsed point, say,
\beq
x_1=\cdots=x_{N-1}=\left(\begin{array}{c}
0\\
\vdots\\
0\\
0
\end{array}\right),\quad
x_N=\left(\begin{array}{c}
0\\
\vdots\\
0\\
v
\end{array}\right).
\eeq
In this case, we must treat the super-Higgs multiplet,
consisting of the right-most off-diagonal vector multiplets 
$\Vct^a_{\,\,\,\,k}$
and the complexified gauge orbit directions $x^a_N$, as one block.
This block gives no contribution to $\widetilde{W}_{\it eff}$
--- the one from vectors and the one from chirals cancel against each
other. The rest is as in $O(k-1)$ or $SO(k-1)$ theory with
$N-1$ fundamentals. Note that $k-1$ is even and the above result 
(\ref{Weffevenk}) can be used. The result is
$\widetilde{W}_{\it eff}=\pi i (N-1)\sum_a\sigma_a$, which is equal to
(\ref{Weffoddk}) modulo $2\pi i$ times $\sigma_a$'s.
Actually, we did not have to do the two computations in view of the 
decoupling between the chiral and twisted chiral multiplets ---
the result for $\widetilde{W}_{\it eff}$ should not depend on where in $H_1$
you do the computation. Nevertheless, the fact that we indeed
obtained the same result is gratifying.

Computation in various mixed branches should be obvious by now,
thanks to the exercise given above that involves 
the super-Higgs multuplet. We obtain
(\ref{Weffevenk}) or (\ref{Weffoddk}) depending on
whether $k$ is even or odd, where the sum over $a$ must be reduced 
appropriately (for example, the sum is over $a\in \{1,\ldots,s\}$ on
$C_s\times H_{k-2s}$).

The result is that we have an effective theta angle
\beq
\theta_{\it eff}=\left\{\begin{array}{ll}
\pi N&\mbox{$k$ even},
\\
\pi (N+1)&\mbox{$k$ odd}.
\end{array}\right.
\eeq
for each $U(1)$ factor on the classical Coulomb or mixed branch.
If it is zero modulo $2\pi$, the energy density is zero and we do have
non-compact Coulomb and mixed branches. 
If it is non-zero modulo $2\pi$, then the energy density is
$e^2_{\it eff}(\sigma)/8$ times the number of unbroken $U(1)$ factors, 
where $e_{\it eff}(\sigma)$ is the
effective gauge coupling at the given value of $\sigma$. 
The latter approaches the classical value $e$ as $|\sigma_a|$
and $|\sigma_a\pm\sigma_b|$ for $a\ne b$ are all much larger compared to
$e$. Thus, the non-compact Coulomb and mixed branches
are all lifted in this case.
To summarize, {\it the theory is regular
if and only if $N-k$ is odd.}

\subsection*{\sl Mod $2$ Theta Angle}

In the analysis so far, we have implicitly assumed that
the mod $2$ theta angle is set equal to zero. Let us now see its effect.
First, the universal cover of the group $SO(k)$ is
$Spin(k)$ which is realized as the subset of the Clifford algebra
${\rm C}(\R^k)$ \footnote{It is generated by $1$ and 
the basis $e_1,\ldots, e_k$
of $\R^k$ which obey the relation $e_ae_b+e_be_a=-2\delta_{a,b}$.
Note that $\exp(te_ae_b)=\cos(t)+e_ae_b\sin(t)$ for $a<b$.
In particular, $\exp(\pi e_ae_b)=-1$ and $\exp(2\pi e_ae_b)=1$.}
generated multiplicatively by elements of
the form $\exp\left(\sum_{a<b}t_{ab}e_ae_b\right)$.
The conjugation 
action of $Spin(k)$ on $\R^k\subset {\rm C}(\R^k)$
induces an isomorphism $SO(k)\cong Spin(k)/\Z_2$
where $\Z_2$ is the subgroup consisting of $\pm 1\in {\rm C}(\R^k)$.
That is why $SO(k)$ has the fundamental group $\Z_2$. 
An example of non-trivial loop in $SO(k)$ is
\beq
t\in\R/2\pi\Z~\longmapsto~g_t=\left(\begin{array}{ccc}
\cos(t)&-\sin(t)&\\
\sin(t)&\cos(t)&\\
&&{\bf 1}_{k-2}
\end{array}\right)\in SO(k).
\eeq
Indeed it lifts to a path
$\widetilde{g}_t=\exp\left({t\over 2}e_1e_2\right)$ from 
$\widetilde{g}_0=1$ to 
$\widetilde{g}_{2\pi}=-1$ in $Spin(k)$.
A topologically non-trivial $SO(k)$ bundle over a closed surface
$\Sigma$ is the one having the 
transition function $g_t$ along a circle in $\Sigma$
parametrized by $t\in \R/2\pi\Z$
that separates $\Sigma$ into two connected components.
The mod $2$ theta angle assigns the phase $(-1)$ to such a
principal $SO(k)$ bundle.
By this exercise, we see that it yields the theta angle
\beq
\theta=\pi,
\eeq 
for the subgroup $U(1)\cong SO(2)\subset SO(k)$ of 2-dimensional
rotations for each orthogonal decomposition
$\R^k\cong \R^2\oplus \R^{k-2}$. 
In particular, it yields a contribution $\pi$ to the theta angle 
for each $U(1)$ factor on the classical Coulomb or mixed branch. 
We can now state the complete criterion:

{\it When $N-k$ is odd ({\rm resp}. even), 
the theory is regular if and only if the tree level mod $2$
theta angle is turned off ({\rm resp}. turned on).}

Thus, the theory with $N-k$ even, which is not by itself
regular, can be made regular by turning on the mod 2 theta angle.
Alternatively, we may consider adding a fundamental chiral multiplet
with a complex mass $m$, as a regulator.
In fact, in the limit $|m|\to \infty$
the effective action for the vector multiplet
obtained by integrating it out is nothing but the mod $2$ theta angle.
This can be seen by noting that it yields the theta angle $\theta=\pi$
for each $SO(2)$ subgroup of 2-dimensional rotation.
When we consider the theory with gauge group
$O(k)$, the regulator field also
has the effect of inverting the definition of the
$\Z_2$ orbifold, i.e., dressing the generator by $(-1)^{F_s}$.
The situation is exactly the same as what we have seen in
the $O(2)$ gauge theory.

In what follows, unless otherwise stated,
we shall always assume that the theory is regular.
When $N-k$ is even, either the mod 2 theta angle is turned on or
a regulator field is introduced.

\subsection{Twisted Masses}
\label{subsec:Oktmass}

Before discussing the theory with massless fundamentals
$x_1,\ldots,x_N$, let us study the theory in which they have
twisted masses $\wtm_1,\ldots,\wtm_N$.
Along the way, we introduce a notation that distinguishes
the two orbifold projections for the case where the gauge group is $O(k)$.
We assume that the masses are generic and in particular satisfy
\beq
~~~~~\wtm_i+\wtm_j\ne 0\quad~\forall (i,j).
\label{genericm}
\eeq
Then, the Higgs branch is lifted, and the theory is 
well behaved in any direction.
Our focus is the spectrum of supersymmetric ground states of 
this regularized system.

We integrate out the fundamentals as they 
are massive in any field configuration.
We also stay in the generic locus on the Coulomb branch
and integrate out the massive off-diagonal components of
the vector multiplet.
The resulting effective twisted superpotential is
\beqa
\widetilde{W}_{\it eff}
&=&
-\sum_{i,a}(\sigma_a-\wtm_i)(\log(\sigma_a-\wtm_i)-1)
-\sum_{i,a}(-\sigma_a-\wtm_i)(\log(-\sigma_a-\wtm_i)-1)\nn\\
&&+\pi i k\sum_a\sigma_a+\pi i (N-k+1)\sum_a\sigma_a.
\label{WeffO}
\eeqa
The first line is from the chiral multiplets, while the term
$\pi i k\sum_a\sigma_a$ on the second line is from the massive vectors
--- as we have seen in the previous section, we have a non-zero
contribution $\pi i \sum_a\sigma_a$ if and only if $k$ is odd.
The last term $\pi i (N-k+1)\sum_a\sigma_a$, which is non-zero
when $N-k$ is even, is from either the mod $2$ theta angle
or the regulator field.
The vacuum equation reads
\beq
\prod_{i=1}^N(\sigma-\wtm_i)=(-1)^{N+1}\prod_{i=1}^N(-\sigma-\wtm_i),
\label{roots}
\eeq
for $\sigma=\sigma_1,\ldots,\sigma_{\ell}$.
The solutions are identified under the action of
the Weyl group:
permutations of $\sigma_a$'s as well as the sign flips
\beq
\sigma_a~\longmapsto\, \epsilon_a\sigma_a,\quad
\left\{\begin{array}{ll}
\epsilon_1\cdots\epsilon_{\ell}=1&\mbox{$SO(k)$, $k$ even},\\
\mbox{no condition}&\mbox{otherwise}.
\end{array}\right.
\eeq
Note that the reflections with $\epsilon_1\cdots\epsilon_{\ell}=-1$
is allowed for $O(k)$, $k$ even as well 
--- they are from the disconnected component
of $O(k)$ and represent the $\Z_2$ orbifold generator.
We require the solutions to obey
\beqa
&&\sigma_a\ne \pm \wtm_i,\nn\\
&&\sigma_a\ne \pm \sigma_b\quad a\ne b,\label{forbidden}\\
&&\sigma_a\ne 0\quad \forall a\quad\mbox{if $k$ is odd}.\nn
\eeqa
In the forbidden region, there are massless degrees of freedom
other than the $U(1)^{\ell}$ vector multiplets, and the effective twisted
superpotential (\ref{WeffO}) can not be trusted.
We do not have to worry about the first condition,
$\sigma_a\ne\pm \wtm_i$, as $\sigma=\pm \wtm_i$ are not among the roots of
(\ref{roots}) thanks to the condition (\ref{genericm}).
We simply ignore solutions violating the other conditions.
Namely, we assume that there is no supersymmetric ground state
supported in the forbidden region.
This point was examined in \cite{HoTo} in a specific class of models
and consistent picture has emerged.

Let us first study the case where $k$ is even ($k=2\ell$) and
$N$ is even. The equation (\ref{roots}) have ${N\over 2}$ pairs
of non-zero roots.
Solutions for $\sigma_1,\ldots,\sigma_{\ell}$ 
obeying the condition (\ref{forbidden}) exists if and only if
${N\over 2}\geq \ell$.
For $O(k)$ gauge group, the number of inequivalent solutions
is ${{N\over 2}\choose \ell}$. 
For $SO(k)$ gauge group, the number is twice as much,
$2{{N\over 2}\choose \ell}$, because of the constraint
$\epsilon_1\cdots\epsilon_{\ell}=1$ on the Weyl group elements.
As the Weyl group is completely broken at each of these solutions,
these are the number of supersymmetric ground states.
For the same reason, for the $O(k)$ theory,
the result does not depend on the choice of the orbifold.

When $k$ is even ($k=2\ell$) and $N$ is odd, 
the equation (\ref{roots}) has ${N-1\over 2}$ pairs of 
non-zero roots and one root at $\sigma=0$.
Solutions for $\sigma_a$'s obeying (\ref{forbidden}) exists 
if and only if ${N-1\over 2}+1\geq \ell$.
The count from solutions for which $\sigma_a$'s are all non-zero
is as in the case above: ${{N-1\over 2}\choose \ell}$
for $O(k)$ and $2{{N-1\over 2}\choose \ell}$ for $SO(k)$.
Let us consider solutions where one $\sigma_a$ vanish, say,
$\sigma_1=0$ (there are ${{N-1\over 2}\choose \ell-1}$ 
inequivalent solutions of this
type).  For $SO(k)$ gauge group, the Weyl group is completely
broken and we obtain ${{N-1\over 2}\choose \ell-1}$ 
as the number of ground states.
For $O(k)$ gauge group, exactly one Weyl group element is unbroken.
It is the one that acts as the sign flip of $\sigma_1$ only, and is
represented by
\beq
\left(\begin{array}{cccc}
-1&&&\\
&1&&\\
&&\ddots&\\
&&&1
\end{array}\right).
\eeq
The spectrum of supersymmetric ground states from this sector
is sensitive to the definition of the orbifold. As in the $O(2)$ theory
discussed in Section~\ref{subsec:O2tmass}, we denote the gauge group by
$\Ost(k)$ if we receive two supersymmetric ground states 
(one twisted and one untwisted),
and by $\Ons(k)$ if we receive one twisted supersymmetric ground state.
The total number of states
of this type is $2{{N-1\over 2}\choose \ell-1}$ for $\Ost(k)$ theory
and ${{N-1\over 2}\choose \ell-1}$ for $\Ons(k)$ theory.
To summarize, the number of ground states for even $k$ and odd $N$ case is
${{N-1\over 2}\choose \ell}+2{{N-1\over 2}\choose \ell-1}$ for $\Ost(k)$,
${{N-1\over 2}\choose \ell}+{{N-1\over 2}\choose \ell-1}
={{N+1\over 2}\choose \ell}$ 
for $\Ons(k)$,
and $2{{N-1\over 2}\choose \ell}+{{N-1\over 2}\choose \ell-1}$ for $SO(k)$.

Let us next study the case where $k$ is odd ($k=2\ell+1$).
Note that there is one component of each vector $x_i$ that is neutral
with respect to the maximal torus.
If we choose the torus as in (\ref{Omaxt}), then it is the last 
($k$-th) component. Thus, the variables $x^k_1,\ldots, x^k_N$ are decoupled
from the rest of the degrees of freedom on the generic locus of the
Coulomb branch.
Note that $O(k)$ and $SO(k)$ differ
in the presence/absence of the group element
\beq
\left(\begin{array}{ccc|c}
1&&&\\
&\ddots&&\\
&&1&\\
\hline
&&&-1
\end{array}\right),
\eeq
that flips the sign of these components.
For $SO(k)$, the last component system is simply the model of
$N$ variables with twisted mass (possibly with one 
additional regulator field). 
This sector hence provides a unique supersymmetric ground state.
For $O(k)$, the last component system is the $\Z_2$ orbifold thereof.
The spectrum in this sector again is sensitive to the choice of
orbifold. We denote the gauge group by $\Ost(k)$ if the number of
supersymmetric ground states is two for even $N$ and one for odd $N$,
and by $\Ons(k)$ if opposite, i.e., one for even $N$ and two for odd $N$.
We now turn to the sector of the first $k-1=2\ell$ components.

We first consider the case where $N$ is even.
The equation (\ref{roots}) has ${N\over 2}$ pairs of non-zero roots.
Solutions obeying (\ref{forbidden}) exist
if and only if ${N\over 2}\geq\ell$, and 
the number of inequivalent ones is ${{N\over 2}\choose \ell}$.
This is for both $O(k)$ and $SO(k)$ since they share the same Weyl group
when $k$ is odd.
Since the Weyl group is completely broken at each of them, 
this is the number of vacuum states from this sector.
Combining with the last component sector, the total number of
supersymmetric ground states is
$2{{N\over 2}\choose\ell}$ for $\Ost(k)$,
${{N\over 2}\choose\ell}$ for $\Ons(k)$,
 and 
${{N\over 2}\choose\ell}$ for $SO(k)$.

Next, we consider the case where $N$ is odd. 
The equation (\ref{roots}) has ${N-1\over 2}$ pairs of non-zero roots
and one root at $\sigma=0$.
According to  (\ref{forbidden}) we need to avoid the one at
$\sigma=0$ when $k$ is odd.
The solutions exist if and only if 
${N-1\over 2}\geq \ell$, and the number of inequivalent ones is
${{N-1\over 2}\choose\ell}$ for both $O(k)$ and $SO(k)$.
Since the Weyl group is completely broken at each of them, 
this is the number of vacuum states from this sector.
Combining with the last component sector, the total number of
supersymmetric ground states is
${{N-1\over 2}\choose\ell}$ for $\Ost(k)$,
$2{{N-1\over 2}\choose\ell}$ for $\Ons(k)$,
 and 
${{N-1\over 2}\choose\ell}$ for $SO(k)$.

The definition of orbifold for $\Ostns(k)$ can be extended by continuity
to the theories where the twisted masses are turned off
and then, possibly, the superpotential is turned on.
For $k$ even, this is defined originally for odd $N$ and then is extended
to the even $N$ case via the regulator field.
For $k$ odd, this is already defined for both even and odd $N$, 
and we would like to check whether the two are continuously connected. 
We can focus on the last 
component sector in which the choice of orbifold is relevant.
Let us start from the even $N$ case, with no mod 2 theta angle
nor the regulator field. The number of ground states is $2$ for $\Ost(k)$, 
and this is the number in the ``standard''
$\Z_2$ orbifold in the sense of Section~\ref{subsec:massive}.
We then turn off the twisted mass for, say, $x_N$,
 and give it a complex mass. Then we have $N_{\it eff}=N-1$ (odd)
fundamentals with a regular field $x_N$.
According to Section~\ref{subsec:massive},
the number of supersymmetric ground states is $1$, which
is indeed the number we are assigning for $\Ost(k)$ with odd $N_{\it eff}$.
This shows the continuity of our definition.

To summarize, for $N\leq k-2$ there is no supersymmetric ground state.
For $N\geq k-1$, the number of supersymmetric ground states is
given by: 
\beq
\begin{array}{c|c|c|c}
\mbox{group}&k&N&\mbox{number}\\
\hline
\Ost(k)&\mbox{even}&\mbox{even}&
{\displaystyle {{N\over 2}\choose {k\over 2}}}\\[0.4cm]
\Ons(k)&\mbox{even}&\mbox{even}&
{\displaystyle {{N\over 2}\choose {k\over 2}}}\\[0.4cm]
SO(k)&\mbox{even}&\mbox{even}&
{\displaystyle 2{{N\over 2}\choose {k\over 2}}}
\\[0.4cm]
\hline
\Ost(k)&\mbox{even}&\mbox{odd}&
{\displaystyle {{N-1\over 2}\choose {k\over 2}}
+2{{N-1\over 2}\choose {k\over 2}-1}}\\[0.4cm]
\Ons(k)&\mbox{even}&\mbox{odd}&
{\displaystyle {{N+1\over 2}\choose {k\over 2}}}\\[0.4cm]
SO(k)&\mbox{even}&\mbox{odd}&
{\displaystyle 2{{N-1\over 2}\choose {k\over 2}}
+{{N-1\over 2}\choose {k\over 2}-1}}
\end{array}
\quad
\begin{array}{c|c|c|c}
\mbox{group}&k&N&\mbox{number}\\
\hline
\Ost(k)&\mbox{odd}&\mbox{even}&
{\displaystyle 2{{N\over 2}\choose {k-1\over 2}}}\\[0.4cm]
\Ons(k)&\mbox{odd}&\mbox{even}&
{\displaystyle {{N\over 2}\choose {k-1\over 2}}}\\[0.4cm]
SO(k)&\mbox{odd}&\mbox{even}&
{\displaystyle {{N\over 2}\choose {k-1\over 2}}}\\[0.4cm]
\hline
\Ost(k)&\mbox{odd}&\mbox{odd}&
{\displaystyle {{N-1\over 2}\choose {k-1\over 2}}}\\[0.4cm]
\Ons(k)&\mbox{odd}&\mbox{odd}&
{\displaystyle 2{{N-1\over 2}\choose {k-1\over 2}}}\\[0.4cm]
SO(k)&\mbox{odd}&\mbox{odd}&
{\displaystyle {{N-1\over 2}\choose {k-1\over 2}}}
\end{array}
\label{WItable}
\eeq

\subsection{$N\leq k-2$: Supersymmetry Breaking}
\label{subsec:SUSYBO}

Let us consider the (regular) theory with massless fundamentals
where the number $N$ is in the range $1\leq N\leq  k-2$.
The observed fact that there is no supersymmetric ground state
when the twisted masses are turn on
implies that there is no normalizable
supersymmetric ground state in the massless theory either. 
This is because \cite{HoTo},
if there were a normalizable zero energy state in the massless
theory, that would stay in the spectrum even if the masses are turned on,
since the masses would only make better the behaviour of states 
at infinity in the field space.

In fact, there is no normalizable supersymmetric ground state
also in irregular theory with $1\leq N\leq k-2$ as well as in
the pure Yang-Mills theory (regular or not) for $k\geq 3$.

To see that, let us continue from the previous subsection and take the limit
where some of the twisted masses are sent to infinity. 
If an odd number of $\wtm_i$'s are sent to infinity,
the behaviour of the superpotential (\ref{WeffO}) at large values of
$\sigma_a$'s is changed and a regular theory becomes an irregular theory.
If an even number of $\wtm_i$'s are sent to infinity, 
the behaviour does not change and a regular theory becomes another 
regular theory.
Note that a pure-Yang-Mills theory is obtained by sending all 
twisted masses to infinity --- the regular one if $N$ is even and 
the irregular one if $N$ is odd.
Let us look closely into the equation (\ref{roots}).
If one twisted mass, say $\wtm_N$, is sent to infinity, then one pair
of non-zero roots go away to infinity. To see that we rewrite the equation 
as 
\beq
\left(-1+{\sigma\over\wtm_N}\right)
\prod_{i=1}^{N-1}(\sigma-\wtm_i)
=\left(-1-{\sigma\over\wtm_N}\right)
\prod_{i=1}^{N-1}(\sigma+\wtm_i).
\eeq
We see that the equation has a limit as $\wtm_N\to\infty$. It is an equation
of order $(N-2)$. Since the order has decreased by $2$,
two roots must have gone away to infinity.
If two twisted masses, say $\wtm_{N-1}$ and $\wtm_N$, are sent to infinity,
the same argument shows that still 
one pair of non-zero roots go away to infinity.

This and the analysis of the previous subsection lead us to conclude
that the irregular theory in the range $1\leq N\leq k-2$
has no supersymmetric ground state, 
when generic twisted masses are turned on and hence
also when they are turned off.
We also find that the pure Yang-Mills
theory with $k\geq 3$, 
whether regular or not, has no supersymmetric ground state.

\subsection{$N=k-1$: Free Conformal Field Theory}
\label{subsec:freeO}

Let us now consider the theory with $N=k-1$ massless fundamentals.
In the regular theory, the Coulomb and mixed branches are lifted
and we are left with the Higgs branch $H_k=H_{k-1}$ only.
As a complex manifold, the Higgs branch is isomorphic to
the affine space $\C^{(k-1)k\over 2}$ since the chiral ring of gauge invariants
is isomorphic to the polynomial ring of the ${k(k-1)\over 2}$
scalar products $(x_ix_j)$ (the ``mesons'') with no relations,
see \cite{Weyl}
\beq
\C[x_1,\ldots,x_{k-1}]^{SO(k,\C)}=\C\Bigl[\,(x_ix_j)\,\Bigl|
\,1\leq i\leq j\leq k-1\,
\Bigr].
\eeq
The classical metric is singular at the roots of Coulomb and mixed branches
where parts of the gauge symmetry is unbroken.
However, the singularity is expected to be smeared as
these branches are lifted by quantum corrections.
We claim that the theory flows in the infra-red limit
to the free theory of the mesons.
This is just as in the $U(1)$ theory discussed in 
Section~\ref{subsec:O2dual} and in the $SU(k)$ theory with
$N=k+1$ massless fundamentals discussed in \cite{HoTo}.
To be precise, this is for the gauge group $SO(k)$.
In the $O(k)$ case, we must take the orbifold with respect to
the $\Z_2$ symmetry that acts trivially on the mesons. 
This will make either two copies or one copy of the Higgs branch.
We claim that the former is the case for the $\Ost(k)$ theory
and the latter is the case for the $\Ons(k)$ theory.
We do not provide a proof of this here, 
but consistency will be seen in what follows.

To summarize, we claim that
{\it the $\Ost(k)$ {\rm resp.} $\Ons(k)$ {\rm resp}. $SO(k)$
 gauge theory with $N=k-1$
massless fundamentals flows in the infra-red limit to
two copies {\rm resp}. one copy {\rm resp}. one copy
of the free conformal field
theory of the ${k(k-1)\over 2}$ mesonic variables.}

\subsection{$N\geq k$: Duality}
\label{subsec:dualityO}

Finally, let us consider the theory with $N\geq k$ massless fundamentals.
We claim that there is a duality, where the correspondence
of the gauge groups is
\beqa
~~~\Ost(k)&\longleftrightarrow& SO(N-k+1)\nn\\
~~~SO(k)&\longleftrightarrow& \Ost(N-k+1)\\
~~~\Ons(k)&\longleftrightarrow& \Ons(N-k+1).\nn
\eeqa
{\it The theory with the gauge group on the left hand side
and $N$ massless fundamentals $x_1,\ldots, x_N$
flows in the infra-red limit to the same fixed point
as the theory with the gauge group on the right hand side
and $N$ fundamentals, $\wtx^1,\ldots,\wtx^N$,
plus ${N(N+1)\over 2}$ singlets, $s_{ij}=s_{ji}$ ($1\leq i,j\leq N$), 
having the superpotential
\beq
W=\sum_{i,j=1}^Ns_{ij}(\wtx^i\wtx^j).
\label{OdualW}
\eeq
The mesons in the original theory correspond to
the singlets in the dual,
\beq
(x_ix_j)=s_{ij}.
\label{Omesons}
\eeq
 The baryons
$[x_{i_1}\cdots x_{i_k}]$ in the original $SO(k)$ theory correspond
to twist operators in the dual $\Ost(N-k+1)$ theory
regarded as a $\Z_2$ orbifold. More fundamentally,
the order $2$ symmetry $O(k)/SO(k)$ of the original theory
corresponds to the quantum symmetry of the dual.
Likewise for the baryons $[\wtx^{i_1}\cdots \wtx^{i_{N-k+1}}]$
and the order $2$ symmetry of the dual $SO(N-k+1)$
theory.
The quantum symmetry of the $\Ons(k)$ theory corresponds to
the quantum symmetry combined with $(-1)^{\bf F}$ in the
dual $\Ons(N-k+1)$ theory, and vice versa.}

The claimed relation is indeed a {\it duality}.
I.e., the dual of the dual is the original.
Let us start from $\Ost(k)$ with $N$ massless fundamentals.
The dual has gauge group $SO(N-k+1)$ and its dual has
$\Ost(N-(N-k+1)+1)=\Ost(k)$.
The latter has
$N$ fundamentals, $\widetilde{\wtx}_i$, and $2{N(N+1)\over 2}$
singlets, $\wts^{ij}$ and $s_{ij}$, having
the superpotential
\beq
W=\sum\wts^{ij}(\widetilde{\wtx}_i\widetilde{\wtx}_j)
+\sum s_{ij}\wts^{ij}.
\eeq
The second term comes from the first dual superpotential
and the relation $(\wtx^i\wtx^j)=\wts^{ij}$.
If $\wts^{ij}$ is integrated out, we obtain the 
constraints $s_{ij}=-(\widetilde{\wtx}_i\widetilde{\wtx}_j)$.
The resulting theory is simply the $\Ost(k)$ gauge theory
with $N$ fundamentals $\widetilde{\wtx}_i$ and no superpotential,
which is indeed the theory we started with.
Note that the constraints on $s_{ij}$
and the meson/singlet correspondence implies the relation
$(x_ix_j)=-(\widetilde{\wtx}_i\widetilde{\wtx}_j)$.
The minus sign is typical for duality.
The case where we start from $SO(k)$ or $\Ons(k)$ is the same.
Another way to see the duality nature is to couple the original system to 
singlets $\wts^{ij}$ via the superpotential $W=\sum\wts^{ij}(x_ix_j)$.
This corresponds in the dual theory to the superpotential
$W=\sum s_{ij}(\wtx^i\wtx^j)+\sum\wts^{ij}s_{ij}$. 
Integrating out $\wts^{ij}$ eliminates $s_{ij}$ and we have the
theory of the fundamentals $\wtx^i$ only. Note that we have
the relation $\wts^{ij}=-(\wtx^i\wtx^j)$, 
which is the singlet/meson correspondence 
(\ref{Omesons}) again up to a sign.

In what follows, we shall provide several evidences of the claimed duality.

\subsection*{\sl Special Cases}

Special cases of the duality for small values of $k$ and $N$
had already been encountered and
established in the earlier sections:
$$
\begin{array}{|c|c|c|c|}
\hline
\mbox{$SO(1)/\Ostns(1)$, $N=1$}&
\mbox{$SO(2)$, $N=2$}&\mbox{$\Ostns(2)$, $N=2$}
&\mbox{$SO(1)/\Ostns(1)$, $N=2$}\\
\hline
\ref{subsec:rr1}&\ref{subsec:rr2}&
\ref{subsec:O2dual}&\ref{subsec:O2rr2}\\
\hline
\end{array}
$$
In fact, the author used these cases as hints to find the general duality.
The top row shows the left hand sides of the duality, 
and the number below
shows the section in which the duality appeared.
We define $\Ost(1)$ {\it resp}. $\Ons(1)$ theories as
the standard {\it resp}. non-standard $\Z_2$ orbifold
when $N-1$ is odd, and the definition is continued to
the $N-1$ even case by introducing a ``regulator'' field.
In other words, by periodicity,
$\Ost(1)=\Z_2$ {\it resp}. $\Z_2(-1)^{F_s}$ 
and $\Ons(1)=\Z_2(-1)^{F_s}$ {\it resp}. $\Z_2$ 
when $N$ is even {\it resp}. odd.

\subsection*{\sl From $\Ost/SO$ Duality To $\Ons$ Duality}

The duality between $\Ons(k)$ and $\Ons(N-k+1)$ can be derived
using the chain of standard relations and the duality
between $\Ost(k)$ and $SO(N-k+1)$, as in (\ref{chain}).
To see this, replace $({\mathcal A}_3,\tau)$ in (\ref{chain})
by the $SO(k)$ gauge theory with $N$ fundamentals
equipped with the $\Z_2$ symmetry to define the $\Ons(k)$ theory.
Then, $({\mathcal A}_2,\tau)$ should be replaced by
the same theory equipped with the $\Z_2$ symmetry to define the $\Ost(k)$
theory. 
$({\mathcal A}_2/\tau,\widehat{\tau})$ is now the
$\Ost(k)$ theory with $N$ fundamentals equipped with the
quantum $\Z_2$ symmetry.
According to the $\Ost/SO$ duality, this is equivalent to the dual
$SO(N-k+1)$ theory equipped with the symmetry 
$\tau\in O(N-k+1)/SO(N-k+1)$, which replaces $({\mathcal B},\tau)$.
Finally, the same theory equipped with $\tau(-1)^{F_s}$
replaces $({\mathcal C},\tau)$.
The end result is the dual $\Ons(N-k+1)$ theory, with the exchange of
the twisted and untwisted sectors in the RR sector.
The exchange shows that the quantum symmetry $\widehat{\tau}$ in the
$\Ons(k)$ theory corresponds to the symmetry $(-1)^{\bf F}\widehat{\tau}$
in the $\Ons(N-k+1)$ theory.

\subsection*{\sl Central Charge}

The dual pair of theories
have the same symmetry other than the $\Z_2$ symmetry which was already
mentioned: the $U(N)$ or $SU(N)\times U(1)_B$ flavor symmetry
and the vector and axial $U(1)$ R-symmetries.
Charge assignment compatible with the duality statement is
\beq
\begin{array}{ccccc}
&SU(N)&U(1)_B&U(1)_V&U(1)_A\\
\hline
x&{\bf N}&1&0&0\\
\hline
\wtx&\overline{\bf N}&-1&1&0\\
s&{\bf S}&2&0&0
\end{array}
\eeq
The R-charges could be modified by the $U(1)_B$ charge.
The above choice is the unique one that assigns vanishing R-charges to $x$,
and only with this choice, the two $U(1)$ R's can
become parts of the $(2,2)$ superconformal symmetry
in the infra-red fixed point of the original theory.
The latter follows from the following argument \cite{WittenH}:
for large values of $x$ where
the semi-classical sigma model analysis is valid,
the R-currents can be chiral only
if $x$ does not rotate under the R-symmetries.

Assuming that these R-symmetries indeed become the parts of the superconformal
symmetry, one can compute the central charge of the fixed point
 \cite{WittenLG,SilWi,HK,HoTo}:
Each Dirac fermion with vector R-charge $q$ and axial R-charge $\mp 1$
contributes by $-q$ to the normalized central charge
$\whc=c/3$. Recall that the fermionic component 
of a chiral multiplet of vector R-charge $Q$ has $q=Q-1$,
and that the gaugino normally has $q=1$.
The central charge of the infra-red fixed point 
of the original theory is
\beq
\whc=kN-{k(k-1)\over 2}.
\label{coriO}
\eeq
Note that this is the dimension of the Higgs branch $H_k$.
This is consistent with the fact that
the sigma model on a K\"ahler manifold of dimension $d$ 
is classically a conformal field theory
with central charge $\whc=d$.
One the other hand, the central charge of the dual theory
reads
\beq
\whc_{\rm dual}={N(N+1)\over 2}-{(N-k+1)(N-k)\over 2}.
\label{cdualO}
\eeq
The two, (\ref{coriO}) and (\ref{cdualO}), indeed agree.
This comparison can be regarded as the 't Hooft anomaly matching 
for $U(1)_R^2$, which is the only non-trivial one.

\subsection*{\sl The Dual Theory In Some Detail}

Let us study the low energy behaviour of the dual theory.
We first look at the classical flat directions.
The D-term equations are as in (\ref{OD}), and we also have
the F-term equations from the superpotential (\ref{OdualW}),
\beqa
(\wtx^i\wtx^j)=0&&\forall (i,j),\label{Feqdu1}\\
s_{ij}\wtx^j_{\tilde{a}}=0&&\forall (i,\tilde{a}).\label{Feqdu2}
\eeqa
The first equations and the D-term equations require $\wtx^1=\cdots =\wtx^N=0$,
which makes the second equation vacuous.
Thus, the space of classical vacua is just the space $S_N$
parametrized by the singlets $s_{ij}$. 
It is the space of all symmetric $N\times N$ matrices and has dimension 
${N(N+1)\over 2}$.

The gauge group is entirely unbroken everywhere in the 
flat directions, and therefore, quantum effects
of gauge interactions should be taken into account.
Let us first consider the case where the original gauge group is
$\Ost(k)$ and the dual gauge group is $SO(\wtk)$, with
$\wtk=N-k+1$.
As we have done many times in Sections~\ref{sec:O1} and \ref{sec:O2},
we work in the Born-Oppenheimer approximation,
treating the singlets $s_{ij}$ as slow variables and
$SO(\wtk)$ gauge fields and the fundamentals $\wtx^i$ as fast variables.
From this view point,
we may regard $s=(s_{ij})$ as a mass matrix for $\wtx^i$'s,
and its corank ($N$ minus its rank)
is the effective number $N_{\it eff}$ of massless fundamentals.
We have learned that, if $N_{\it eff}\leq \wtk-2$, 
the supersymmetry is spontaneously broken, i.e.,
there is no zero energy state.
Thus, unless the supersymmetry is entirely broken,
the low energy dynamics will concentrate on the locus of
$s_{ij}$'s where the matrix $s$ has corank $\wtk-1$ or higher. That is,
rank $N-(\wtk-1)=k$ or lower,
\beq
S_{N,\leq k}=\Bigl\{ \,s\in S_N\,\Bigr|\,\,{\rm rank}(s)\leq k\,\,\Bigr\}.
\label{defSNk}
\eeq
Let us look at the behaviour of the theory near such a locus.
For concreteness, let us look at the region of $S_N$ where
the last $k\times k$ block of $(s_{ij})$ has rank $k$. We separate 
$\wtx^i$'s into two groups: the first $N-k$ of them,
$\wtx^{\alpha}$ for $\alpha=1,\ldots,N-k$,
and the last $k$ of them, $\wtx^{\mu}$ for $\mu=N-k+1,\ldots, N$.
Fields from the latter group are massive and can be integrated out.
This leaves us with the superpotential
\beq
W=\sum_{\alpha,\beta=1}^{N-k}
\left(s_{\alpha\beta}-\sum_{\mu,\nu=N-k+1}^N
s_{\alpha \mu}s^{\mu\nu}s_{\nu \beta}\right)(\wtx^{\alpha}
\wtx^{\beta})
\label{OdSup}
\eeq
In the above,
$(s^{\mu\nu})$ is the inverse of the last $k\times k$ block $(s_{\mu\nu})$
of $(s_{ij})$.
At this point, we again use what we have learned:
The $SO(\wtk)$ theory with $N_{\it eff}=N-k=\wtk-1$ massless fundamentals
is the free theory of the mesons at low energies. 
Then, the composites $(\wtx^{\alpha}\wtx^{\beta})$ in (\ref{OdSup})
should be regarded as elementary fields and can be integrated out.
This yields the constraints
$s_{\alpha\beta}=\sum s_{\alpha\mu}s^{\mu\nu}s_{\nu\beta}$, which
means that $s$ is of rank $k$ since
\beq
A=BC^{-1}B^T
\,\Longrightarrow\,
\left(\begin{array}{c|c}
A&B\\
\hline
\!\!B^T\!&C
\end{array}\right)
=\left(\begin{array}{c|c}
{\bf 1}_{N-k}&BC^{-1}\!\!\!\\
\hline
&{\bf 1}_k
\end{array}\right)
\left(\begin{array}{c|c}
{\bf 0}_{N-k}&\\
\hline
&C
\end{array}\right)
\left(\begin{array}{c|c}
{\bf 1}_{N-k}&\\
\hline
\!\!\!C^{-1}B^T&{\bf 1}_k
\end{array}\right).
\label{Oasasas}
\eeq
Therefore, the low energy theory is the sigma model whose target space
is the submanifold $S_{N,\leq k}$,
in the region of the field space where the rank of $s$ is at least $k$.
The space $S_{N,\leq k}$ has codimension
${(N-k)(N-k+1)\over 2}$ in $S_N$,\footnote{The subspace of $S_N$ 
consisting of matrices  of corank $i$ or higher
has codimension ${i(i+1)\over 2}$:
to choose such a matrix,
we first choose a subspace of codimension $i$
and then choose a symmetric bilinear form in that subspace.
The first choice involves $i(N-i)$ parameters, as it
corresponds to choosing a point of the Grassmannian
$G(N-i,N)$, and the second choice involves
${(N-i)(N-i+1)\over 2}$ parameters. Therefore the codimension is
${N(N+1)\over 2}-\{i(N-i)+{(N-i)(N-i+1)\over 2}\}={i(i+1)\over 2}$.}
and that explains the central charge (\ref{cdualO}).
It can be regarded as the same space as the the Higgs branch $H_k=H_{O(k),N}$ 
of the original theory, in the sense that both spaces are parametrized by
$N\times N$ symmetric matrices of rank $k$ or less: $(x_ix_j)$ for
$H_{O(k),N}$ and $s_{ij}$ for $S_{N,\leq k}$, which indeed 
correspond to each other under the duality (\ref{Omesons}).

The analysis for the case where the original gauge group
is $\Ons(k)$ and the dual group is
$\Ons(\wtk)$ proceeds in the same way.
We find that the singlet $s_{ij}$ is constrained to
be in the subspace $S_{N,\leq k}$ and the dual theory reduces
to the sigma model on $S_{N,\leq k}$ in the open domain of rank exactly $k$. 
Of course, this dual pair is different from the one above.
In the original side, they differ in the orbifold projections 
in the twisted NSNS and untwisted RR sectors.
In the dual side, the non-standard orbifold should be at work in 
the $\Ons(\wtk)$ theory.

Finally, let us study the case where the original gauge group
is $SO(k)$ and the dual group is $\Ost(\wtk)$.
The analysis of the dual theory proceeds in the same way
until the point where we use the
low energy description of the theory with $N_{\it eff}=\wtk-1$. 
In the present case, where the gauge group
is $\Ost(\wtk)$, there are two copies of the free theory
of invariants $(\wtx^{\alpha}\wtx^{\beta})$.
Therefore, we have a {\it double cover} 
of $S_{N,\leq k}$ at least over the open subset consisting
of matrices $s$ of maximal rank $k$. The two sheets are exchanged under 
the $\Z_2$ quantum symmetry of the orbifold.
Let us compare it with the Higgs branch of the original theory,
$H_k=H_{SO(k),N}$. Since the Higgs branch for
the $O(k)$ theory is obtained by a $\Z_2$ quotient
of the one for the $SO(k)$ theory, $H_{SO(k),N}$ is indeed a double cover of
$H_{O(k),N}$.
The baryons $[x_{i_1}\cdots x_{i_k}]$ are the ones that distinguish
the two sheets above $H_{O(k),N}$, and 
they are indeed claimed to be twist fields in the dual theory.

\subsection*{\sl Flow By Complex Mass}

Let us consider the $\Ost(k)$ {\it resp}. $SO(k)$ {\it resp}. $\Ons(k)$ 
gauge theory
with $N$ fuandamental matter fields with a superpotential 
mass term for one of them, say the last one,
$W=(x_Nx_N)$.
This introduces a term $s_{NN}$ in the dual superpotential,
\beq
W=\sum_{i,j=1}^Ns_{ij}\wtx^i\wtx^j+s_{NN}.
\eeq
If we integrate out $s_{NN}$, we obtain the constraint $(\wtx^N)^2+1=0$
which can be solved by
\beq
\wtx^N=\left(\begin{array}{c}
0\\
\vdots\\
0\\
\pm i
\end{array}\right).
\label{wtxN}
\eeq
It breaks the dual gauge group to the subgroup
$SO(\wtk-1)$ {\it resp}. $\Ost(\wtk-1)$ {\it resp}. $\Ons(\wtk-1)$.
Note that the solution is unique
except in the case where
$\wtk=1$ and the dual gauge group is $SO(1)=\{1\}$, i.e.,
$N=k$ and the original gauge group is $\Ost(k)$,
in which the two solutions $+i$ and $-i$ are inequivalent.
Plugging (\ref{wtxN}) to the 
superpotential we have terms of the form
$\pm 2i s_{j'N}\wtx_{\wtk}^{j'}$ for
$j'=1,\ldots,N-1$. Integrating out $s_{j'N}$, we obtain the constraint
$\wtx_{\wtk}^{j'}=0$. Thus, we are left with
the $SO(\wtk-1)$ {\it resp}. $\Ost(\wtk-1)$ {\it resp}. $\Ons(\wtk-1)$ 
gauge theory with
$N-1$ fundamentals $\wtx^{\prime 1},\ldots,\wtx^{\prime N-1}$
and ${(N-1)N\over 2}$ singlets $s_{i'j'}$,
having the remaining superpotential.
This is indeed the dual of the 
$\Ost(k)$ {\it resp}. $SO(k)$ {\it resp}. $\Ons(k)$ theory
with $N-1$ massless fuandamentals.

For the case where the starting point is $N=k$, the dual gauge group
$SO(1)$ or $\Ostns(1)$ is trivial or completely broken by (\ref{wtxN}) and
the ``fundamentals'' $\wtx^{j'}$'s are completely gone.
If the original gauge group is $SO(k)$ or $\Ons(k)$ 
(the dual group $\Ost(1)$ or $\Ons(1)$),
we have the free theory of only the singlets $s_{i'j'}$, which
correspond to the mezons $(x_{i'}x_{j'})$ in the original.
If the original gauge group is $\Ost(k)$ (the dual group $SO(1)$),
 then, since the two solutions (\ref{wtxN}), i.e.,
$\wtx^N=\pm i$, are inequivalent, we have two copies of
the free theory of the singlets.
To summarize, the duality reproduces the effective theory for
the $N=k-1$ theory obtained in Section~\ref{subsec:freeO}.

\subsection*{\sl Vacuum Counting With Twisted Mass}

As another test of duality, let us compare the number of
supersymmetric ground states, or more precisely the Witten index,
of the dual pair perturbed by twisted masses.
The counting for the original theory,
where $x_1,\ldots,x_N$ are given 
twisted masses $\wtm_1,\ldots,\wtm_N$, has been done in 
Section~\ref{subsec:Oktmass} under the genericity assumption 
including (\ref{genericm}).
As this is associated with the $U(1)^N\subset U(N)$ global symmetry,
this corresponds in the dual side to giving twisted mass
$\wtm_i+\wtm_j$ to $s_{ij}$ and $-\wtm_i$ to $\wtx^i$.
Note that the masses for $s_{ij}$ are all non-zero 
under (\ref{genericm}).
We discussed in Section~\ref{subsec:rr1}
the vacuum counting problem in such a system.
As argued there, the spectrum of supersymmetric ground states,
or at least the Witten index,
is expected not to change if we turn off the superpotential,
since no vacuum runs off to nor 
come in from infinity.
(This was confirmed in a simple example by an exact analysis in 
Appendix~\ref{app:SUSYQM}.)
Then, since the singlet sector provides ``one'' as the number
of ground states, the total number is the same as the theory
of $\wtx$'s only. For this the result of Section~\ref{subsec:Oktmass}
is applicable, though of course for the dual group.
Thus, the comparison is a straightforward task --- 
just stare the table (\ref{WItable}).
A complete match!

\subsection{Chiral Rings}
\label{subsec:ringO}

We discuss the chiral rings, both $(c,c)$ and $(a,c)$ rings, of the models
we are studying.
In some cases, the duality can be used to determine them.
In some other cases, we can determine the rings in
both sides of the dual pair and the result can be used to test the duality.

\subsection*{\sl The $(c,c)$ Ring}

The classical $(c,c)$ ring 
is the ring of gauge invariant polynomials of the
chiral multiplet fields.
For the $O(k)$ theory (in the untwisted sector) and for 
the $SO(k)$ theory, it is respectively \cite{Weyl}
\beq
\C[x_1,\ldots,x_N]^{O(k,\C)}=\C\bigl[\,(x_ix_j)\,\bigr]
\bigr/J_1,
\label{HringO}
\eeq
and
\beq
\C[x_1,\ldots,x_N]^{SO(k,\C)}=\C\bigl[\,(x_ix_j),\,
[x_{i_1}\cdots x_{i_k}]\,\bigr]
\bigr/(J_1,J_2,J_3),
\label{HringSO}
\eeq
where $J_1, J_2, J_3$ denote relations of the form
\beqa
J_1:&&\det\left(\begin{array}{ccc}
(x_{i_0}x_{j_0})&\cdots&(x_{i_0}x_{j_{k}})\\
\vdots&&\vdots\\
(x_{i_k}x_{j_0})&\cdots&(x_{i_k}x_{j_k})
\end{array}\right)
=0,\\
J_2:&&[x_{i_1}\cdots x_{i_k}][x_{j_1}\cdots x_{j_k}]
=\det\left(\begin{array}{ccc}
(x_{i_1}x_{j_1})&\cdots&(x_{i_1}x_{j_{k}})\\
\vdots&&\vdots\\
(x_{i_k}x_{j_1})&\cdots&(x_{i_k}x_{j_k})
\end{array}\right),\\
J_3:&&\sum_{p=0}^k(-1)^p[x_{i_0}\cdots\widehat{x_{i_p}}
\cdots x_{i_k}](x_{i_p}x_j)=0.
\eeqa
These relations must be satisfied in the semiclassical regime where
the gauge group is completely broken, and hence must be the exact 
chiral ring relations, as a potential parameter of correction
does not exist. 
In the dual side, the corresponding relations are not visible in
the classical theory and appear only in the infra-red limit of 
the quantum theory.
Indeed the relations $J_1$ are consistent with
the constraint ${\rm rank} (s)\leq k$ obtained 
in the paragraph including (\ref{defSNk})-(\ref{Oasasas}).
In the case of $O(k)$ gauge group, we also have $(c,c)$ ring elements
from the twisted sector. For the $\Ost(k)$ theory, 
the twist fields correspond to $O(\wtk)/SO(\wtk)$ anti-invariants
in the dual theory and are generated by the baryons
$[\wtx^{i_1}\cdots \wtx^{i_{\wtk}}]$.
We can find the relations involving these fields using the above relations 
$J_2$ as well as the classical constraint from the F-term equations 
(\ref{Feqdu1})-(\ref{Feqdu2}).
These lead to the following conclusions. 
We write $s_{ij}=(x_ix_j)$, $b_{i_1\cdots i_k}=[x_{i_1}\cdots x_{i_k}]$,
$\wtb^{i_1\cdots i_{\wtk}}=[\wtx^{i_1}\cdots \wtx^{i_{\wtk}}]$
whenever appropriate.

The $(c,c)$ ring of the $\Ost(k)$ theory is the polynomial
ring of $s_{ij}$ and $\wtb^{i_1\cdots i_{\wtk}}$
modulo the relations
\beq
\det\left(\begin{array}{ccc}
s_{i_0j_0}&\cdots&s_{i_0j_k}\\
\vdots&&\vdots\\
s_{i_kj_0}&\cdots&s_{i_kj_k}
\end{array}\right)
=0,\qquad
\wtb^{i_1\cdots i_{\wtk}}\,\wtb^{j_1\cdots j_{\wtk}}=0,
\qquad
\sum_{j_1=1}^Ns_{ij_1}\,\wtb^{j_1j_2\cdots j_{\wtk}}=0.
\eeq
The $(c,c)$ ring of the $SO(k)$ theory is the polynomial ring
of $s_{ij}$ and $b_{i_1\cdots i_k}$ modulo the relations
\beqa
&\det\left(\begin{array}{ccc}
s_{i_0j_0}&\cdots&s_{i_0j_k}\\
\vdots&&\vdots\\
s_{i_kj_0}&\cdots&s_{i_kj_k}
\end{array}\right)
=0,\qquad
b_{i_1\cdots i_{\wtk}}\,b_{j_1\cdots j_{\wtk}}=
\det\left(\begin{array}{ccc}
s_{i_1j_1}&\cdots&s_{i_1j_k}\\
\vdots&&\vdots\\
s_{i_kj_1}&\cdots&s_{i_kj_k}
\end{array}\right),\nn\\
&
\displaystyle 
\sum_{p=0}^k(-1)^p\, s_{ij_p}\,b_{j_0\cdots\widehat{j_p}\cdots j_k}=0.
\eeqa
The $(c,c)$ ring of the untwisted elements
of the $\Ons(k)$ theory is the polynomial
ring of $s_{ij}$
modulo the relations
\beq
\det\left(\begin{array}{ccc}
s_{i_0j_0}&\cdots&s_{i_0j_k}\\
\vdots&&\vdots\\
s_{i_kj_0}&\cdots&s_{i_kj_k}
\end{array}\right)
=0.
\eeq

\subsection*{\sl The $(a,c)$ Ring}

Classically, the $(a,c)$ ring is the ring of 
gauge invariant polynomials of the scalar components
$\sigma$ of the gauge multiplets, which is isomorphic to the ring of 
Weyl invariant polynomials of the components 
$\sigma_1,\ldots,\sigma_{\ell}$ for the maximal torus.
Examples of elements are
\beq
c_{2i}={\rm tr}(\sigma^{2i})=\sum_{a=1}^{\ell}\sigma_a^{2i},\qquad
p_{\ell}
={\rm Pf}(\sigma)=\sigma_1\cdots\sigma_{\ell} \,\,\,\mbox{(for $SO(2\ell)$).}
\eeq
As the generators, we can take
$c_2,\ldots,c_{2\ell}$ for $SO(2\ell+1)$, $O(2\ell+1)$ and $O(2\ell)$, and
$c_2,\ldots,c_{2\ell-2},p_{\ell}$ for $SO(2\ell)$.
There are no relations among them and thus the ring is
the polynomial ring of these generating variables.
The underlying vector space is of course infinite dimensional.

The story is different in the quantum theory.
Coulomb branch is lifted by quantum corrections if the theory is regular.
This implies that the underlying vector space of
 the $(a,c)$ ring of the infra-red conformal field theory
is finite dimensional.
Thus, we expect to have quantum relations among the generators,
$c_2,\ldots, c_{2\ell}$, or $c_2,\ldots,c_{2\ell-2},p_{\ell}$.
In addition, we also have $(a,c)$ ring elements from the twisted sector
in the theories with gauge group $O(k)$ or $O(\wtk)$.
In the $SO(k)$ and $\Ost(k)$ theories, where we have
the spectral flow between $(a,c)$ ring elements and RR ground states,
the dimension is expected to be equal to the number (\ref{WItable}) of
supersymmetric ground states in the model perturbed by twisted masses.

The key to find the quantum relations is the relations
(\ref{roots}) for the mass deformed system. In the massless limit, 
$\wtm_i\to 0$, they become
\beq
(\sigma_a)^N=0\qquad a=1,\ldots, \ell.
\label{relrel}
\eeq
From these we would like to extract relations among the gauge 
invariants. Analogous problem has been discussed in \cite{CV}
(see also \cite{WVerlinde,HV}). We apply the ``change of variables method''
from that reference to the case when $N$ is odd, 
which proceeds as follows.
The relations (\ref{relrel}) are the Jacobi relations of
the function
${1\over N+1}(\sigma_1^{N+1}+\cdots+\sigma_{\ell}^{N+1})$,
which is invariant under the $SO(k)$ and $O(k)$ Weyl group
when $N$ is odd. We express this function
in terms of the generators of the $SO(k)$ Weyl invariants, and call it 
$W$;
\beq
{1\over N+1}(\sigma_1^{N+1}+\cdots+\sigma_{\ell}^{N+1})=\left\{
\begin{array}{ll}
W(c_2,\ldots,c_{2\ell-2},p)&\mbox{$k$ even},\\
W(c_2,\ldots,c_{2\ell})&\mbox{$k$ odd},
\end{array}\right.
\eeq
For the $SO(k)$ theory, the $(a,c)$ ring is identified
with the Jacobi ring of $W$, that is, the $(c,c)$ ring of the 
Landau-Ginzburg model with the superpotential $W$.
For the $\Ostns(k)$ theory, the $(a,c)$ ring is identified
with the $(c,c)$ ring of the 
Landau-Ginzburg {\it orbifold} with the superpotential $W$
with respect to $\Ostns(k)/SO(k)\cong \Z_2$.
We are not claiming that the conformal field theory
is dual, or mirror to be precise, 
to the Landau-Ginzburg model/orbifold. 
We are simply identifying the $(a,c)$ ring of our theory
with the $(c,c)$ ring of the LG.
The two theories cannot be mirror to each other
as the $(c,c)$ ring of our theory is infinite dimensional
while the $(a,c)$ ring of the LG is finite dimensional.

This applies both to the original $O(k)$ or
$SO(k)$ theory and to the dual $O(\wtk)$ or $SO(\wtk)$ theory.
Let us compute the ring in some examples, and check against the duality.

\noindent
\underline{$SO(2), N=5$ versus $\Ost(4), N=5$}

\noindent
The $(a,c)$ ring of the $SO(2)$ theory is the Jacobi ring of 
$W={1\over 6}p_1^6$, i.e. the polynomial ring of $p_1$
modulo the relation
\beq
p_1^5=0.
\eeq
Under the $\Z_2=O(2)/SO(2)$,
three elements, $1,p^2,p^4$, are even and two elements, $p,p^3$, are odd.
The $(a,c)$ ring of the $\Ost(4)$ theory is the $(c,c)$ ring of 
a LG orbifold with the superpotential $W={1\over 6}c_2^3-{1\over 2}p_2^2c_2$
with respect to $\Z_2:(c_2,p_2)\to (c_2,-p_2)$.
The spectra of $(c,c)$ elements and RR ground states for the two cases are
as follows (We follow the notation of \cite{IV}. In particular, 
$K_{\tau}\in \Z/2\Z$ is the parameter that distinguishes two possible
orbifold projections):
For $K_{\tau}=1$, the states surviving the orbifold projection are
\beqa
(c,c):&&\left\{\begin{array}{ll}
|0\rangle^{1}_{(c,c)},\,\,c_2|0\rangle^{1}_{(c,c)},\,\,
c_2^2|0\rangle^{1}_{(c,c)}&\mbox{from untwisted}\\
|0\rangle^{\tau}_{(c,c)},\,\,c_2|0\rangle^{\tau}_{(c,c)}
&\mbox{from twisted}
\end{array}\right.
\\
{\rm RR}:&&\left\{\begin{array}{ll}
|0\rangle^{1}_{\rm R},\,\,c_2|0\rangle^{1}_{\rm R},\,\,
c_2^2|0\rangle^{1}_{\rm R}&\mbox{from untwisted}\\
|0\rangle^{\tau}_{\rm R},\,\,c_2|0\rangle^{\tau}_{\rm R}
&\mbox{from twisted}.
\end{array}\right.
\eeqa
For $K_{\tau}=0$, the surviving states are
\beqa
(c,c):&&\left\{\begin{array}{ll}
|0\rangle^{1}_{(c,c)},\,\,c_2|0\rangle^{1}_{(c,c)},\,\,
c_2^2|0\rangle^{1}_{(c,c)}&\mbox{from untwisted}\\
\mbox{~~~~~~~~~~~ none}
&\mbox{from twisted}
\end{array}\right.
\label{nonefromtwisted}
\\
{\rm RR}:&&\left\{\begin{array}{ll}
p_2|0\rangle^{1}_{\rm R}&\mbox{from untwisted}\\
|0\rangle^{\tau}_{\rm R},\,\,c_2|0\rangle^{\tau}_{\rm R}
&\mbox{from twisted}.
\end{array}\right.
\eeqa
We see that we need to take $K_{\tau}=1$ for the $\Ost(4)$ theory,
in order to have $3$ untwisted and $2$ twisted
$(c,c)$ ring elements,
as expected from
the vacuum counting in Section~\ref{subsec:Oktmass}.
(Then $K_{\tau}=0$ should correspond to the $\Ons(4)$ theory.
Note that the number of RR ground states also matches with
(\ref{WItable}) also for $\Ons(4)$.)
The ring relation is standard for the untwisted sector elements.
Relations involving twist operators can be found
using the recent result by Krawitz \cite{Krawitz}.\footnote{We thank
Tyler Jarvis for instruction on the ring structure 
and also pointing out that the correct version
is in his Ph D Thesis.}
Let $1_{\tau}$ be an element corresponding to the state
$|0\rangle^{\tau}_{(c,c)}$. The ring relation is then
\beq
\quad {1}_{\tau}\cdot 1_{\tau}=-c_2,\quad c_2^2\cdot 1_{\tau}=0,
\eeq
in addition to $c_2^3=0$ that comes from
 the Jacobi relations $c_2^2=p_2^2$, $p_2c_2=0$.
The rings for the dual pair 
are indeed isomorphic under the correspondence
\beq
1,p_1,p_1^2,p_1^3,p_1^4\longleftrightarrow 
1,1_{\tau},-c_2,-c_2\cdot 1_{\tau}, c_2^2.
\eeq
We can also perform the test for the other versions of dual pair: 
$\Ost(2)$ versus $SO(4)$ as well as
$\Ons(2)$ versus $\Ons(4)$.
The ring of $\Ost(2)$ is again found from \cite{Krawitz}. It is
generated by $1_{\tau}$  and $p_1^2$ which obey
the relations $p_1^2\cdot 1_{\tau}=0$ and $1_{\tau}\cdot 1_{\tau}=p_1^4$.
The ring of $SO(4)$ is the Jocobi ring of
$W=c_2^3/6-p_2^2c_2/2$, i.e., the polynomial ring of
$c_2$ and $p_2$ modulo the relation $c_2^2=p_2^2,p_2c_2=0$.
The two rings are isomorphic under
\beq
1,p_1^2,1_{\tau}, p_1^4\longleftrightarrow 
1, c_2,p_2,c_2^2.
\eeq
The ring of $\Ons(2)$ is the ring of even polynomials of $p_1$ modulo $p_1^5=0$
while the ring of $\Ons(4)$ is the ring of polynomials of $c_2$ with $c_2^3=0$.
They are obviously isomorphic.

\noindent
\underline{$SO(3), N=7$ versus $\Ost(5), N=7$}

\noindent
The $(a,c)$ ring of the $SO(3)$ theory is the Jacobi ring
of $W={1\over 8}c_2^4$, that is, the polynomial ring of $c_2$ modulo 
the relation
$c_2^3=0$.
The $(a,c)$ ring of the $\Ost(5)$ theory
is the $(c,c)$ ring of the orbifold of the LG model with
$W={1\over 8}c_2^2c_4+{1\over 16}c_4^2-{1\over 16}c_2^4$
by the $\Z_2$ that acts trivially on the variables.
In order to be consistent with (\ref{WItable}), 
the orbifold must be the one which is isomorphic to the model
without the orbifold. The ring is therefore the polynomial
ring of $c_2$ and $c_4$ modulo the relation
$c_2^2+c_4=c_2^3-c_2c_4=0$, that is, the polynomial ring
of $c_2$ modulo the relation $c_2^3=0$.
The two rings are indeed isomorphic to each other.
Nothing changes for the dual pair
$\Ost(3)$ versus $SO(5)$.
For the pair $\Ons(3)$ versus $\Ons(5)$, the ring is doubled on both 
sides --- for each untwisted element, there is a copy in the twisted
sector, and the ring relation is the obvious one. The two rings are
isomorphic to each other.

In this paper, we do not try to give full rational for the above
procedure to determine the $(a,c)$ ring for odd $N$ case, nor
even to propose the ring for even $N$ case. Also,
we do not attempt to prove the isomorphism for general
dual pair. It is possible that the would be proven isomorphism
is related to the level-rank duality for the fusion ring of
the Wess-Zumino-Witten models or for the chiral ring of
Kazama-Suzuki models. We postpone the full account on these for future work.

\section{Symplectic Groups}
\label{sec:Sp}

In this section, we study low energy behaviour of theories with
the symplectic gauge group $USp(k)$ with $N$ chiral multiplets,
$x_1,\ldots,x_N$ in the
fundamental representation ${\bf k}$.
Here $k$ is an even integer,  $k=2\ell$, for $\ell=1,2,3,\ldots$.
We recall that the group $USp(k)$ is the group of $k\times k$ unitary matrix
that preserves the symplectic structure defined by
the matrix
\beq
J_k=\left(\begin{array}{c|c}
&-{\bf 1}_{\ell}\\
\hline
{\bf 1}_{\ell}&
\end{array}\right).
\eeq
That is, $USp(k)$ consists of $k\times k$ matrix $g$ satisfying
$g^{\dag}g={\rm 1}_k$ and $g^TJ_kg=J_k$.
It is simply connected and hence any principal $USp(k)$ bundle
on a closed surface is topologically trivial.
In particular, there is no room for theta angle.

Note that $USp(k)$ and $SU(k)$ coincide at $k=2$.
Some of the results below for the $k=2$ case
had been obtained in \cite{HoTo} as results for $SU(2)$ gauge theories.

\subsection{The Space Of Classical Vacua}

Let us first describe the space of classical vacua. 
We denote the scalar component of the vector multiplet
by $\sigma$. It is a $k\times k$ matrix such that
$J_k\sigma$ is symmetric.
We write $x$ for the $k\times N$ matrix $(x^a_i)$,
and denote by $x_{\uparrow}$ and $x_{\downarrow}$ its
upper and lower $\ell\times N$ submatrices.
The vacuum equation reads
\beqa
&&[\sigma,\sigma^{\dag}]=0,\nn\\
&&x_{\uparrow}x_{\uparrow}^{\dag}=(x_{\downarrow}x_{\downarrow}^{\dag})^T,
\quad
x_{\uparrow}x_{\downarrow}^{\dag}=-(x_{\uparrow}x_{\downarrow}^{\dag})^T,
\label{SpD}
\\
&&\sigma x=\sigma^{\dag} x=0.\nn
\eeqa
The first equation means that, up to gauge transformations, 
$\sigma$ is of the form
\beq
\left(\begin{array}{ccc|ccc}
\sigma_1&&&&&\\
&\ddots&&&&\\
&&\sigma_{\ell}&&&\\
\hline
&&&-\sigma_1&&\\
&&&&\ddots&\\
&&&&&-\sigma_{\ell}
\end{array}\right)
\eeq
Let us introduce $\ell\times N$ quaternion matrix
${\bf x}=x_{\uparrow}+jx_{\downarrow}$. The combination ${\bf x}{\bf x}^{\dag}$
is a self-adjoint quaternion matrix and can be diagonalized using a
$U({\bf H}^k)\cong USp(k)$ conjugation. By the equations on the second
line, this means that (after the gauge rotation)
$x_{\uparrow}x_{\uparrow}^{\dag}=(x_{\downarrow}x_{\downarrow}^{\dag})^T$
is a real diagonal matrix and 
$x_{\uparrow}x_{\downarrow}^{\dag}$ vanishes.
This implies that, with an appropriate $U(N)$ flavor rotation,
the solution can be made into the form
\beq
x=\left(\begin{array}{ccc|ccc|c}
a_1&&&&&&\\
&\ddots&&&&&\\
&&a_m&&&&\\
&&&&&&\\
\hline
&&&a_1&&&\\
&&&&\ddots&&\\
&&&&&a_m&\\
&&&&&&~
\end{array}\right),
\eeq
where $r=2m$ can range over even numbers from $0$ to ${\rm min}\{k,N\}$
({\it resp.} ${\rm min}\{k,N-1\}$) for even ({\it resp.} odd) $N$.
The final equation requires that if the number of non-zero
$\sigma_a$'s is $s$, then the number of non-zero $a_i$'s is at most $\ell-s$.
Let $C_s$ ({\it resp.} $H_r$)
be the set of gauge equivalence class of solutions for
$\sigma$ of rank $2s$ or less ({\it resp}. $x$ of rank $r$ or less).
It has complex dimension $s$ ({\it resp.} $Nr-{r(r+1)\over 2}$).
The space of classical vacua is
\beq
{\mathcal M}=\bigcup_{s=s_{\rm min}}^{\ell}\Bigl(C_s\times H_{k-2s}\Bigr),
\label{SpMc}
\eeq
where $s_{\rm min}=0$ if $N\geq k$
and $s_{\rm min}=[{k-N+1\over 2}]$ if $N\leq k-1$. 
When $N\geq k$ there is a Higgs branch $C_0\times H_k$.
There is always a Coulomb branch $C_{\ell}\times H_0$.
Other components are the mixed Coulomb-Higgs branches.

\subsection{Regularity}

We are interested in regular theory where
all the Coulomb and mixed branches are lifted by quantum corrections.
The computation of the effective twisted superpotential is very simple
compared to the orthogonal groups.
Let us consider the classical Coulomb branch, $C_{\ell}\times H_0$.
The massive vector mutiplets give no contribution to the twisted
superpotential --- there are four multiplets that are charged
under $U(1)_a\times U(1)_b$ and yields $\pi i (\sigma_a+\sigma_b)+\pi i
(\sigma_a-\sigma_b)\equiv 0$.
The massive chiral multiplets give the usual contribution,
and the result is
\beq
\widetilde{W}_{\it eff}
=-\sum_{i,a}\sigma_a(\log\sigma_a-1)
-\sum_{i,a}(-\sigma_a)(\log(-\sigma_a)-1)
=\pi i N\sum_{a=1}^{\ell}\sigma_a.
\eeq
Computation on the mixed branches gives the same result except that the sum
is over $a=1,\ldots,s$ for $C_s\times H_{k-2s}$.
The conclusion is that {\it the theory is regular if and only if
$N$ is odd}.

The theory with even $N$ is not regular. Unlike in the orthogonal
groups, the symplectic group is simply connected and does not
allow any theta angle. Also,
the trick using complex mass does not work here ---
the gauge invariants $[x_ix_j]$ are antisymmetric in 
$i\leftrightarrow j$ and any mass reduces the degrees of freedom by
{\it even} number.

\subsection{Twisted Masses}
\label{subsec:Sptmass}

Let us give twisted masses $\wtm_1,\ldots,\wtm_N$
to the chiral matter fields $x_1,\ldots,x_N$, and study the spectrum of
supersymmetric ground states.
We assume genericity of $\wtm_i$'s including
$\wtm_i+\wtm_j\ne 0$ so that the Higgs branch is lifted everywhere
on the $\sigma$-space.
The effective twisted superpotential is
\beq
\widetilde{W}_{\it eff}
=
-\sum_{i,a}(\sigma_a-\wtm_i)(\log(\sigma_a-\wtm_i)-1)
-\sum_{i,a}(-\sigma_a-\wtm_i)(\log(-\sigma_a-\wtm_i)-1).
\label{WeffSp}
\eeq
and the vacuum equation for $\sigma=\sigma_1,\ldots,\sigma_{\ell}$
reads
\beq
\prod_{i=1}^N(\sigma-\wtm_i)=(-1)^{N+1}\prod_{i=1}^N(-\sigma-\wtm_i).
\label{rootsSp}
\eeq
The solutions are identified under the action of
the Weyl group: permutations and independent sign flips of
$\sigma_a$'s.
We require the solutions to obey
\beqa
&&\sigma_a\ne \pm \wtm_i,\nn\\
&&\sigma_a\ne\sigma_b\quad a\ne b,\label{forbiddenSp}\\
&&\sigma_a+\sigma_b\ne 0\quad \forall (a,b).\nn
\eeqa
The same remark made for the $O(k)$ or $SO(k)$ gauge theory
after (\ref{forbidden}) applies here without modification.

The equation (\ref{rootsSp})
has a single root at $\sigma=0$ and ${N-1\over 2}$ pairs of non-zero roots.
There are solutions for $\sigma_a$'s obeying (\ref{forbiddenSp})
if and only if ${N-1\over 2}\geq\ell$, i.e., $N\geq k+1$.
The number of inequivalent solutions is
\beq
{{N-1\over 2}\choose {k\over 2}}.
\label{WISp}
\eeq
The Weyl group is completely broken at each solution, and hence this is
the number of supersymmetric ground states.

\subsection{$N\leq k$: Supersymmetry Breaking}
\label{subsec:SUSYBSp}

By the result of the previous subsection and applying
 the same argument as in the orthogonal gauge groups,
we conclude the following: There is no normalizable
supersymmetric ground state in pure $USp(k)$ Yang-Mills theory
(an irregular theory)
as well as in the theory with $N\leq k$ massless fundamentals,
for both $N$ odd (regular theory) and $N$ even (irregular theory).

\subsection{$N=k+1$: Free Conformal Field Theory}
\label{subsec:freeSp}

Let us now consider the theory with $N=k+1$ massless fundamentals.
In the regular theory, the Coulomb and mixed branches are lifted
and we are left with the Higgs branch $H_k$ only.
As a complex manifold, the Higgs branch is isomorphic to
the affine space $\C^{(k+1)k\over 2}$ since the chiral ring of gauge invariants
is isomorphic to the polynomial ring of the ${k+1\choose 2}$
symplectic products $[x_ix_j]$ (the ``mesons'') with no relations,
\beq
\C[x_1,\ldots,x_{k+1}]^{Sp(k,\C)}=\C\Bigl[\,[x_ix_j]\,\Bigl|
\,1\leq i < j\leq k+1\,\Bigr]
\eeq
The metric is singular at the roots of Coulomb and mixed branches,
but the singularity is expected to be smeared as these
branches are lifted.
We claim that 
{\it the $USp(k)$ gauge theory with $N=k+1$
massless fundamentals flows in the infra-red limit to
the free conformal field
theory of the ${(k+1)k\over 2}$ mesonic variables.}

\subsection{$N\geq k+3$: Duality}
\label{subsec:dualitySp}

Finally, let us consider the theory with an odd
$N\geq k+3$ massless fundamentals.
We claim that there is a duality:

{\it The $USp(k)$ gauge theory with $N$ fundamentals
$x_1,\ldots, x_N$
flows in the infra-red limit to the same fixed point
as the $USp(N-k-1)$ gauge theory
with $N$ fundamentals, $\wtx^1,\ldots,\wtx^N$,
and ${N(N-1)\over 2}$ singlets, $a_{ij}=-a_{ji}$ ($1\leq i,j\leq N$), 
having the superpotential
\beq
W=\sum_{i,j=1}^Na_{ij}[\wtx^i\wtx^j].
\label{SpdualW}
\eeq
The mesons in the original theory correspond to
the singlets in the dual,}
\beq
[x_ix_j]=a_{ij}.
\label{Spmesons}
\eeq

It is a duality ---
the dual of the dual is the same as the original.
We omit the detail here, except showing the equality,
$N-(N-k-1)-1=k$.

\subsection*{\sl The Central Charge}

The two theories 
has the same symmetry: the $SU(N)\times U(1)_B$ flavor symmetry
and the vector and axial $U(1)$ R-symmetries:
\beq
\begin{array}{ccccc}
&SU(N)&U(1)_B&U(1)_V&U(1)_A\\
\hline
x&{\bf N}&1&0&0\\
\hline
\wtx&\overline{\bf N}&-1&1&0\\
s&{\bf A}&2&0&0
\end{array}
\eeq
We have assigned vanishing R-charges to $x$,
so that the two $U(1)$ R's can
become parts of the $(2,2)$ superconformal symmetry
in the infra-red fixed point of the original theory.
Assuming that they indeed do correspond 
to the parts of the superconformal symmetry,
let us compute the central charge of the original theory.
The one for the original theory is
\beq
\whc=kN-{k(k+1)\over 2},
\label{coriSp}
\eeq
which is also the dimension of the Higgs branch $H_k$.
The one for the dual theory is
\beq
\whc_{\rm dual}={N(N-1)\over 2}-{(N-k-1)(N-k)\over 2}.
\label{cdualSp}
\eeq
The two, (\ref{coriSp}) and (\ref{cdualSp}), indeed agree.

\subsection*{\sl The Dual Theory In Some Detail}

Let us study the low energy behaviour of the dual theory
with gauge group $USp(\wtk)$, with $\wtk=N-k-1$.
The D-term equations are like (\ref{SpD}) and the F-term equations from 
(\ref{SpdualW}) are
\beqa
[\wtx^i\wtx^j]=0&&\forall (i,j),\\
a_{ij}\wtx^j_{\tilde{a}}=0&&\forall (i,\tilde{a}).
\eeqa
They force $\wtx^1=\cdots =\wtx^N=0$ but no condition on the singlets 
$a_{ij}$. The space of classical vacua is
the space $A_N=\{(a_{ij})\}
\cong \C^{N(N-1)\over 2}$ of $N\times N$ antisymmetric matrices.
The gauge group is unbroken everywhere, and
quantum effects of gauge interactions must be taken into account.
When we view $a=(a_{ij})$ as the mass matrix for $\wtx^i$ and study 
the gauge sector first, the nature of the low energy theory
depends very much on the corank of $a$, as it is equal to
the effective number $N_{\it eff}$ of massless fundamentals.
If $N_{\it eff}\leq \wtk-1$, there is no zero energy state.
Thus, the low energy dynamics will concentrate on the locus 
where the matrix $a$ has corank $\wtk+1$ or higher, i.e.,
rank $N-(\wtk+1)=k$ or lower,
\beq
A_{N,\leq k}=\Bigl\{\,\,a\in A_N\,\,\Bigr|
\,\,{\rm rank}(a)\leq k\,\,\Bigr\}.
\label{ANdef}
\eeq
Let us look at the behaviour of the theory near such a locus.
For concreteness, let us look at the region of $A_N$ where
the last $k\times k$ block of $(a_{ij})$ has rank $k$. We separate 
$\wtx^i$'s into two groups: the first $N-k$ and the last $k$ of them,
$\wtx^{\alpha}$ and $\wtx^{\mu}$.
Integrating out the massive fields from the latter,
we obtain the superpotential
\beq
W=\sum_{\alpha,\beta=1}^{N-k}
\left(a_{\alpha\beta}-\sum_{\mu,\nu=N-k+1}^N
a_{\alpha \mu}a^{\mu\nu}a_{\nu \beta}\right)[\wtx^{\alpha}
\wtx^{\beta}],
\label{SpdSup}
\eeq
where
$(a^{\mu\nu})$ is the inverse of the last $k\times k$ block $(a_{\mu\nu})$
of $(a_{ij})$.
The $USp(\wtk)$ theory with $N_{\it eff}=N-k=\wtk+1$ massless fundamentals
is the free theory of the mesons at low energies. 
Then, the composites $[\wtx^{\alpha}\wtx^{\beta}]$ in (\ref{SpdSup})
can be integrate them out as elementary fields and we obtain
the constraints
$a_{\alpha\beta}=\sum a_{\alpha\mu}a^{\mu\nu}a_{\nu\beta}$.
This means that $a$ is of rank $k$ since
\beq
A=\!-BC^{-1}B^T
\Longrightarrow
\left(\begin{array}{c|c}
A&B\\
\hline
\!\!\!-B^T\!&C
\end{array}\right)
=\left(\begin{array}{c|c}
\!\!{\bf 1}_{N-k}\!&BC^{-1}\!\!\!\\
\hline
&{\bf 1}_k
\end{array}\right)
\left(\begin{array}{c|c}
\!\!{\bf 0}_{N-k}\!&\\
\hline
&C
\end{array}\right)
\left(\begin{array}{c|c}
{\bf 1}_{N-k}&\\
\hline
\!\!\!-C^{-1}B^T&{\bf 1}_k
\end{array}\right).
\label{Spasasas}
\eeq
Therefore, the low energy theory is the sigma model whose target space
is the submanifold $A_{N,\leq k}$,
in the region of the field space where
the rank of $a$ is at least $k$.
The space $A_{N,\leq k}$ has codimension
${(N-k)(N-k-1)\over 2}$ in $A_N$,\footnote{The subspace of $A_N$ 
consisting of matrices  of an odd corank $i$ or higher
has codimension ${i(i-1)\over 2}$:
to choose such a matrix,
we first choose a subspace of codimension $i$
and then choose an antisymmetric bilinear form in that subspace.
The first choice involves $i(N-i)$ parameters, as it
corresponds to choosing a point of the Grassmannian
$G(N-i,N)$, and the second choice involves
${(N-i)(N-i-1)\over 2}$ parameters. Therefore the codimension is
${N(N-1)\over 2}-\{i(N-i)+{(N-i)(N-i-1)\over 2}\}={i(i-1)\over 2}$.}
which explains the central charge (\ref{cdualSp}).
It can be regarded as the same space
as the Higgs branch $H_k=H_{USp(k),N}$ 
of the original theory, in the sense that
both spaces are parametrized by $N\times N$ antisymmetric matrices of
rank $k$ or less:
$[x_ix_j]$ for $H_{USp(k),N}$ and 
$a_{ij}$ for $A_{N,\leq k}$, which indeed correspond to each other
under the duality (\ref{Spmesons}).

\subsection*{\sl Flow By Complex Mass}

Let us consider the theory with a superpotential 
mass term for two of the $N$ fundamentals, say the last two,
$W=[x_{N-1}x_{N}]$. This introduces a term $a_{(N-1)\,N}$
in the dual superpotential,
\beq
W=\sum_{i,j=1}^{N}a_{ij}[\wtx^i\wtx^j]+a_{(N-1)\, N}.
\eeq
If we integrate out $a_{(N-1)\, N}$, we obtain the constraint 
$[\wtx^{N-1}\wtx^{N}]+1=0$, which can be solved by
\beq
\wtx^{N-1}=\left(\begin{array}{c}
{\bf e}_{\ell}\\
\hline
{\bf 0}_{\ell}
\end{array}\right),\quad
\wtx^{N}=\left(\begin{array}{c}
{\bf 0}_{\ell}\\
\hline
{\bf e}_{\ell}
\end{array}\right)
\quad\mbox{where}\quad
{\bf e}_{\ell}=\left(\begin{array}{c}
0\\
\vdots\\
0\\
1
\end{array}\right).
\eeq
They break the dual gauge group to the subgroup
$USp(N-k-3)$. Plugging them back to the superpotential,
we have terms of the form
$a_{j'N}\wtx^{j'}_{\ell}-a_{j'(N-1)}\wtx^{j'}_{2\ell}$ for
$j'=1,\ldots,N-2$. Integrating out $a_{j'(N-1)}$ and $a_{j'N}$, 
we obtain the constraint $\wtx^{j'}_a=0$ for $j'=1,\ldots, N-2$ and 
$a=\ell,2\ell$. 
Thus, we are left with
the $USp(N-k-3)$ gauge theory with
$N-2$ fundamentals $\wtx^{\prime 1},\ldots,\wtx^{\prime N-2}$
and ${(N-2)(N-3)\over 2}$ singlets $a_{i'j'}$,
having the remaining superpotential.
This is indeed the dual of the 
$USp(k)$ theory
with $N-2$ massless fundamentals.

If the starting point was $N=k+3$, the dual gauge group $USp(2)$
is completely broken and the fundamentals are all gone.
What remains is 
the free theory of the singlets $a_{i'j'}$ for $i,j=1,\ldots, k+1$,
which correspond to the mesons $[x_ix_j]$.
The duality indeed reproduces the effective theory for
the $N=k+1$ theory obtained in Section~\ref{subsec:freeSp}.

\subsection*{\sl Vacuum Counting With Twisted Mass}

Let us compare the Witten index of the dual pair perturbed by twisted masses.
The counting for the original theory,
where $x_1,\ldots,x_N$ are given 
twisted masses $\wtm_1,\ldots,\wtm_N$, has been done in 
Section~\ref{subsec:Sptmass} under a certain genericity assumption.
This corresponds, in the dual side, to non-zero and generic
twisted masses
$\wtm_i+\wtm_j$ for $a_{ij}$ and $-\wtm_i$ for $\wtx^i$.
We expect that the Witten index
does not change as we turn off the superpotential,
since no vacuum runs off to nor come in from infinity,
Then, it is the same as the theory of $\wtx$'s only. 
For this the result of Section~\ref{subsec:Sptmass}
is applicable, though of course for the dual group.
Thus, we only have to compare (\ref{WISp}) for $USp(k)$ and for $USp(N-k-1)$. 
They indeed agree with each other.

\subsection{Chiral Rings}
\label{subsec:ringSp}

\subsection*{\sl The $(c,c)$ Ring}

The classical (c,c) ring is the
ring of gauge invariant polynomials of the chiral multiplet fields, 
which is known to be \cite{Weyl}
\beq
\C[x_1,\ldots,x_N]^{Sp(k,\C)}=\C\Bigl[\,[x_ix_j]\,\Bigr]
\bigr/(J_0,J_1,\ldots,J_{\ell})
\label{HringSp}
\eeq
where the relations are
\beqa
J_0:&&\sum_{\sigma\in\mathfrak{S}_{k+1}}(-1)^{l(\sigma)}
[x_{j_0}x_{i_{\sigma(0)}}][x_{i_{\sigma(1)}}x_{i_{\sigma(2)}}]\cdots
[x_{i_{\sigma(k-1)}}x_{i_{\sigma(k)}}]=0,\nn\\
J_1:&&\sum_{\sigma\in\mathfrak{S}_{k+1}}(-1)^{l(\sigma)}
[x_{j_0}x_{i_{\sigma(0)}}][x_{j_1}x_{i_{\sigma(1)}}]
[x_{j_2}x_{i_{\sigma(2)}}][x_{i_{\sigma(3)}}x_{i_{\sigma(4)}}]
\cdots
[x_{i_{\sigma(k-1)}}x_{i_{\sigma(k)}}]=0,\nn\\
&&\cdots\\
J_{\ell}:&&\sum_{\sigma\mathfrak{S}_{k+1}}(-1)^{l(\sigma)}
[x_{j_0}x_{i_{\sigma(0)}}][x_{j_1}x_{i_{\sigma(1)}}]\cdots
[x_{j_k}x_{i_{\sigma(k)}}]=0.\nn
\eeqa
As these must be satisfied in the semiclassical regime and
as there is no parameter for corrections,
this must be the $(c,c)$ ring of the theory.
In the dual side, the corresponding relations for $a_{ij}$ appear only 
in the infra-red limit. These are consistent with
the constraint ${\rm rank}(a)\leq k$ obtained in the paragraph
including (\ref{ANdef})-(\ref{Spasasas}).

\subsection*{\sl The $(a,c)$ Ring}

The $(a,c)$ ring of the classical theory is the ring of 
gauge invariant polynomials of $\sigma$, or equivalently, the ring of 
Weyl invariant polynomials of
$\sigma_1,\ldots,\sigma_{\ell}$.
It is the polynomial ring of the invariants
$c_2,\ldots,c_{2\ell}$, where
\beq
c_{2i}={\rm tr}(\sigma^{2i})=\sum_{a=1}^{\ell}\sigma_a^{2i}.
\eeq
The underlying vector space is of course infinite dimensional. 
In the quantum theory, since the Coulomb branch is lifted 
if it is regular (i.e. for odd $N$),
we expect to have relations among $c_2,\ldots,c_{2\ell}$,
so that the underlying vector space has a finite dimension which 
is equal to the number (\ref{WISp}).
The relations are found via the $\wtm_i\to 0$ limit
of (\ref{rootsSp}),
\beq
(\sigma_a)^N=0
\qquad a=1,\ldots, \ell,
\eeq
which are the Jacobi relations of the function
${1\over N+1}(\sigma_1^{N+1}+\cdots+\sigma_{\ell}^{N+1})$. 
We express this function,
which is invariant under the Weyl group when $N$ is odd, 
in terms of $c_2,\ldots,c_{2\ell}$ and call it $W$.
Then, the $(a,c)$ ring is the Jacobi ring of $W(c_2,\ldots,c_{2\ell})$.

Notice that it is isomorphic to the ring for
the $SO(2\ell+1)$ or $\Ost(2\ell+1)$ theory with the same $N$.
Of course this does not mean that the theory is equivalent to
the $SO$ or $\Ost$ theories. They are different in  many other ways,
such as the central charge and
the $(c,c)$ ring.

This holds for both the original $USp(k)$ theory and for the
dual $USp(\wtk)$ theory. It is straightforward to
check that the rings for the dual pair are isomorphic to each other,
at least for low values of $(k,N)$. We do not try to give a proof
for general $(k,N)$ in this paper. (In fact,
a proof in this case is equivalent to a proof in 
 the $SO({\rm odd})$ or $\Ost({\rm odd})$ theories with odd $N$, 
by the isomorphism mentioned above.)
It would be interesting to see if there is a relation to level-rank duality
in Wess-Zumino-Witten fusion rings or in the chiral rings 
in Kazama-Suzuki models.

\section{Motivation, Test, And Application}

\subsection{A Linear Sigma Model Including $\Ost(2)$}
\label{subsec:LSM}

The present work started as an attempt to understand, from
the quantum field theory point of view, the relation
discussed in \cite{HoTa} between two different Calabi-Yau manifolds.
One of the two naturally leads us to consider the following linear sigma model.

It is the theory with the gauge group
$G=(U(1)\times O(2))/\{(\pm 1,\pm {\bf 1}_2)\}$, with the matter fields
\beq
\begin{array}{cc}
\underbrace{p^1~~p^2~~p^3~~p^4~~p^5}&
\underbrace{x_1~~x_2~~x_3~~x_4~~x_5}\\
(-2,{\bf 1})&(1,{\bf 2})
\end{array}
\label{LSM1}
\eeq
and the superpotential
\beq
W=\sum_{i,j,k=1}^5S^{ij}_kp^k(x_ix_j).
\label{WLSM1}
\eeq
$S^{ij}_k=S^{ji}_k$ are complex numbers which
are generic in the sense specified soon.
The theta parameter for the $SO(2)\subset O(2)$
is turned off for the theory to be regular, as we have 5 (odd) doublets.
We often use the parametrization introduced in Section~\ref{subsec:O2tmass}:
\beqa
G={U(1)\times U(1)\over\{(\pm 1,\pm 1)\}}\rtimes\Z_2
&\cong& (U(1)_1\times U(1)_2)\rtimes\Z_2
\\
({[}(g,h){]},*)~~&\longmapsto&~~(gh,gh^{-1},*).\nn
\eeqa
The FI-theta parameters for $U(1)_1$ are equal to those of $U(1)_2$
and are denoted by $(r,\theta)$, or $t=r-i\theta$.
The matter fields are $p^k$'s of charge $(-1,-1)$,
$u_i$'s of charge $(1,0)$ and $v_i$'s of charge $(0,1)$ with respect to
$U(1)_1\times U(1)_2$, and the superpotential (\ref{WLSM1})
reads as
\beq
W=\sum_{i,j,k=1}^5S_k^{ij}p^k(u_iv_j+v_iu_j).
\eeq
The generator $\tau$ of $\Z_2$ 
exchanges $U(1)_1$ and $U(1)_2$ as well as $u_i$ and $v_i$.
The D-term equations read
\beq
-|p|^2+|u|^2=-|p|^2+|v|^2=r,
\label{Deq}
\eeq
and the F-term equations are
\beqa
&&\sum_{ij}S_k^{ij}u_iv_j=0,\quad k=1,\ldots,5,\label{Feq1}\\
&&\sum_{j,k}S^{ij}_kp^ku_j=\sum_{j,k}S^{ij}_kp^kv_j=0,\quad
i=1,\ldots,5.\label{Feq2}
\eeqa

\subsection*{\sl The Low Energy Theory At $r\gg 0$}

Let us analyze the low energy theory at $r\gg 0$.
The D-term equations require $u\ne 0$ and $v\ne 0$, 
thus $U(1)_1\times U(1)_2$ is completely broken. 
The space of $(u,v)$ can be identified as $\CP^4\times\CP^4$ 
on which $\tau$ acts by the exchange of the
two $\CP^4$ factors.
Let $\wtX_S$ be the subspace of $\CP^4\times\CP^4$
consisting of $(u,v)$ satisfying 
the first set of F-term equations, (\ref{Feq1}).
We assume that $\wtX_S$ is a smooth submanifold 
of $\CP^4\times \CP^4$ of codimension $5$.
Namely, we require that the differential of the five equations
has the maximal rank,
$$
\mbox{(C):~ {\it 
If $(u,v)$ represents a point of $\wtX_S$,
then the $5\times 10$ matrix 
$(Su,Sv)$ is of rank $5$.}}
$$
Here,
$Su$ is the square matrix whose $(k,i)^{\rm th}$ entry is
$\sum_jS^{ij}_ku_j$. This condition also
forbids the exchange $\tau$ to have a fixed point: 
A fixed point would be represented by $(u,u)$
where $u\ne 0$ satisfies $\sum_{i,j}S_k^{ij}u_iu_j=0$,
but the matrix $(Su,Su)$
has rank 4 or less, as $Su$ has $u$ in the kernel, contradicting (C).
In particular, the gauge group is completely broken.
Again by the condition (C), the second set of F-term equations
(\ref{Feq2}) requires that all $p^k$'s vanish.
We conclude that the vacuum manifold is
the free quotient $X_S=\wtX_S/\Z_2$, which may also be written simply as
\beq
X_S=\left\{\,\,x\in (\C^2)^{\oplus 5}\,\,\Biggr|\,\,x\ne 0,\,~
\sum_{i,j}S_k^{ij}(x_ix_j)=0~\,\forall k\,\,\right\}\Biggl/
(\C^{\times}\times O(2,\C))/\Z_2.
\label{defXS}
\eeq
$\wtX_S$ and $X_S$ are three dimensional Calabi-Yau manifolds,
with $h^{1,1}(\wtX_S)=2$, $h^{2,1}(\wtX_S)=52$
and $h^{1,1}(X_S)=1$, $h^{2,1}(X_S)=26$ \cite{HoTa}.
The modes transverse to $\wtX_S$ are all massive and hence
the low energy theory is the non-linear sigma model whose target space is
$X_S$.

As always, we need to make a choice of the $\Z_2$ orbifold, which
is a part of the definition of the $O(2)$ gauge theory.
We fix this by requiring the sigma model on $X_S$
to be the standard one, where RR ground states are 
in one to one correspondence with
the cohomology classes of $X_S$, i.e., $\Z_2$ invariant 
(rather than anti-invariant) cohomology classes of $\wtX_S$, 
with Hodge diamond 
\beq
\begin{array}{c}
1\\[-0.2cm]
0~~~~0\\[-0.2cm]
0~~~~1~~~~0\\[-0.2cm]
1~~~26~~~26~~~1\\[-0.2cm]
0~~~~1~~~~0\\[-0.2cm]
0~~~~0\\[-0.2cm]
1
\end{array}
\eeq
For this we need to choose the $\Z_2$ orbifold to be the standard one. 
This is achieved if we choose the $O(2)$ factor of the gauge group
to be the $\Ost(2)$.

\subsection*{\sl Singularity}

Let us find the location of the singular points.
We denote by $\sigma_1$ and $\sigma_2$ the fieldstrength for the 
groups $U(1)_1$ and $U(1)_2$.
They are exchanged by the symmetry $\tau$,
\beq
\sigma_1\,\stackrel{\tau}{\longleftrightarrow}\, \sigma_2.
\label{WeylO2}
\eeq
The effective twisted superpotential is
$\widetilde{W}_{\it eff}
=5(\sigma_1+\sigma_2)(\log(-\sigma_1-\sigma_2)-1)
-5\sigma_1(\log\sigma_1-1)
-5\sigma_2(\log\sigma_2-1)
-t(\sigma_1+\sigma_2)$.
The theory is singular if there is a non-compact Coulomb branch
determined by the equations
$\partial_{\sigma_1}\widetilde{W}_{\it eff}=
\partial_{\sigma_2}\widetilde{W}_{\it eff}=0$ mod $2\pi i\Z$,
i.e.,
\beq
\e^{t}={(-\sigma_1-\sigma_2)^5\over \sigma_1^5}
={(-\sigma_1-\sigma_2)^5\over \sigma_2^5}.
\eeq
There are three inequivalent solutions,
$(\sigma_1,\sigma_2)\propto (1,1)$, $(1,\e^{2\pi i\over 5})$,
$(1,\e^{4\pi i\over 5})$, for the values of $t$ given respectively by
\beq
\e^t~=\,-2^5,~ -(1+\e^{2\pi i\over 5})^5,~
-(1+\e^{4\pi i\over 5})^5.
\label{singuLSM}
\eeq
Note that the first point is special in that
the symmetry (\ref{WeylO2}) is unbroken.
There we must take into account the $\Z_2$ orbifold that acts trivially. 
Since we choose the gauge group to be $\Ost(2)$,
the $\Z_2$ orbifold yields
{\it two copies} of the Coulomb branch at $\e^t=-2^5$.
At each of the symmetry breaking points,
there is just one copy of Coulomb branch.

Location of singular points agree with the result of \cite{HoTa}
(under the relation $x=-\e^{-t}$ to the parameter $x$ of \cite{HoTa}).
In that work, the monodromies of the Picard-Fuchs system are also computed
--- at the three points they are conjugate to
\beq
\left(\begin{array}{cccc}
1&&&2\\
&1&&\\
&&1&\\
&&&1
\end{array}\right),
\quad
\left(\begin{array}{cccc}
1&&&1\\
&1&&\\
&&1&\\
&&&1
\end{array}\right),
\quad
\left(\begin{array}{cccc}
1&&&1\\
&1&&\\
&&1&\\
&&&1
\end{array}\right).
\eeq
The difference is clear. 
A spacetime interpretation would be that there are 
{\it two} massless hypermultiplets of charge $1$
at the first point,
while there is just one massless hypermultiplet of charge $1$ at 
each of the latter two.
This observation tempts us to make a

\noindent
{\bf Conjecture:}~ {\it Suppose a linear sigma model
gives rise to a family of $4d$ ${\mathcal N}=2$ compactifications
of Type II superstrings.
If a disjoint union of $n$ copies of Coulomb branch is supported
at a locus of the moduli space,
then there are $n$ massless hypermultiplets of the same charge
along the locus.}

\subsection*{\sl The Low Energy Theory At $r\ll 0$}

Let us now study the low energy theory at $r\ll 0$.
This time the D-term equations require $p\ne 0$.
We then find that 
the equations $|u|^2=|v|^2$ and $p\cdot (Su,Sv)=0$
force $u=v=0$ under the condition (C).
The space of classical vacua is therefore the space of $p$
obeying $|p|^2=|r|$ modulo phase rotations, that is, a $\CP^4$.
However,
the low energy theory is not just the sigma model
on this vacuum manifold, since a non-trivial subgroup of the gauge group 
is unbroken:
Non-zero values of $p$ break the $U(1)$ factor to $\{\pm 1\}$
and hence the unbroken gauge group is
$(\{\pm 1\}\times \Ost(2))/\{(\pm 1,\pm {\bf 1}_2)\}\cong \Ost(2)$.
We may consider the theory as the $\Ost(2)$ gauge theory with five doublets,
$x_1,\ldots, x_5$, fibred over the $\CP^4$, with the
superpotential (\ref{WLSM1}), which may be rewritten as
\beq
W=\sum_{i,j=1}^5S^{ij}(p)(x_ix_j),
\eeq
where $S^{ij}(p)=\sum_{k=1}^5S_k^{ij}p^k$.
The low energy behaviour
of the $\Ost(2)$ gauge theory for a given value of $p$
depends very much on the rank of the mass
matrix $S(p)=(S^{ij}(p))$.
It follows from the condition (C) that $S(p)$ has at least rank $3$
if $p\ne 0$. That is, a possible rank of $S(p)$ is $3$, $4$ or $5$.
Let us put
\beqa
Y_S&=&\Bigl\{\,[p]\in\CP^4\,\Bigl|\,{\rm rank} S(p)\leq 4\,\Bigr\},
\label{defYS}
\\
C_S&=&\Bigl\{\,[p]\in\CP^4\,\Bigl|\,{\rm rank} S(p)=3\,\Bigr\}.
\label{defCS}
\eeqa
$Y_S$ is a hypersurface of $\CP^4$ given by the equation $\det S(p)=0$.
$C_S$ is of codimension $3$ in $\CP^4$ and hence is a curve.
$Y_S$ has $A_1$ singularity along $C_S$ (see
Section~\ref{subsec:O2rr2}).
The mass matrix $S(p)$ has corank $1$ along
$Y_S\setminus C_S$ and corank $2$ along $C_S$.
In this situation, we may apply
the analysis of Section~\ref{subsec:O2rr1} and \ref{subsec:O2rr2}
concerning the $\Ost(2)$ gauge theories with doublets having
 superpotential with
corank $1$ and $2$ degenerations.
The result there implies that, at least locally, the low energy theory
is the sigma model whose target space is
a double cover of $Y_S$ ramified along $C_S$ --- the cover is of the type
$\C^2\to \C^2/\Z_2$ as in (\ref{A1quot}),
 in the direction of $Y_S$ transverse to $C_S$.
The question is whether such a ramified double cover of $Y_S$ exists
globally. This is in fact proven to be the case in \cite{HoTa}. 
Thus, we can say that the low energy theory is indeed the sigma model
on that double cover.

However, it is quite unsatisfactory in that the local understanding 
cannot tell anything about the existence of the global cover,
let alone the construction of such a cover. 
It would have been disastrous if there were more than one covers 
or if there were none (though neither is the case fortunately). 
This was the actual motivation to look for the dual description of the
$O(2)$ theory with more than two massless flavors, 
which resulted in the discovery of the non-Abelian duality.

\subsection{The Dual Linear Sigma Model Including $SO(4)$}
\label{subsec:DLSM}

As described in Section~\ref{subsec:O2dual},
 the $\Ost(2)$ gauge theory with five doublets
$x_1,\ldots, x_5$ is dual to 
the $SO(4)$ gauge theory with five quartets $\wtx^1,\ldots,\wtx^5$
and fifteen singlets $s_{ij}$ with the superpotential
$W=\sum_{i,j}s_{ij}(\wtx^i\wtx^j)$. 
The singlets are related to the gauge invariants by
$s_{ij}=(x_ix_j)$. 
This duality can be incorporated into the full linear sigma model
and leads us to consider the following theory.

It is the $(U(1)\times SO(4))/\{(\pm 1,\pm {\bf 1}_4)\}$ gauge theory
with the following field content
\beq
\begin{array}{ccc}
\underbrace{p^1~~p^2~~p^3~~p^4~~p^5}&
\underbrace{\wtx^1~~\wtx^2~~\wtx^3~~\wtx^4~~\wtx^5}&
\underbrace{(s_{ij})_{1\leq i\leq j\leq 5}}\\
(-2,{\bf 1})&(-1,{\bf 4})&(2,{\bf 1})
\end{array}
\eeq
and the superpotential
\beq
W=\sum_{i,j}s_{ij}(\wtx^i\wtx^j)+\sum_{i,j,k}S^{ij}_kp^ks_{ij}.
\eeq
The sum of $U(1)$ charges vanish,
$-2\times 5-1\times 20+2\times 15=0$. Hence, 
the axial $U(1)$ R-symmetry is not anomalous and the FI-theta
parameters $(\wtr,\wttheta)$ is a free parameter of the theory.
As we will show later,
we can normalize the latter so that the D-term equation for the $U(1)$ subgroup
reads as
\beq
-2|p|^2-|\wtx|^2+2|s|^2=2\wtr.
\label{DeqdLSM}
\eeq
Let us consider the theory at $\wtr\ll 0$.
The D-term and F-term equations require $p\ne 0$, which breaks
the gauge group to the subgroup
$(\{\pm 1\}\times SO(4))/\{(\pm 1,\pm {\bf 1}_4)\}\cong SO(4)$.
Thus, we have an $SO(4)$ gauge theory fibred over $\CP^4=\{p\}$.
Since this fibration of $SO(4)$ theories is dual to the fibration of
$\Ost(2)$ theories discussed in the previous subsection, we conclude
that this linear sigma model at $\wtr\ll 0$ 
is dual to the original one
(\ref{LSM1})-(\ref{WLSM1}) at $r\ll 0$.
Comparing (\ref{DeqdLSM}) and (\ref{Deq}),
we expect that $\wtr$ agrees with $r$ in the limit $r\to-\infty$. 
The precise relation will be determined momentarily.

\subsection*{\sl The Low Energy Theory At $\wtr\ll 0$}

Let us further study the low energy behaviour of
this theory at $\wtr\ll 0$.
Integrating out the fields $s_{ij}$ we obtain the constraints
\beq
(\wtx^i\wtx^j)+S^{ij}(p)=0\quad \forall (i,j).
\label{Feq5}
\eeq
Since $S(p)$ has rank at least three for $p\ne 0$ by the condition (C),
we find that the $4\times 5$ matrix $(\wtx^i_a)$ has rank at least $3$ 
if it obeys the constraints (\ref{Feq5}). 
This completely breaks the residual $SO(4)$ gauge group.
We find that the manifold of classical vacua is the free quotient
\beq
\wtY_S=\Bigl\{ (p,\wtx)\in \C^{\oplus 5}\oplus (\C^4)^{\oplus 5}
\,\Bigl|\, p\ne 0,\,\,\mbox{(\ref{Feq5})}\,\Bigr\}
\Bigl/{\C^{\times}\times SO(4,\C)\over
\{(\pm 1,\pm {\bf 1}_4)\}}.
\label{thecover}
\eeq
If it is a smooth manifold, we identify the low energy theory
as the non-linear sigma model with this target space.
Since our theory is dual to (\ref{LSM1})-(\ref{WLSM1}), 
this must be the double cover of $Y_S$ which we were
longing for!
This is indeed the case, as we show now.

Since $\wtx^i$ are in the quartet ${\bf 4}$, the $5\times 5$ matrix
$(\wtx^i\wtx^j)$ has at most rank $4$. Thus, (\ref{Feq5}) yields
the constraint on  $p$ that $S(p)$ must have rank at most $4$. 
I.e., we obtained the constraint that $p$ must represent
a point of $Y_S$, this time, by a completely classical argument.
Therefore, $(p,\wtx)\mapsto p$ defines a map
\beq
f:\wtY_S\,\longrightarrow\, Y_S.
\label{forget}
\eeq
Let us find the fibre of this map. A choice of 
$p\ne 0$ that represents a point of $Y_S$ breaks
the group of the quotient (\ref{thecover}) to the subgroup
$(\{\pm 1\}\times SO(4,\C))/\Z_2\cong SO(4,\C)$.
The symmetric matrix $S(p)$ can be diagonalized using
the $GL(5,\C)$ coordinate change and we may assume
$S(p)=-{\rm diag}(c_1,c_2,c_3,c_4,0)$, where three of $c_1,\ldots,c_4$
must be non-zero. 
A solution to the equation (\ref{Feq5}) for such $S(p)$ 
can be made into the following form using the $SO(4,\C)$ symmetry,
\beq
(\wtx^1,\ldots, \wtx^5)=\left(\begin{array}{ccccc}
\pm\sqrt{c_1}&&&&\\
&\pm\sqrt{c_2}&&&\\
&&\pm\sqrt{c_3}&&\\
&&&\pm\sqrt{c_4}&0
\end{array}\right).
\eeq
Even number of sign flip of the non-zero entries
is a part of the $SO(4,\C)$ symmetry, while
odd number of sign flip is not a part of it
if $c_1,\ldots, c_4$ are all non-zero ---
$\det(\wtx^1\cdots \wtx^4)$ distinguishes the orbits.
If one of $c_1,\ldots, c_4$ vanishes, any number of sign flip is a part of
$SO(4,\C)$.
Therefore, the fibre $f^{-1}([p])$ consists of two
points if ${\rm rank} S(p)=4$
while it is a single point if ${\rm rank} S(p)=3$ (i.e. if
$[p]$ belongs to $C_S$).
Let us look at the behaviour of $f$ near the curve $C_S$.
For example, take a point $p_*$ where $S(p_*)=-{\rm diag}(0,0,1,1,1)$.
If we assume that $p_*$ is a smooth point of $C_S$, we may assume
that we can find three coordinates $(a,b,c)$ 
of $\CP^4=\{p\}$ transverse to $C_S$ so that
\beq
S(p)=-\left(\begin{array}{ccccc}
a&c&&&\\
c&b&&&\\
&&1&&\\
&&&1&\\
&&&&1
\end{array}\right).
\eeq
$[p]$ belongs to $Y_S$ if and only if $ab=c^2$.
The equation (\ref{Feq5}) is solved by
\beq
(\wtx^1,\ldots, \wtx^5)=\left(\begin{array}{ccccc}
\xi&\eta&&&\\
&&1&&\\
&&&1&\\
&&&&1
\end{array}\right),\qquad
\xi^2=a,\,\,\,\eta^2=b,\,\,\,\xi\eta=c.
\label{wtxina}
\eeq
This is indeed the expected local behaviour.
Namely, we found that the forgetful map (\ref{forget})
is a double cover of $Y_S$ ramified along the curve $C_S$.

We confirmed that the space $\wtY_S$ in (\ref{thecover})
is the double cover of $Y_S$ which we were looking for.
Note that we were able to obtain the target space $\wtY_S$
by a completely classical analysis
in this dual model, since the gauge group is completely Higgsed.
This construction of $\wtY_S$ 
is strikingly explicit compared to the one given in
\cite{HoTa} based on the relative spectrum of a sheaf of algebra over
$Y_S$.

\subsection*{\sl Smoothness of $\wtY_S$}

Let us discuss the condition for smoothness of $\wtY_S$.
In view of the fact that it is a double cover of $Y_S$ ramified over $C_S$,
with the local structure as above,
it is smooth if $C_S$ is a smooth submanifold of $\CP^4$. 
The curve $C_S$ is locally an intersection of three hypersurfaces.
To see this, take a point $[p_*]\in C_S$ and choose a coordinate system
such that $S(p_*)=-{\rm diag}(0,0,1,1,1)$. Then, $C_S$ can be defined
by $\Delta_{11}(p)=\Delta_{12}(p)=\Delta_{22}(p)=0$ in the region where
the matrix $S(p)_{345}=(S^{ij}(p))_{3\leq i,j\leq 5}$ has rank $3$. Here 
$\Delta_{ij}(p)$
is the determinant of the $4\times 4$ obtained by deleting $i$-th raw and
$j$-the column of $S(p)$. The curve $C_S$ is smooth if and only if 
the $5\times 3$ matrix of the differentials of these three equations has 
rank $3$. The matrix is equal to
$(S_k^{11}\,S_k^{12}\,S_k^{22})$ times $\det S(p)_{345}\ne 0$.
Therefore, the condition for the smoothness of the curve $C_S$ is
$$
\mbox{(D):~ {\it If $S(p_*)$ is of the form
$\left(\begin{array}{c|c}
{\bf 0}_2&\\
\hline
&*_3
\end{array}\right)$,
the $5\times 3$ matrix
$(S_k^{11}\,S_k^{12}\,S_k^{22})$ has rank $3$.}}
$$
The condition (D) is also a necessary
condition for the smoothness of $\wtY_S$. 
To see this, 
let us recall that $\wtY_S$ is defined by the free quotient (\ref{thecover}).
The number of variables is $5+20=25$ while the number of equations is $15$
and the dimension of the group is $1+6=7$. And the difference
$25-15-7=3$ matches the dimension of $\wtY_S$. Thus, $\wtY_S$
is smooth if and only if the $25\times 15$ matrix of
differentials of the equations has rank $15$.
At the point $p_*$ in the discussion above, and for $\wtx$ given by
(\ref{wtxina}) with $\xi=\eta=0$,
the matrix takes the form
\beq
\left(\begin{array}{cc}
{\bf 0}_{20\times 3}&*_{20\times 12}\\
S_k^{11}\,S_k^{12}\,S_k^{22}&*_{5\times 12}
\end{array}\right).
\eeq
It has rank $15$ only if the $5\times 3$ part
$(S_k^{11}\,S_k^{12}\,S_k^{22})$ has rank $3$.
In summary, (D) is the condition for smoothness of $\wtY_S$.

This condition follows from the condition (C) for smoothness of $\wtX_S$.
To see this, suppose (D) fails. That is, there is some $p_*$ such that
$S(p_*)=-{\rm diag}(0,0,1,1,1)$ but 
$(S_k^{11}\,S_k^{12}\,S_k^{22})$ has rank $2$ or less, i.e.,
there is some $(\alpha,\beta,\gamma)\ne (0,0,0)$ such that
$\alpha S_k^{11}+\beta S_k^{22}+\gamma S_k^{12}=0$ for all $k$.
We can find $(u_1,u_2)\ne (0,0)$ and $(v_1,v_2)\ne (0,0)$ such that
$\alpha=u_1v_1$, $\beta=u_2v_2$ and $\gamma=u_1v_2+u_2v_1$.
Then, $u=(u_1,u_2,0,0,0)$ and $v=(v_1,v_2,0,0,0)$ represent a point
of $\wtX_S$. Note that $(Su,Sv)$ is annihilated by $p_*\ne 0$,
which means that $(Su,Sv)$ has rank $4$ or less. I.e., (C) fails.

The converse also holds if we assume that the $\tau$ action on $\wtX_S$
has no fixed point.
Suppose that (C) fails under that assumption. 
Then, there are linearly independent two
5-vectors $u$ and $v$ satisfying $\sum_{i,j}S^{ij}_ku_iv_j=0$ for all $k$
such that $(Su,Sv)$ has rank $4$ or less. That means that there is some
$p_*$ that annihilates this $5\times 10$ matrix.
With a choice of coordinates, we may assume
$u=(1,0,0,0,0)$ and $v=(v_1,1,0,0,0)$.
That $p_*$ annihilates $(Su,Sv)$ means that the matrix
$S(p_*)$ is of the form in the set-up of (D).
However, the equation $\sum_{i,j}S^{ij}_ku_iv_j=0$
reads $S^{11}_kv_1+S^{12}_k=0$, which means that 
$(S_k^{11}\,S_k^{12}\,S_k^{22})$ cannot have rank $3$.
Thus, (D) fails.

To summarize, 
``$\wtX_S$ is smooth'' is equivalent to
``$\wtY_S$ is smooth and the $\tau$ action on $\wtX_S$ is fixed point 
free''. $\wtX_S$, $X_S$ and $\wtY_S$ are all smooth under the condition (C).

\subsection*{\sl The FI-Theta Parameters}

Let us now carry out the promise concerning 
the FI-theta parameters $(\wtr,\wttheta)$.
The Lie algebra of the gauge group is the direct sum of
$\mathfrak{u}(1)$ and $\mathfrak{so}(4)$, where 
$\alpha\in \mathfrak{u}(1)$, regarded as a real number,
generates the one parameter subgroup $\{[(\e^{it\alpha},{\bf 1}_4)]\}_{t\in \R}$.
We denote by $F_{\mathfrak{u}(1)}$ the $\mathfrak{u}(1)$
component of the curvature. On a closed worldsheet $\Sigma$, the flux
$\int_{\Sigma}F_{\mathfrak{u}(1)}$ obeys a certain quantization
condition. For the usual $U(1)$ gauge group the condition is that
the flux can take all values of $2\pi\Z$.
For the present gauge group, it can take all values of
$\pi\Z$. For example, we may decompose $\Sigma$ into two parts by a circle
$S^1$ parametrized by $t\in \R/2\pi\Z$ and consider the principal
bundle determined by the transition function along $S^1$, given by
$[(\e^{it\over 2},h_t)]$, where $h_t$ is some
$SO(4)$-valued function such that $h_{t+2\pi}=-h_t$.
Then, the flux for any connection of this bundle is $\pm \pi$ (the sign
depends on the orientation of $\Sigma$ versus that of $S^1$).
Therefore, the theta term must be of the form
$\int_{\Sigma}{2\wttheta\over 2\pi}F_{\mathfrak{u}(1)}$,
rather than the usual $\int_{\Sigma}{\wttheta\over 2\pi}F_{\mathfrak{u}(1)}$,
in order to have the periodicity $\wttheta\equiv\wttheta+2\pi$.
The corresponding twisted superpotential is
\beq
\widetilde{W}_{\it tree} =-2\,\wtt\,\sigma_{\mathfrak{u}(1)},
\label{tWdLSM}
\eeq
for $\wtt=\wtr-i\wttheta$.
The FI parameter $\wtr$
enters into the D-term equation indeed as (\ref{DeqdLSM}).

Let us find the singular points in the parameter space, in order
to find the precise relation between $(\wtr,\wttheta)$ and
$(r,\theta)$. A maximal torus of the gauge group is
$(U(1)\times SO(2)\times SO(2))/\{(\pm 1,\pm {\bf 1}_2, \pm {\bf 1}_2)\}$
and we identify it as $U(1)_0\times U(1)_1\times U(1)_2$,
where the element $[(z,h_1,h_2)]$ of the former group
is identified with the element $(g_0,g_1,g_2)$ of the latter group by
\beq
g_0=z^2,\quad g_1=zh_1,\quad g_2=zh_2.
\eeq
Here we abuse the notation: the rotation of $\R^2$ 
by an angle $\alpha$ is identified 
with the element $\e^{i\alpha}\in U(1)$.
The $SO(4)$ Weyl group action $(h_1,h_2)\mapsto (h_2,h_1)$,
$(h_1^{-1},h_2^{-1})$ corresponds to
\beq
(g_0,g_1,g_2)\mapsto (g_0,g_2,g_1),~
(g_0,g_0g_1^{-1},g_0g_2^{-1}).
\label{WeylSO4}
\eeq
By $z^2=g_0$,
the tree level twisted superpotential (\ref{tWdLSM}) is written as
$\widetilde{W}_{\it tree}=-\wtt\sigma_0$.
The fields have the following charges under 
$U(1)_0\times U(1)_1\times U(1)_2$:
\beq
\begin{array}{ccc}
\underbrace{p^1~~p^2~~p^3~~p^4~~p^5}&
\underbrace{\wtx^1~~\wtx^2~~\wtx^3~~\wtx^4~~\wtx^5}&
\underbrace{(s_{ij})_{1\leq i\leq j\leq 5}}\\
(-1,0,0)&{{(-1,1,0)\atop (0,-1,0)}\atop {(-1,0,1)\atop (0,0,-1)}}
&(1,0,0)
\end{array}
\eeq
Writing down the effective twisted superpotential and extremizing it, we 
obtain the equations determining the Coulomb branch,
\beq
{(-\sigma_0)^5(-\sigma_0+\sigma_1)^5(-\sigma_0+\sigma_2)^5
\over (\sigma_0)^{15}}=-\e^{\wtt},\quad
{(-\sigma_1)^5\over (-\sigma_0+\sigma_1)^5}
={(-\sigma_2)^5\over (-\sigma_0+\sigma_2)^5}
=1.
\eeq
The sign in $-\e^{\wtt}$ comes from integrating out the off diagonal
components of the vector multiplet, as in the $\pi i\sigma_a$ shift
(\ref{Weffoddk}) in Section~\ref{subsec:Oreg}.
We need to avoid solutions such that
$\sigma_1=\sigma_2$ and $\sigma_0=\sigma_1+\sigma_2$ 
at which the unbroken subgroup is bigger than the maximal torus.
By the second set of equations, we find
$\sigma_a={\sigma_0\over 1+\omega_a}$, with $\omega_a^5=1$, for $a=1,2$,
where $\omega_1=\omega_2^{\pm 1}$ needs to be avoided.
The Weyl group action (\ref{WeylSO4})
is translated into
$(\omega_1,\omega_2)\mapsto (\omega_2,\omega_1)$, 
$(\omega_1^{-1},\omega_2^{-1})$.
The are four inequivalent possibilities,
$(\omega_1,\omega_2)=(\e^{2\pi i\over 5},\e^{4\pi i\over 5})$,
$(\e^{-{2\pi i\over 5}},\e^{4\pi i\over 5})$,
$(1,\e^{2\pi i\over 5})$,
$(1,\e^{4\pi i\over 5})$.
Correspondingly, $\wtt$ has values
\beq
\e^{\wtt}\,=\,-1,~\,-1,~\,
-2^{-5}(1+\e^{4\pi i\over 5})^5,~\,
-2^{-5}(1+\e^{2\pi i\over 5})^5.
\eeq
The Weyl group is completely broken at each of 
the four Coulomb branches. Therefore, there are two copies of
one-dimensional Coulomb branch at $\e^{\wtt}=-1$, while 
there is one at each of the other two values of $\e^{\wtt}$.
Comparing it with (\ref{singuLSM}), we find the relation
between $t$ and $\wtt$: It is
$\e^t=2^5\e^{\wtt}$, i.e.,
\beq
r=\wtr+5\log 2,\quad\theta=\wttheta.
\eeq
We have reproduced, in a very different way, the observation 
that $\e^t=-2^5$ has a double Coulomb branch.

\subsection*{\sl The Low Energy Theory At $\wtr\gg 0$}

Nothing stops us from studying the low energy behaviour of
the model in the opposite regime
$\wtr\gg 0$. 
The D-term equation (\ref{DeqdLSM}) requires that $s=(s_{ij})$ is non-zero.
This breaks the gauge group to $SO(4)$.
We have an $SO(4)$ gauge theory fibred over $\CP^{14}=\{s\}$. 
The $5$ quartets of this theory has mass matrix
$s_{ij}$ and the nature of the low energy theory depends on its rank.
Analysis of such a system has been carried out in 
Section~\ref{subsec:dualityO}, which can be applied here without 
modification. By the supersymmetry breaking for the $SO(4)$ theory with
$N_{\it eff}\leq 2$, we find that the low energy dynamics concentrates
near the locus of $\CP^{14}$ where $s_{ij}$ is of rank $2$ or less. 
The low energy description of the theory with $N_{\it eff}=3$ 
in terms of composite mesons tells us that,
inside the open subset of $\CP^{14}$ where $s$ has rank at least $2$,
the theory reduces to the sigma model whose target space is the
locus of rank exactly $2$.
In the present case, we also have the F-term constraints
\beq
\sum_{i,j}S_k^{ij}s_{ij}=0,\quad k=1,\ldots,5.
\label{Feq6}
\eeq
Thus, the low energy theory is a simple non-linear sigma model
whose target space is
\beq
\Bigl\{\,[s]\in \CP^{14}\,\Bigl|\,\, {\rm rank} \,s= 2,\,\,
(\ref{Feq6})\,\Bigr\}.
\label{spsp}
\eeq
To be precise, we may need to worry about the rank 1 locus.
Here we simply assume genericity of $S^{ij}_k$ so that
there is no rank 1 solution to (\ref{Feq6}).
This is equivalent to the assumption that 
no $u\ne 0$ solves $\sum_{i,j}S_k^{ij}u_iu_j=0$, i.e.,
the $\tau$ action on $\wtX_S$ is free, which is guaranteed 
by the condition (C).
The space (\ref{spsp}) is isomorphic to $X_S=\wtX_S/\Z_2$ 
via the correspondence $s_{ij}\propto (x_ix_j)$
--- compare (\ref{spsp}) with (\ref{defXS}).
We have reproduced the low energy theory of the
original linear sigma model at $r\gg 0$.

\subsection*{\sl Summary}

In the original $(U(1)\times \Ost(2))/\Z_2$ theory,
the $\Ost(2)$ gauge symmetry is completely
Higgsed and the classical analysis suffices
in the $r\gg 0$ phase while it is entirely unbroken and its strong
quantum effect is essential in the $r\ll 0$ phase.
In the dual $(U(1)\times SO(4))/\Z_2$ theory,
the $SO(4)$ gauge symmetry is unbroken in the $r\gg 0$ phase while it is
completely Higgsed in the $r\ll 0$ phase. 
The exchange between Higgs/weak/classical
and confinement/strong/quantum is a typical feature of duality.

\subsection{$SO(2)$ And $\Ons(2)$ Versions}
\label{subsec:SOOns}

Purely from curiosity, we study the linear sigma model 
of Section~\ref{subsec:LSM} in which the $\Ost(2)$ factor 
of the gauge group is replaced by $SO(2)$ or $\Ons(2)$.

\subsubsection*{$SO(2)$}

Let us consider the model with the gauge group 
$(U(1)\times SO(2))/\{(\pm 1,\pm {\bf 1}_2)\}\cong U(1)_1\times U(1)_2$.
The FI-theta parameters $t_1$ and $t_2$ of the two $U(1)$ factors
are independent in this model.
The D-term equations read
\beq
-|p|^2+|u|^2=r_1,\quad -|p|^2+|v|^2=r_2,
\eeq
and the F-term equations remain the same as (\ref{Feq1})-(\ref{Feq2}).
The model has three classical phases:
(I) $r_1>0$, $r_2>0$, (II) $r_2<0$, $r_1>r_2$,
and (III) $r_1<0$, $r_1<r_2$.
The D-term equations require $u\ne 0$ and $v\ne 0$ in Phase I,
$u\ne 0$ and $p\ne 0$ in Phase II, and 
$v\ne 0$ and $p\ne 0$ in Phase III.
The low energy theory in the respective phase
is the sigma model with the target space
\beqa
X_{\rm I}&=&\Bigl\{\,(u,v)\in \CP^4\times \CP^4\,\Bigr|
\sum_{i,j}S_k^{ij}u_iv_j=0~\forall k\,\,\Bigr\},\nn\\
X_{\rm II}&=&\Bigl\{\,(u,p)\in \CP^4\times \CP^4\,\Bigr|
\sum_{i,j}S_k^{ij}p^ku_i=0~\forall j\,\,\Bigr\},\\
X_{\rm III}&=&\Bigl\{\,(v,p)\in \CP^4\times \CP^4\,\Bigr|
\sum_{i,j}S_k^{ij}p^kv_j=0~\forall i\,\,\Bigr\}.\nn
\eeqa
Note that $X_{\rm I}=\wtX_S$ and $X_{\rm II}\cong X_{\rm III}$.
Hodge diamond of RR ground states is
\beq
\begin{array}{c}
1\\[-0.2cm]
0~~~~0\\[-0.2cm]
0~~~~2~~~~0\\[-0.2cm]
1~~~52~~~52~~~1\\[-0.2cm]
0~~~~2~~~~0\\[-0.2cm]
0~~~~0\\[-0.2cm]
1
\end{array}
\eeq
The quantum K\"ahler moduli space can be found by looking for Coulomb branch
vacua.
Writing down the effective twisted superpotential and extremizing it,
we obtain the vacuum equations for $\sigma_1$ and $\sigma_2$, which read for
$z:=\sigma_2/\sigma_1$ as
\beq
\e^{t_1}=-(1+z)^5,\qquad \e^{t_2}=-(1+z^{-1})^5.
\eeq
This provides a parametric representation of the singular locus.
We indeed see the three phase boundaries:
the I-II phase boundary corresponds to $z\to \infty$ 
where $\e^{t_1}\to \infty$ and $\e^{t_2}\to -1$,
the I-III boundary corresponds to $z\to 0$ where $\e^{t_2}\to \infty$ and 
$\e^{t_1}\to -1$,
and the II-III boundary corresponds to $z\to -1$ where 
$\e^{t_1}\to 0$, $\e^{t_2}\to 0$
and $\e^{t_1-t_2}\to -1$.

Let us look at the locus
\beq
r_1=r_2\ll 0,\quad\mbox{and}\quad \theta_1=\theta_2.
\eeq
It is on the classical II-III phase boundary but avoids the
quantum phase boundary which is the line on the opposite side
$\theta_1-\theta_2=\pi$ (mod $2\pi \Z$).
The theory is regular and we may apply our understanding
of the low energy behaviour of the $SO(2)$ gauge theory or its dual
$\Ost(4)$ gauge theory.
This tells us that the low energy theory is the orbifold conformal field 
theory
\beq
\wtY_S/\Z_2,
\eeq
where $\wtY_S$ is defined by (\ref{thecover}) and
$\Z_2$ is the symmetry associated with $\Ost(4)/SO(4)$.
Note that the map
$f:\wtY_S\to Y_S$ in (\ref{forget}) is the mathematical
quotient with respect to this $\Z_2$.

We would like to make two remarks
concerning the dual model with gauge group 
$(U(1)\times \Ost(4))/\{(\pm 1,\pm {\bf 1}_4)\}$.
The center of this group is $U(1)$ and it may appear that the model 
has just one K\"ahler parameter.
This corresponds to the $t_1=t_2$ subspace of
the original quantum K\"ahler moduli space. 
The missing exactly marginal $(a,c)$
operator should come from the twisted sector with respect to
the $\Z_2\cong \Ost(4)/SO(4)$. Indeed, this must be the twist operator
denoted by $1_{\tau}$ in the study of
the $(a,c)$ ring of the relevant
$\Ost(4)$ theory in Section~\ref{subsec:ringO}.
Another remark is about the local analysis of the dual in
the $\wtr\gg 0$ phase.
The difference from the $SO(4)$ case occurs at the point where we use
the low energy description of the theory with 
$N_{\it eff}=3$ massless fundamentals:
For $SO(4)$ we had one copy of the free theory of composite mesons
but for $\Ost(4)$ we have two copies.
This gives us a double cover of (\ref{spsp}),
and that must be the double cover
$\wtX_S$ of $X_S$.

\subsubsection*{$\Ons(2)$}

Next, we consider the model with the gauge group 
$(U(1)\times \Ons(2))/\{(\pm 1,\pm {\bf 1}_2)\}\cong 
(U(1)_1\times U(1)_2)\rtimes \Z_2(-1)^{F_s}$.
This is a one parameter model, $t_1=t_2=t$, with
the same D- and F-term equations as in Section~\ref{subsec:LSM}.
The difference is that the orbifold group is
the non-standard one.
Accordingly the low energy theory is the non-standard orbifold:
At $r\gg 0$ we have
\beq
\wtX_S/\Z_2(-1)^{F_s},
\eeq
and at $r\ll 0$ we have
\beq
\wtY_S/\Z_2(-1)^{F_s}
\eeq
Hodge diamond of RR ground states is
\beq
\begin{array}{c}
0\\[-0.2cm]
0~~~~0\\[-0.2cm]
0~~~~1~~~~0\\[-0.2cm]
0~~~26~~~26~~~0\\[-0.2cm]
0~~~~1~~~~0\\[-0.2cm]
0~~~~0\\[-0.2cm]
0
\end{array}
\eeq
The singular points are $\e^t=-2^5$, $-(1+\e^{2\pi i\over 5})^5$
and $-(1+\e^{4\pi i\over 5})^5$ as in (\ref{singuLSM}) but this time, 
there is only {\it one} copy of Coulomb branch at the first point
as well as in the other two.

The dual model with gauge group
$(U(1)\times \Ons(4))/\{(\pm 1,\pm {\bf 1}_4)\}$
has no other exactly marginal
operator than the one that generates $\delta \wtt$.
The orbifold with respect to
$\Ons(4)/SO(4)$ is opposite to that of $\Ost(4)/SO(4)$
in the twisted NSNS sector, and we have indeed seen in 
(\ref{nonefromtwisted}) that there is no twisted $(a,c)$ ring element
in the relevant $\Ons(4)$ theory.
In the $\wtr\gg 0$ phase, we find that the target space
is (\ref{spsp}), i.e.
$X_S$, rather than the double cover $\wtX_S$,
since the $\Ons(4)$ theory with
$N_{\it eff}=3$ massless fundamentals flows to
{\it one} copy of the free theory
of composite mesons.
It is not the standard sigma model as it is obtained after
the non-standard orbifold operated locally.

\subsection{R\o dland's Example --- $USp(2)$ Versus $USp(4)$}
\label{subsec:Rodland}

The linear sigma model studied in \cite{HoTo} has gauge group
$(U(1)\times USp(2))/\{(\pm,\pm{\bf 1}_2)\}\cong U(2)$
and the following matter fields,
superpotential and twisted superpotential:
\beq
\begin{array}{cc}
\underbrace{p^1~\cdots~p^7}&
\underbrace{x_1~\cdots~x_7}\\
\det^{-1}&{\bf 2}
\end{array}
\eeq
\beq
W=\sum_{i,j,k=1}^7A_k^{ij}p^k[x_ix_j],
\eeq
\beq
\widetilde{W}=-t\, {\rm tr}_{\bf 2}^{}\sigma.
\eeq
$A^{ij}_k=-A^{ji}_k$ are complex numbers which are generic in a suitable sense.
The low energy theory at $r\gg 0$ is the non-linear sigma model
whose target space is the complete intersection of
seven planes in the Grassmannian $G(2,7)$,
\beq
X_A=\left\{\,\, [x]\in G(2,7)\,\,\Biggl|\,\,
\sum_{i,j}A_k^{ij}[x_ix_j]=0~\,\forall k\,\,\right\}.
\eeq
The low energy theory at $r\ll 0$ is the non-linear sigma model
whose target space is the Pfaffian Calabi-Yau manifold,
\beq
Y_A=\left\{\,\,[p]\in\CP^6\,\,\Bigl|\,\,
{\rm rank} A(p)=4\,\,\right\},
\label{defYA}
\eeq
where $A^{ij}(p)=\sum_kA_k^{ij}p^k$. This was demonstrated in
\cite{HoTo} by finding and employing the low energy description of
$USp(2)$ gauge theories
with $1$ or $3$ massless fundamentals.
Both $X_A$ and $Y_A$ have Hodge diamond
\beq
\begin{array}{c}
1\\[-0.2cm]
0~~~~0\\[-0.2cm]
0~~~~1~~~~0\\[-0.2cm]
1~~~50~~~50~~~1\\[-0.2cm]
0~~~~1~~~~0\\[-0.2cm]
0~~~~0\\[-0.2cm]
1
\end{array}
\eeq
but they are topologically and birationally inequivalent.
Finally, the theory is singular at
\beq
\e^t\,=\,(1+\omega)^7,~\,
(1+\omega^2)^7,~\,(1+\omega^3)^7,
\label{singuSp1}
\eeq
where $\omega:=\e^{2\pi i\over 7}$.
There is a single one dimensional Coulomb branch at each point.
Note that (\ref{singuSp1}) differs from \cite{HoTo} by a sign.
This is because \cite{HoTo} missed the possible effect of
integrating out the off-diagonal components of the
vector multiplet discussed in Section~\ref{subsec:Oreg}.
In the present case, the effect is
the $\pi$ shift of the theta angle of the central $U(1)$,
which contributes to the sign change.\footnote{This
effect was also missed in the first version of the present paper.}

As in the model including $O(2)$, we may employ the duality
for the $USp(2)$ part, and consider the dual linear sigma model.
It has gauge group 
$(U(1)\times USp(4))/\{(\pm,\pm{\bf 1}_4)\}$ and the following
matter fields,
superpotential and twisted superpotential:
\beq
\begin{array}{ccc}
\underbrace{p^1~\cdots~p^7}&
\underbrace{\wtx^1~\cdots~\wtx^7}&
\underbrace{(a_{ij})_{1\leq i<j\leq 7}}\\
(-2,{\bf 1})&(-1,{\bf 4})&(2,{\bf 1})
\end{array}
\eeq
\beq
W=\sum_{i,j=1}^7a_{ij}[\wtx^i\wtx^j]+\sum_{i,j,k=1}^7A_k^{ij}p^ka_{ij},
\eeq
\beq
\widetilde{W}=-2\,\wtt\,\sigma_{\mathfrak{u}(1)}.
\eeq
At $\wtr\ll 0$, the gauge group is completely Higgsed.
We have F-term equations
\beq
[\wtx^i\wtx^j]+A^{ij}(p)=0\quad\forall (i,j).
\label{Feq7}
\eeq
The low energy theory is the non-linear sigma model whose target space is
the free quotient
\beq
\left\{\,\,(p,\wtx)\in \C^{\oplus 7}\oplus (\C^4)^{\oplus 7}\,\,\Biggl|
\,\,p\ne 0,\,\,
(\ref{Feq7})\,\,\right\}
\Biggl/{\C^{\times}\times Sp(4,\C)\over
\{(\pm 1,\pm {\bf 1}_4)\}}.
\label{YAA}
\eeq
There is a map to $Y_A$ given by $(p,\wtx)\mapsto p$.
Indeed if $(p,\wtx)$ solves (\ref{Feq7}), then
$A(p)$ has rank $4$. In fact it is an isomorphism.
To see that, let us choose a point
$[p]\in Y_A$ and use the $GL(7,\C)$ coordinate change to make
$A(p)$ into the form
\beq
\left(\begin{array}{ccc}
J_2&&\\
&J_2&\\
&&{\bf 0}_3
\end{array}\right).
\eeq
A solution to (\ref{Feq7}) is given by
\beq
\wtx=\left({\bf 1}_4,{\bf 0}_{4\times 3}\right),
\eeq
and any solution is in its $Sp(4,\C)$ orbit.
Thus, the fibre of the map
at each point of $Y_A$ consists of a unique point.
We have seen that (\ref{YAA}) is another representation
of the Pfaffian variety $Y_A$.
It is obtained in a completely classical manner
in this dual model, while (\ref{defYA}) is obtained by a very
non-trivial analysis of the quantum theory in the original model.

At $\wtr\gg 0$, the D-term equation requires $a\ne 0$ and the gauge group is
broken to $USp(4)$. This leaves us with a $USp(4)$ gauge theory fibred over
$\CP^{20}=\{a\}$. The 7 quartets have mass matrix $a_{ij}$ and the nature
of the theory depends on its rank. 
Analysis of such a system has been carried out in 
Section~\ref{subsec:dualitySp}, which can be applied here without 
modification: By the supersymmetry breaking for the $USp(4)$ theory with
$N_{\it eff}\leq 3$ and by the low energy description of the theory
with $N_{\it eff}=5$ in terms of composite mesons,
we find that the theory reduces at low energies to the
sigma model on the locus of $\CP^{20}$ where
$a_{ij}$ has rank $2$. (Note that $a$ cannot have
rank $0$ as that would violate the D-term equation for the $U(1)$.)
We also have the F-term constraints,
\beq
\sum_{i,j}A_k^{ij}a_{ij}=0,\qquad k=1,\ldots,7.
\label{Feq7b}
\eeq
The low energy theory is a simple non-linear sigma model
with the target space
\beq
\Bigl\{\,\,[a]\in\CP^{20}\,\,\Bigl|\,\,
{\rm rank} \, a=2,\,\,\,(\ref{Feq7b})\,\,\Bigr\}.
\label{XAan}
\eeq
This is isomorphic to $X_A$ under the correspondence
$a_{ij}\propto [x_ix_j]$.
We have reproduced the low energy theory of the original
linear sigma model at $r\gg 0$.

Finally, let us identify the singular points and find the relation
to the FI-theta parameters of the original model.
Identification and parametrization of the maximal torus can be done
in almost the same way as in the model including $SO(4)$.
One difference is that the Weyl group is slightly bigger,
$(h_1,h_2)\mapsto (h_2,h_1)$, $(h_1^{-1},h_2)$, $(h_1,h_2^{-1})$,
and there are more forbidden loci: 
$\sigma_0=2\sigma_1$ and
$\sigma_0=2\sigma_2$ in addition to
$\sigma_1=\sigma_2$ and $\sigma_0=\sigma_1+\sigma_2$.
The equation determining the Coulomb branch is
\beq
{(-\sigma_0)^7(-\sigma_0+\sigma_1)^7(-\sigma_0+\sigma_2)^7
\over (\sigma_0)^{21}}=-\e^{\wtt},\quad
{(-\sigma_1)^7\over (-\sigma_0+\sigma_1)^7}
={(-\sigma_2)^7\over (-\sigma_0+\sigma_2)^7}
=1.
\eeq
The sign in $-\e^{\wtt}$ comes from the effect of integrating out the
off diagonal components of the vector multiplet.
We find $\sigma_a={\sigma_0\over 1+\omega_a}$, with
$\omega_a^7=1$, for $a=1,2$, where
we need to avoid
$\omega_1=1$ and $\omega_2=1$ in addition to $\omega_1=\omega_2^{\pm 1}$.
The Weyl group action becomes
$(\omega_1,\omega_2)\mapsto (\omega_2,\omega_1)$,
$(\omega_1^{-1},\omega_2)$, $(\omega_1,\omega_2^{-1})$. There are
three inequivalent possibilities
$(\omega_1,\omega_2)=(\omega,\omega^2)$, $(\omega,\omega^3)$
and $(\omega^2,\omega^3)$ (again, $\omega:=\e^{2\pi i\over 7}$), 
for which
$\e^{\wtt}=(1+\omega^3)^7$, $(1+\omega^2)^7$ and 
$(1+\omega)^7$ respectively.
The Weyl group is completely broken and there is a single Coulomb branch
at each of these points.
Comparing with (\ref{singuSp1}), we may set
\beq
\wtr=r,\quad\wttheta=\theta.
\eeq

To summarize, we obtained completely consistent results from the
dual pair of linear sigma models. The two play complementary r\^oles.
If the gauge symmetry is unbroken and a non-trivial quantum analysis is needed
in one theory, the gauge group is Higgsed and the result is obtained by 
purely classical analysis in the dual. This happens both at $r\gg 0$
and $r\ll 0$.

\subsection{Intersection Of Quadrics}
\label{subsec:IQ}

Let $S_1(x),\ldots,S_M(x)$ be quadratic polynomials of $N$ variables
$x=(x_1,\ldots, x_N)$. We denote by $Q_S$ the intersection
of $M$ quadrics in $\CP^{N-1}$
\beq
S_1(x)=\cdots=S_M(x)=0.
\label{iq}
\eeq
We assume that $Q_S$ is a smooth submanifold of dimension $N-1-M$.

A linear sigma model for $Q_S$ is the
$U(1)$ gauge theory with fields $p^1,\ldots, p^M$ of charge $-2$
and fields $x_1,\ldots,x_N$ of charge $1$, with the
superpotential
\beq
W=p^1S_1(x)+\cdots p^MS_M(x).
\label{Wiq}
\eeq
For large positive values of the FI parameter, $r\gg 0$, 
the D-term equation forces $x$ to have non-zero values and the 
gauge group is completely broken. The 
theory reduces at low energies
to the non-linear sigma model whose target space is $Q_S$.
For $r\ll 0$, on the other hand, $p=(p^1,\ldots,p^M)$ must 
have non-zero values,
and  the $U(1)$ gauge group is broken to the $\Z_2$ subgroup.
The low energy theory is the so called hybrid model.
It is a Landau-Ginzburg model on the $\Z_2$ orbibundle 
${\mathcal O}(-{1\over 2})^{\oplus N}$ over $\CP^{M-1}$.
The equation for Coulomb branch vacua is
\beq
(-2\sigma)^{2M}/\sigma^N=\e^t.
\eeq
When $N=2M$, where $Q_S$ is a Calabi-Yau manifold, 
the axial $U(1)$ R symmetry is anomaly free and
$t$ is a parameter of the theory.
There is a singular point at $\e^t=2^N$.
When $N>2M$, the theory is a flow from the sigma model on $Q_S$
to the $r\ll 0$ hybrid model or one of the 
$(N-2M)$ Coulomb branch vacua.
When $N<2M$, the theory is a flow from the $r\ll 0$ hybrid model
to the sigma model on $Q_S$ or one of the $(2M-N)$ Coulomb branch vacua.

Our main interest is the nature of the hybrid model at $r\ll 0$.
(See \cite{CDHPS} for an earlier study.)
We may rewrite the superpotential (\ref{Wiq}) as
\beq
W=\sum_{i,j=1}^NS^{ij}(p)x_ix_j,
\eeq
where $S^{ij}(p)=\sum_kS_k^{ij}p^k$ for $S_k(x)=\sum_{i,j}S_k^{ij}x_ix_j$.
The model can be regarded as the $\Z_2$ Landau-Ginzburg orbifold
of $x$ with this quadratic superpotential fibred over
${\bf P}:=\CP^{M-1}$. 
We denote by ${\bf P}_{(i)}$ the locus of $p\in {\bf P}$
 where $S(p)$ has rank $N-i$.
It has codimension ${i(i+1)\over 2}$.
Over the generic locus ${\bf P}_{(0)}$, the fields $x_i$ are all massive
and can be integrated out.
The $\Z_2$ orbifold is the standard one, 
so that the number of zero energy states
in the $x$ sector is two {\it resp}. one if $N$ is even {\it resp}. odd
(Section~\ref{subsec:massive}).
Thus, we have a double {\it resp}. single cover over ${\bf P}_{(0)}$.
Near the first degeneration locus ${\bf P}_{(1)}$ (codimension $1$),
the result of Section~\ref{subsec:rr1} can be applied:
The double cover for the $N$ even case is branched along ${\bf P}_{(1)}$,
while the cover for the odd $N$ case is of the form of the orbifold
$\C/\Z_2$ in the transverse direction to ${\bf P}_{(1)}$.
Near the second degeneration locus ${\bf P}_{(2)}$ (codimension $3$),
the result of Section~\ref{subsec:rr2} and \ref{subsec:O2dual} 
can be applied:
In the transverse direction with coordinate
$(a,b,c)$, the double cover for the $N$ even case is the conifold $c^2-ab=d^2$
with $\theta=\pi$ where the $\Z_2$ deck transformation is $d\to \pm d$,
while the cover for the $N$ odd case is the $\Z_2(-1)^{F_s}$ orbifold thereof.
We would like to see how such local behaviour may be glued together
and find a global picture. 
For this purpose we turn to the dual model.

A key to find the dual is to rewrite the gauge group as
\beq
U(1)={U(1)\times O(1)\over\{(\pm 1,\pm 1)\}},
\eeq
and apply the duality to the $O(1)$ sector that appears in the
$r\ll 0$ hybrid model. Since we have the standard $\Z_2$ orbifold
with $N$ fields that transform by sign flip, the $O(1)$
is $\Ost(1)$ when $N$ is even while it is $\Ons(1)$ when $N$ is odd
(see {\sl Special Cases} in Section~\ref{subsec:O2dual}).
For even {\it resp}. odd $N$, the dual model has gauge group
\beq
{U(1)\times SO(N)\over \{(\pm 1,\pm {\bf 1}_N)\}}\quad
{\it resp}.\quad
{U(1)\times \Ons(N)\over \{(\pm 1,\pm {\bf 1}_N)\}},
\eeq
the matter fields
\beq
\begin{array}{ccc}
\underbrace{~p^1~\cdots~p^M}&
\underbrace{~\wtx^1~\cdots~\wtx^N}&
\underbrace{(s_{ij})_{1\leq i\leq j\leq N}}\\
(-2,{\bf 1})&(-1,{\bf N})&(2,{\bf 1})
\end{array}
\eeq
and the superpotential
\beq
W=\sum_{j,j=1}^Ns_{ij}(\wtx^i\wtx^j)+\sum_{i,j=1}^NS^{ij}(p)s_{ij}.
\eeq
The mod 2 theta angle for the $SO(N)$ factor is turned on (as
$N-k=0$ (even)).
We write $\wtt_{\mathfrak{u}(1)}=\wtr_{\mathfrak{u}(1)}-i\wthu$ 
for the FI-theta parameter 
for the $U(1)$ factor.

Let us be more precise about the theta angle.
There is no subtlety for odd $N$ since the gauge group is
simply isomorphic to $U(1)\times SO(N)$,
as the element $(-1,-{\bf 1}_N)$ identifies the two connected components.
In particular, the theta parameter has the standard periodicity
$\wthu\equiv \wthu+2\pi$.
For even $N$, the fundamental group of the gauge group is
isomorphic to $\Z\oplus \Z_2$ but not canonically so.
For example, a loop associated to $(n,0)\in \Z\oplus \Z_2$
may be chosen as $t\in\R/2\pi\Z\mapsto g_t=[(\e^{int\over 2},h_t)]$ where
$h_t$ is represented by $\widetilde{h}_t=\exp\left({nt\over 4}(e_1e_2+\cdots
+e_{N-1}e_N)\right)])$ in $Spin(N)$, 
but we could equally well choose the one
where the sign of $e_ae_{a+1}$  in the exponent is flipped.
To be specific, let us define $\wthu$ so that
the path-integral weight is given the phase 
$\e^{{in\over 2}\wthu}$ 
for the gauge bundle defined by this particular transition function $g_t$.
It has the extended periodicity
$\wthu\equiv \wthu+4\pi$.
Recall that $2\wttheta$ in (\ref{tWdLSM}) also has the extended periodicity
for the same reason.
However, unlike in that case, the theories with $\wthu$
and $\wthu+2\pi$ are equivalent
--- the symmetry $\tau\in O(N)/SO(N)$ makes the shift
\beq
\wthu~\longrightarrow~
\wthu+2\pi.
\label{theshift}
\eeq
To see this, note that conjugation by $\tau$ changes $h_t$ by multiplication of
a non-contractible loop in $SO(N)$.
For example, if $\tau$ is represented by ${\rm diag}(-1,1,\ldots,1)$,
then
$\tau \widetilde{h}_t\tau^{-1}=
\exp\left({nt\over 4}(-e_1e_2+\cdots+e_{N-1}e_N)\right)
=\exp\left(-{nt\over 2}e_1e_2\right) \widetilde{h}_t$.
Since we have a non-trivial mod 2 theta angle for $SO(N)$,
the path-integral weight changes by $(-1)^n$. This change is nothing but
the shift (\ref{theshift}).
In the model of (\ref{tWdLSM}), the symmetry $\tau\in O(4)/SO(4)$ exists
but does not shift 
$\wthu=2\wttheta$ since the mod 2 theta angle for
 $SO(4)$ is turned off.
Note that the shift (\ref{theshift}) 
is related to the ambiguity in the choice of isomorphism
of the fundamental group to $\Z\oplus \Z_2$.

Let us now analyze the theory at $\wtr_{\mathfrak{u}(1)}\ll 0$.
The D- and F-term equations require $p$ to have non-zero values, 
and the gauge group is broken to $SO(N)$ (even $N$)
or $\Ons(N)$ (odd $N$).
Integrating out the fields $s_{ij}$ we obtain the constraints
\beq
(\wtx^i\wtx^j)+S^{ij}(p)=0\quad\forall (i,j).
\label{eqnq}
\eeq
Suppose $S(p)$ has rank at least $N-1$ for all $p\ne 0$, i.e.,
${\bf P}={\bf P}_{(0)}\cup{\bf P}_{(1)}$,
which would be the case if $\dim{\bf P}\leq 2$.
Then, $\wtx$ has rank at least $N-1$ for every solution to (\ref{eqnq}).
For even $N$ the residual gauge group $SO(N)$ is completely broken
at any solution to (\ref{eqnq}).
Therefore, the low energy theory is the sigma model whose target space
is the free quotient
\beq
\widetilde{\bf P}_S=\Bigl\{(p,\wtx)\in \C^M\oplus (\C^N)^{\oplus N}\,
\Bigl|\,p\ne 0,~(\ref{eqnq})\,\Bigl\}\Bigl/{\C^{\times}\times SO(N,\C)
\over \{(\pm 1,\pm {\bf 1}_N)\}}.
\label{qwtPS}
\eeq
The map $(p,\wtx)\in \widetilde{\bf P}_S\mapsto p\in {\bf P}$ is a double 
cover that is branched along the degeneration locus ${\bf P}_{(1)}$.
This can be seen as in the argument to show that (\ref{forget}) is a
 ramified double cover.
Indeed, if $S(p)=-{\rm diag}(z,1,\ldots,1)$, the solution to
(\ref{eqnq}) is given by
$\wtx={\rm diag}(\wtz,1,\ldots, 1)$ with $z=\wtz^2$.
For non-zero $z$, the two solutions with opposite signs of $\wtz$
are distinct.
$\widetilde{\bf P}_S$ provides an explicit global realization
of the branched double cover that is expected in the local analysis
of the original linear sigma model.
For odd $N$, the residual gauge group $\Ons(N)$ is completely
broken at maximal rank solutions over ${\bf P}_{(0)}$
but a $\Z_2$ subgroup remains unbroken at corank $1$ solutions
over ${\bf P}_{(1)}$.
The low energy theory is the sigma model on an orbifold
\beq
{\bf P}_S=\Bigl\{(p,\wtx)\in \C^M\oplus (\C^N)^{\oplus N}\,
\Bigl|\,p\ne 0,~(\ref{eqnq})\,\Bigl\}\Bigl/{\C^{\times}\times \Ons(N,\C)
\over \{(\pm 1,\pm {\bf 1}_N)\}}.
\label{qPS}
\eeq
For $S(p)=-{\rm diag}(z,1,\ldots,1)$ and
$\wtx={\rm diag}(\wtz,1,\ldots, 1)$ with $z=\wtz^2$,
the relevant unbroken gauge group is
$\Ons(1)$ that acts on a single variable $\wtz$ as $\wtz\to -\wtz$. 
Therefore, the orbifold is the standard one $\C/\Z_2$
in the direction transverse to ${\bf P}_{(1)}$.
Again, ${\bf P}_S$ provides an explicit global realization
of the orbifold that is expected in the local analysis of the
 original linear sigma model.
In a general case, especially when ${\bf P}$ has dimension three or higher,
${\bf P}_{(i)}$ with $i\geq 2$ are non-empty.
At solutions to (\ref{eqnq}) over such higher degeneration locus,
 continuous subgroups of the
gauge group remain unbroken.
In particular, the quotients (\ref{qwtPS}) and (\ref{qPS}) 
are not smooth manifolds nor orbifolds,
and we no longer have a sigma model description of the
low energy theory. 
But we do see what we have locally in the direction transverse to
${\bf P}_{(i)}$: For even {\it resp}. odd $N$ it is the 
$SO(i)$ {\it resp}. $\Ons(i)$ gauge theory
with $i$ massless fundamentals, 
which flows to a superconformal field theory with central charge
$\whc={i(i+1)\over 2}$.
The case $i=2$ is indeed (orbifold of) the conifold at $\theta=\pi$.
It would be interesting to understand
the total system better, using the original and the dual models.

The model at $\wtr_{\mathfrak{u}(1)}\gg 0$ can be analyzed as follows.
The D-term equation requires $s\ne 0$ and the gauge group is broken to
$SO(N)$ (even $N$) or $\Ons(N)$ (odd $N$).
By the supersymmetry breaking for the theory with $N_{\it eff}\leq N-2$
and the low energy description of the theory with $N_{\it eff}= N-1$
in terms of the composite mesons, we find that the theory reduces
to the sigma model on the locus of $s\in {\bf CP}^{N(N+1)\over 2}$
of corank $N-1$, i.e., rank $1$, so that one may write
\beq
s_{ij}=x_ix_j.
\eeq 
We also have the F-term constraints
\beq
\sum_{ij}S^{ij}_ks_{ij}=0,\quad k=1,\ldots,M.
\eeq
Namely, we have the sigma model whose target space is
the complete intersection of the quadrics (\ref{iq}).

Finally, let us analyze the Coulomb branch vacua.

\subsubsection*{\sl Even $N$}

We use the following parametrization of the maximal torus
of the gauge group;
\beqa
{U(1)\times SO(2)_1\times\cdots\times SO(2)_{N\over 2}\over
\{(\pm 1,\pm {\bf 1}_2,\ldots,\pm {\bf 1}_2)\}}
&\cong&U(1)_0\times U(1)_1\times\cdots \times U(1)_{N\over 2}
\label{maxT}\\
(z,h_1,\ldots, h_{N\over 2})~~~~~~~~&\longmapsto&~~~~~~~~
(z^2,zh_1,\ldots, zh_{N\over 2}).
\nn
\eeqa
For the theta parameter $\wthu$ (period $4\pi$)
as defined in the paragraph including 
(\ref{theshift}), the tree level twisted superpotential
on the Coulomb branch is
\beq
\widetilde{W}_{\it tree}=
-\wtt_{\mathfrak{u}(1)}
\sigma_{\mathfrak{u}(1)}+\pi i \sum_{a=1}^{N\over 2}
(\sigma_{\mathfrak{so}(2)_a}-\sigma_{\mathfrak{u}(1)})
=-t_0\sigma_0+\pi i \sum_{a=1}^{N\over 2}\sigma_a.
\label{Wtreee}
\eeq
Here we write $\sigma_0=2\sigma_{\mathfrak{u}(1)}$
and $\sigma_a=\sigma_{\mathfrak{u}(1)}+\sigma_{\mathfrak{so}(2)_a}$
following (\ref{maxT}) and we put
$t_0:={1\over 2}\wtt_{\mathfrak{u}(1)}+{N\over 2}\pi i$ (period $2\pi i$). 
The $\pi i$ terms in (\ref{Wtreee})
come from the mod $2$ theta angle for the $SO(N)$ gauge group.
The effective twisted superpotential is
\beqa
\widetilde{W}_{\it eff}\!\!&=&\!\!
-{{N\over 2}({N\over 2}-1)\over 2}\pi i \sigma_0
-M(-\sigma_0)(\log(-\sigma_0)-1)\nn\\
&&-N\sum_{a=1}^{N\over 2}\Bigl\{
(-\sigma_0+\sigma_a)(\log(-\sigma_0+\sigma_a)-1)
+(-\sigma_a)(\log(-\sigma_a)-1)\Bigr\}\nn\\
&&-{N(N+1)\over 2}\sigma_0(\log\sigma_0-1)
-t_0\sigma_0+\pi i \sum_{a=1}^{N\over 2}\sigma_a,
\eeqa
where the first term results from integrating out the off diagonal components
of the vector multiplet. The vacuum equations read
\beq
{(-\sigma_0)^M\prod_{a=1}^{N\over 2}(-\sigma_0+\sigma_a)^N
\over \sigma_0^{N(N+1)\over 2}}=(-1)^{{N\over 2}({N\over 2}-1)\over 2}
\e^{t_0},\qquad
{\sigma_a^N\over (\sigma_0-\sigma_a)^N}=-1\quad\forall a.
\eeq
Writing $\sigma_a=\sigma_0\left({1\over 2}+u_a\right)$ 
(i.e., $u_a:=\sigma_{\mathfrak{so}(2)_a}/\sigma_0$), we see that
each $u_a$ must solve
\beq
\left({1\over 2}+u\right)^N+\left({1\over 2}-u\right)^N=0
\label{eqnfu}
\eeq
and $\sigma_0$ is then determined by
\beq
\sigma_0^{M-{N\over 2}}(-1)^M\prod_{a=1}^{N\over 2}
\left({1\over 2}-u_a\right)^N=(-1)^{{N\over 2}({N\over 2}-1)\over 2}\e^{t_0}.
\label{eqnfs0}
\eeq
The equation (\ref{eqnfu}) has ${N\over 2}$ pairs of non-zero 
roots, and we must find solutions such that
$u_a\ne \pm u_b$ ($a\ne b$) modulo the $SO(N)$ Weyl group action ---
permutations and sign flips of $u_a$'s preserving
the product $u_1\cdots u_{N\over 2}$.
There are two inequivalent solutions.
When $N\ne 2M$, the equation (\ref{eqnfs0})
for given $u_a$'s has $\left|M-{N\over 2}\right|$
solutions for $\sigma_0$. Thus, there are total of 
$|2M-N|$ solutions, matching with the result in the original
linear sigma model.
When $N=2M$, the equation (\ref{eqnfs0}) should be regarded as
the one that determines the location of singular points in the 
parameter space. Corresponding to the two solutions for $u_a$'s,
we have {\it two} singular points, related by
$\e^{t_0}\to -\e^{t_0}$, i.e., $t_0\to t_0+\pi i$.
Thus our one parameter family of theories cannot be the same as
the one parameter family from the original model which has one
singular point ($\e^t=2^N$). At this point, we recall that
we have a symmetry $\tau\in O(N)/SO(N)$ which shifts $\wtt_{\mathfrak{u}(1)}$
by $2\pi i$ (\ref{theshift}), that is,
$t_0\to t_0+\pi i$. Thus, our dual family can be regarded as a double cover
of the original family.
A precise covering map is given by
\beq
\e^t=2^N\prod_{a=1}^{N\over 2}\left(1+\e^{\pi i(2a-1)\over N}\right)^{2N}
\cdot\e^{2t_0}.
\eeq
This also relates $t$ and $\wtt_{\mathfrak{u}(1)}$ via
$\e^{2t_0}=\e^{\wtt_{\mathfrak{u}(1)}}$.

\subsubsection*{\sl Odd $N$}

This case is more straightforward as the gauge group
is simply isomorphic to $U(1)\times SO(N)$. 
We denote the scalar components of the vector multuplet by
 $\sigma$ for the $U(1)$ part and by $\sigma_1,\ldots,\sigma_{N-1\over 2}$
for the maximal torus of $SO(N)$.
The tree level twisted superpotential is
\beq
\widetilde{W}_{\it tree}=
-\wtt_{\mathfrak{u}(1)}\sigma+\pi i (\sigma_1+\cdots+\sigma_{N-1\over 2}).
\eeq
Again, the $\pi i$ terms come from the mod $2$ theta angle of the $SO(N)$
gauge group.
Computation of the effective twisted superpotential is straightforward and 
the vacuum equation reads
\beq
{(-2\sigma)^{2M}(-\sigma)^N
\prod_{a=1}^{N-1\over 2}(-\sigma+\sigma_a)^N(-\sigma-\sigma_a)^N\over
(2\sigma)^{N(N+1)}}=\e^{\wtt_{\mathfrak{u}(1)}},\qquad
{(-\sigma-\sigma_a)^N\over (-\sigma+\sigma_a)^N}=1\quad\forall a.
\eeq
We see that each $z_a=\sigma_a/\sigma$ must solve the equation
\beq
(1+z)^N-(1-z)^N=0,
\label{eqfz}
\eeq
and $\sigma$ is then determined by
\beq
(-\sigma)^{2M-N}2^{2M-N(N+1)}\prod_{a=1}^{N-1\over 2}(1-z_a^2)^N
=\e^{\wtt_{\mathfrak{u}(1)}}.
\label{eqfsa}
\eeq
The equation (\ref{eqfz}) has one root at $z=0$ and ${N-1\over 2}$ pairs
of non-zero roots. We look for solutions such that
$z_a\ne \pm z_b$ ($a\ne b$) and $z_a\ne 0$ 
modulo the $SO(N)$ Weyl group action ---
permutations and independent sign flips of $z_a$'s.
There is a unique solution. Thus, we find $|2M-N|$ Coulomb branch vacua,
matching with the result in the original linear sigma model.

\subsection{Equivalences Of D-Brane Categories}

As mentioned earlier, the present work is motivated by recent development
in mathematics concerning equivalences of derived categories of
certain pairs of algebraic varieties.
Such categories are realized as the categories of B-branes
in the supersymmetric non-linear sigma models.
If two varieties $X$ and $Y$ sits on a common quantum K\"ahler moduli space,
it is expected from the general principle of
$(2,2)$ supersymmetry that $X$ and $Y$ have equivalent derived categories.
Our task was to promote the equivalences found in mathematics to
statements in quantum field theories.
Here we summarize the relevant equivalences, give some references,
and make some comments.

The study of Section~\ref{subsec:LSM} and \ref{subsec:DLSM}
is directly related to the work \cite{HoTa} by Hosono and Takagi.
The relevant equivalence is
\beq
D^b(X_S)\,\cong\, D^b(\wtY_S).
\eeq
The proof is being done by the authors of \cite{HoTa}.

From the linear sigma model
with gauge group $(U(1)\times SO(2))/\{(\pm 1,\pm {\bf 1}_2)\}$
studied in Section~\ref{subsec:SOOns}, we have equivalences
\beq
\begin{array}{ccccc}
&&\!\!\!\!\!D^b(X_{\rm II})\!\!\!\!\!&&\\[-0.1cm]
{}^{\cong}\!\!\!\!\!\!\!\!\!\!\!\!\!\!\!\!\!\!\!\!
&\!\!\!\!\!\!\!\!\!\!\mbox{\large $\nearrow$}\!\!\!\!\!\!\!\!\!\!
&&\!\!\!\!\!\!\!\!\!\!\mbox{\large $\searrow$}\!\!\!\!\!\!\!\!\!\!
&\!\!\!\!\!\!\!\!\!\!\!\!\!\!\!\!\!\!\!\!{}^{\cong}\\[-0.1cm]
D^b(\wtX_S)&&&&D^b_{\Z_2}(\wtY_S)\\[-0.1cm]
{}_{\cong}\!\!\!\!\!\!\!\!\!\!\!\!\!\!\!\!\!\!\!\!
&\!\!\!\!\!\!\!\!\!\!\mbox{\large $\searrow$}\!\!\!\!\!\!\!\!\!\!
&&\!\!\!\!\!\!\!\!\!\!\mbox{\large $\nearrow$}\!\!\!\!\!\!\!\!\!\!
&\!\!\!\!\!\!\!\!\!\!\!\!\!\!\!\!\!\!\!\!{}_{\cong}\\[-0.1cm]
&&\!\!\!\!\!D^b(X_{\rm III})\!\!\!\!\!&&
\end{array}
\eeq
The two arrows on the left (i.e. those not involving
the orbifold category $D^b_{\Z_2}(\wtY_S)$)
as well as the unwritten vertical arrow in the middle
are already mentioned in \cite{HoTa}.
They can be promoted to the relations
between ${\mathcal N}=2$ theories with boundaries \cite{HHP}.

From the linear sigma model
with gauge group $(U(1)\times \Ons(2))/\{(\pm 1,\pm {\bf 1}_2)\}$
studied also in Section~\ref{subsec:SOOns}, we have
\beq
D^b_{\Z_2(-1)^{F_s}}(\wtX_S)\cong D^b_{\Z_2(-1)^{F_s}}(\wtY_S).
\eeq
We invented a notation $D^b_{\Z_2(-1)^{F_s}}(-)$ for the
category of the non-standard $\Z_2$ orbifold,
not knowing the well accepted notation in mathematics.

The equivalence relevant for
Section~\ref{subsec:Rodland}
is
\beq
D^b(X_A)\cong D^b(Y_A).
\eeq
This was first pointed out by E. Witten
as a consequence of the work \cite{HoTo}.
Proofs are given by Borisov-Caldararu \cite{BC}
and Kuznetsov \cite{KPf}.

The study of Section~\ref{subsec:IQ} is related to the following
equivalences, found by Bondal-Orlov
\cite{BOso,BOrev,KiQ}:
\beq\begin{array}{cl}
D^b(Q_S)~\cong~\left\langle {\mathcal C}_S,{\mathcal O}_1,\ldots,
{\mathcal O}_{N-2M}\right\rangle
&\mbox{if $N\geq 2M$},
\\[0.2cm]
\left\langle {\mathcal B}_{N-2M},\ldots,
{\mathcal B}_{-1},D^b(Q_S)\right\rangle~\cong~ {\mathcal C}_S
&\mbox{if $N\leq 2M$,}
\end{array}
\label{sod}
\eeq
where ${\mathcal C}_S$ is a category that corresponds to
the linear sigma model at $\sigma_{\mathfrak{u}(1)}=0$ for $r\ll 0$.
For $M\leq 3$,
${\mathcal C}_S=D^b(\widetilde{\bf P}_S)$ for even $N$
and ${\mathcal C}_S=D^b({\bf P}_S)$ for odd $N$.
${\mathcal O}_i$ and ${\mathcal B}_{-j}$ are ``exceptional objetcs'' of 
the category on the other side of the equivalence.
$\langle -,-,\ldots,-\rangle$ stands for
 ``semi-orthogonal decomposition''.
In fact, this equivalence for odd $N$ was the motivation to refine
our understanding of $\Z_2$ orbifolds at an earlier stage, and the refinement
resulted in finding the $\Ons$ duality
(the earlier understanding gave us only the
$SO$/$\Ost$ duality).
Linear sigma models relevant for the equivalence for the case $N=2M$
were studied earlier in \cite{CDHPS}.

A point of view that seems to
underlie all of these equivalences is {\it projective duality} and
its categorical counterpart proposed by A.~Kuznetsov, 
called Homological Projective Duality \cite{HPD}.
Indeed, comparison of the presentation (\ref{defYS}) of $Y_S$
and (\ref{spsp}) of $X_S$ suggests projective duality.
Note that these two presentations have been found
in two linear sigma models which are dual to each other.
The same applies to (\ref{defYA}) of $Y_A$ and (\ref{XAan}) of $X_A$.
It would be interesting to see if there is
a relation between the gauge theory duality and projective duality
at a more fundamental level.

\section*{Acknowledgement}

We would like to thank Alexey Bondal, Mike Douglas,
Simeon Hellerman, Shinobu Hosono, Minxin Huang,
Johanna Knapp, Alexander Kuznetsov,
Todor Milanov, Dave Morrison, Yongbin Ruan and David Tong for useful
discussions and instructions.

This work is supported by
JSPS Grant-in-Aid for Scientific Research No. 21340109
and WPI Initiative, MEXT, Japan at IPMU, the University of Tokyo.

\appendix{Supersymmetric Quantum Mechanics With Both
Superpotential And Twisted Masses}
\label{app:SUSYQM}

We study the spectrum of supersymmetric ground states of the theory of
two variables $z$ and $x$
having the superpotential
\beq
W=zx^2,
\label{Wapp}
\eeq
and twisted masses associated with the symmetry
\beq
x\longmapsto \lambda^{-1} x,\quad
z\longmapsto \lambda^2 z.
\label{prot}
\eeq
That is, $x$ and $z$ have twisted masses $-\wtm$ and $2\wtm$ respectively.

For the purpose of finding the ground states, we may only consider
the zero mode sector.
The supercharges are given by
\beqa
\oQ_+&=&\bpsi^x_+{\partial\over \partial\bx}
+\bpsi^z_+{\partial\over\partial\bz}
-\left(2zx\psi^x_-+x^2\psi^z_--\wtm x\bpsi^x_-+2\wtm z\bpsi^z_-\right)
\\
\oQ_-&=&\bpsi^x_-{\partial\over \partial\bx}
+\bpsi^z_-{\partial\over\partial\bz}
+\left(2zx\psi^x_++x^2\psi^z_++\overline{\wtm} x\bpsi^x_+
-2\overline{\wtm} z\bpsi^z_+\right)
\\
Q_+&=&-\psi^x_+{\partial\over \partial x}
-\psi^z_+{\partial\over\partial z}
-\left(2\bz\bx\bpsi^x_-+\bx^2\bpsi^z_--\overline{\wtm} \bx\psi^x_-
+2\overline{\wtm} \bz\psi^z_-\right)
\\
Q_-&=&-\psi^x_-{\partial\over \partial x}
-\psi^z_-{\partial\over\partial z}
+\left(2\bz\bx\bpsi^x_++\bx^2\bpsi^z_++\wtm \bx\psi^x_+
-2\wtm \bz\psi^z_+\right).
\eeqa
Non-zero anticommutators are
\beqa
&\{\oQ_+,Q_+\}=\{\oQ_-,Q_-\}=H,\\
&\{\oQ_+,Q_-\}=\wtm J,\quad \{\oQ_-,Q_+\}=\overline{\wtm} J,
\eeqa
where $H$ and $J$ are the Hamiltonian and the generator of (\ref{prot})
respectively,
\beqa
H\!\!&=&\!\!-{\partial^2\over \partial x\partial \bx}
-{\partial^2\over \partial z\partial \bz}
+|2zx|^2+|x^2|^2+|\wtm x|^2+|2\wtm z|^2\\
&&\!\!
+\left[\Bigl(2z\psi^x_+\psi^x_-+2x\psi^x_+\psi^z_-+2x\psi^z_+\psi^x_-
\Bigr)+
\mbox{\it h.c.}\,\right]
+\left[\Bigl(-\wtm\psi^x_+\bpsi^x_-+2\wtm\psi^z_+\bpsi^z_-\Bigr)+
\mbox{\it h.c.}\,\right],\nn
\\
J\!\!&=&\!\!-\left(x{\partial\over \partial x}-\bx{\partial \over\partial \bx}
-\bpsi^x_+\psi^x_++\psi^x_-\bpsi^x_-\right)
+2\left(z{\partial\over \partial z}-\bz{\partial \over\partial \bz}
-\bpsi^z_+\psi^z_++\psi^z_-\bpsi^z_-\right).
\eeqa
The space of states can be chosen to be the space
\beq
{\mathcal H}=\bigoplus_{p,q=0}^2\,\Omega^{0,p}(\C^2,\wedge^q T_{\C^2})
\label{defH}
\eeq 
of differential forms on $\C^2=\{(x,z)\}$ with values in polyvector fields.
The fermions are represented on it as
\beqa
&\displaystyle \bpsi^x_++\bpsi^x_-=\dd \bx\wedge,\quad \psi^x_++\psi^x_-
=\imath\left({\partial\over\partial \bx}\right),\nn\\
&\displaystyle \psi^x_+-\psi^x_-=\imath\left(\dd x\right),\quad 
\bpsi^x_+-\bpsi^x_-={\partial\over \partial x}\wedge,
\nn
\eeqa
and similarly for $\psi^z_{\pm}$ and $\bpsi^z_{\pm}$.

If we choose $\wtm$ to be pure imaginary, $\wtm+\overline{\wtm}=0$, 
the operator
$Q=\oQ_++\oQ_-$ obey the relation,
\beq
Q^2=0,\quad\{Q,Q^{\dag}\}=2H.
\eeq
In particular, there is a one to one correspondence between
supersymmetric ground states and $Q$-cohomology classes.
Under the same condition, $\wtm+\overline{\wtm}=0$, 
$Q$ is represented by the operator
$Q=\overline{\partial}+Q_{\it hol}$ on the space 
${\mathcal H}$ in (\ref{defH}), where
\beq
Q_{\it hol}=\imath(\dd W)-\wtm K\wedge,
\eeq
with $dW=2zx\dd x+x^2\dd z$ and $K=-x{\partial\over\partial x}
+2z{\partial\over\partial z}$.
Using the standard argument, one can show that $Q$-cohomology classes are
in one to one correspondence with the $Q_{\it hol}$-cohomology classes
where $Q_{\it hol}$ is regarded as the differential on
the space of holomorphic polyvector fields,
\beq
{\mathcal H}_{\it hol}=\bigoplus_{q=0}^2
\Gamma_{\it hol}(\C^2,\wedge^q T_{\C^2}).
\eeq
Note that
\beqa
&\displaystyle 
Q_{\it hol}(1)
=-\wtm \left(-x{\partial\over\partial x}+2z{\partial\over \partial z}
\right),\nn\\
&\displaystyle 
Q_{\it hol}
\left({\partial\over\partial x}\right)=2zx+2\wtm z{\partial\over\partial x}
\wedge{\partial\over\partial z},\qquad
Q_{\it hol}
\left({\partial\over\partial z}\right)=x^2+\wtm x{\partial\over\partial x}
\wedge{\partial\over\partial z},\nn\\
&\displaystyle 
Q_{\it hol}
\left({\partial\over\partial x}\wedge{\partial\over\partial z}\right)
=x\left(-x{\partial\over\partial x}+2z{\partial\over\partial z}\right).\nn
\eeqa
It is easy to see that there is only one cohomology class,
which is represented by
\beq
x+\wtm {\partial\over\partial x}\wedge{\partial\over\partial z}.
\eeq
This proves that the system has a unique supersymmetric ground state.

\end{document}